
\input fontch.tex

%
%
%
\def\unredoffs{} \def\redoffs{\voffset=-.31truein\hoffset=-.48truein}
\def\speclscape{}
%
%
%
%
%
\newbox\leftpage \newdimen\fullhsize \newdimen\hstitle \newdimen\hsbody
\tolerance=1000\hfuzz=2pt
\catcode`\@=11 
\ifx\hyperdef\UNd@FiNeD\def\hyperdef#1#2#3#4{#4}\def\hyperref#1#2#3#4{#4}\fi
\def\bigans{b }
\def\answ{b }
%
\ifx\answ\bigans\message{(This will come out unreduced.}
\magnification=1200\unredoffs\baselineskip=16pt plus 2pt minus 1pt
\hsbody=\hsize \hstitle=\hsize 
\else\message{(This will be reduced.} \let\l@r=L
\magnification=1000\baselineskip=16pt plus 2pt minus 1pt \vsize=7truein
\redoffs \hstitle=8truein\hsbody=4.75truein\fullhsize=10truein\hsize=\hsbody
\output={\ifnum\pageno=0 
  \shipout\vbox{\speclscape{\hsize\fullhsize\makeheadline}
    \hbox to \fullhsize{\hfill\pagebody\hfill}}\advancepageno
  \else
  \almostshipout{\leftline{\vbox{\pagebody\makefootline}}}\advancepageno
  \fi}
\def\almostshipout#1{\if L\l@r \count1=1 \message{[\the\count0.\the\count1]}
      \global\setbox\leftpage=#1 \global\let\l@r=R
 \else \count1=2
  \shipout\vbox{\speclscape{\hsize\fullhsize\makeheadline}
      \hbox to\fullhsize{\box\leftpage\hfil#1}}  \global\let\l@r=L\fi}
\fi
%
\newcount\yearltd\yearltd=\year\advance\yearltd by -2000

\def\Title#1#2{\nopagenumbers\abstractfont\hsize=\hstitle\rightline{#1}%
\vskip 1in\centerline{\titlefont #2}\abstractfont\vskip .5in\pageno=0}
\def\Date#1{\vfill\leftline{#1}\tenpoint\supereject\global\hsize=\hsbody%
\footline={\hss\tenrm\hyperdef\hypernoname{page}\folio\folio\hss}}%
%

\def\draftmode{\message{ DRAFTMODE }\def\draftdate{{\rm preliminary draft:
\number\month/\number\day/\number\yearltd\ \ \hourmin}}%
\headline={\hfil\draftdate}\writelabels\baselineskip=20pt plus 2pt minus 2pt
 {\count255=\time\divide\count255 by 60 \xdef\hourmin{\number\count255}
  \multiply\count255 by-60\advance\count255 by\time
  \xdef\hourmin{\hourmin:\ifnum\count255<10 0\fi\the\count255}}}
\def\nolabels{\def\wrlabeL##1{}\def\eqlabeL##1{}\def\reflabeL##1{}}
\def\writelabels{\def\wrlabeL##1{\leavevmode\vadjust{\rlap{\smash%
{\line{{\escapechar=` \hfill\rlap{\sevenrm\hskip.03in\string##1}}}}}}}%
\def\eqlabeL##1{{\escapechar-1\rlap{\sevenrm\hskip.05in\string##1}}}%
\def\reflabeL##1{\noexpand\llap{\noexpand\sevenrm\string\string\string##1}}}
\nolabels
%
\global\newcount\secno \global\secno=0
\global\newcount\meqno \global\meqno=1
\def\s@csym{}
\def\newsec#1{\global\advance\secno by1%
{\toks0{#1}\message{(\the\secno. \the\toks0)}}%
\global\subsecno=0\eqnres@t\let\s@csym\secsym\xdef\secn@m{\the\secno}\noindent
{\bf\hyperdef\hypernoname{section}{\the\secno}{\the\secno.} #1}%
\writetoca{{\string\hyperref{}{section}{\the\secno}{\the\secno.}} {#1}}%
\par\nobreak\medskip\nobreak}
\def\eqnres@t{\xdef\secsym{\the\secno.}\global\meqno=1\bigbreak\bigskip}
\def\sequentialequations{\def\eqnres@t{\bigbreak}}\xdef\secsym{}
\global\newcount\subsecno \global\subsecno=0
\def\subsec#1{\global\advance\subsecno by1%
{\toks0{#1}\message{(\s@csym\the\subsecno. \the\toks0)}}%
\ifnum\lastpenalty>9000\else\bigbreak\fi
\noindent{\it\hyperdef\hypernoname{subsection}{\secn@m.\the\subsecno}%
{\secn@m.\the\subsecno.} #1}\writetoca{\string\quad
{\string\hyperref{}{subsection}{\secn@m.\the\subsecno}{\secn@m.\the\subsecno.}}
{#1}}\par\nobreak\medskip\nobreak}
\def\appendix#1#2{\global\meqno=1\global\subsecno=0\xdef\secsym{\hbox{#1.}}%
\bigbreak\bigskip\noindent{\bf Appendix \hyperdef\hypernoname{appendix}{#1}%
{#1.} #2}{\toks0{(#1. #2)}\message{\the\toks0}}%
\xdef\s@csym{#1.}\xdef\secn@m{#1}%
\writetoca{\string\hyperref{}{appendix}{#1}{Appendix {#1.}} {#2}}%
\par\nobreak\medskip\nobreak}
%
%
\def\checkm@de#1#2{\ifmmode{\def\f@rst##1{##1}\hyperdef\hypernoname{equation}%
{#1}{#2}}\else\hyperref{}{equation}{#1}{#2}\fi}
\def\eqnn#1{\DefWarn#1\xdef #1{(\noexpand\relax\noexpand\checkm@de%
{\s@csym\the\meqno}{\secsym\the\meqno})}%
\wrlabeL#1\writedef{#1\leftbracket#1}\global\advance\meqno by1}
\def\f@rst#1{\c@t#1a\em@ark}\def\c@t#1#2\em@ark{#1}
\def\eqna#1{\DefWarn#1\wrlabeL{#1$\{\}$}%
\xdef #1##1{(\noexpand\relax\noexpand\checkm@de%
{\s@csym\the\meqno\noexpand\f@rst{##1}}{\hbox{$\secsym\the\meqno##1$}})}
\writedef{#1\numbersign1\leftbracket#1{\numbersign1}}\global\advance\meqno by1}
\def\eqn#1#2{\DefWarn#1%
\xdef #1{(\noexpand\hyperref{}{equation}{\s@csym\the\meqno}%
{\secsym\the\meqno})}$$#2\eqno(\hyperdef\hypernoname{equation}%
{\s@csym\the\meqno}{\secsym\the\meqno})\eqlabeL#1$$%
\writedef{#1\leftbracket#1}\global\advance\meqno by1}
\def\xeqn{\expandafter\xe@n}\def\xe@n(#1){#1}
\def\xeqna#1{\expandafter\xe@n#1}
\def\eqns#1{(\e@ns #1{\hbox{}})}
\def\e@ns#1{\ifx\UNd@FiNeD#1\message{eqnlabel \string#1 is undefined.}%
\xdef#1{(?.?)}\fi{\let\hyperref=\relax\xdef\next{#1}}%
\ifx\next\em@rk\def\next{}\else%
\ifx\next#1\xeqn#1\else\def\n@xt{#1}\ifx\n@xt\next#1\else\xeqna#1\fi
\fi\let\next=\e@ns\fi\next}

\def\DefWarn#1{\ifx\UNd@FiNeD#1\else
\immediate\write16{*** WARNING: the label \string#1 is already defined ***}\fi}
%
\newskip\footskip\footskip14pt plus 1pt minus 1pt 
\def\footnotefont{\ninepoint}\def\f@t#1{\footnotefont #1\@foot}
\def\f@@t{\baselineskip\footskip\bgroup\footnotefont\aftergroup\@foot\let\next}
\setbox\strutbox=\hbox{\vrule height9.5pt depth4.5pt width0pt}
\global\newcount\ftno \global\ftno=0
\def\foot{\global\advance\ftno by1\def\foot@rg{\hyperref{}{footnote}%
{\the\ftno}{\the\ftno}\xdef\foot@rg{\noexpand\hyperdef\noexpand\hypernoname%
{footnote}{\the\ftno}{\the\ftno}}}\footnote{$^{\foot@rg}$}}
%
\newwrite\ftfile
\def\footend{\def\foot{\global\advance\ftno by1\chardef\wfile=\ftfile
\hyperref{}{footnote}{\the\ftno}{$^{\the\ftno}$}%
\ifnum\ftno=1\immediate\openout\ftfile=\jobname.fts\fi%
\immediate\write\ftfile{\noexpand\smallskip%
\noexpand\item{\noexpand\hyperdef\noexpand\hypernoname{footnote}
{\the\ftno}{f\the\ftno}:\ }\pctsign}\findarg}%
\def\footatend{\vfill\eject\immediate\closeout\ftfile{\parindent=20pt
\centerline{\bf Footnotes}\nobreak\bigskip\input \jobname.fts }}}
\def\footatend{}
%
%
\global\newcount\refno \global\refno=1
\newwrite\rfile
\def\ref{[\hyperref{}{reference}{\the\refno}{\the\refno}]\nref}
\def\nref#1{\DefWarn#1%
\xdef#1{[\noexpand\hyperref{}{reference}{\the\refno}{\the\refno}]}%
\writedef{#1\leftbracket#1}%
\ifnum\refno=1\immediate\openout\rfile=\jobname.refs\fi
\chardef\wfile=\rfile\immediate\write\rfile{\noexpand\item{[\noexpand\hyperdef%
\noexpand\hypernoname{reference}{\the\refno}{\the\refno}]\ }%
\reflabeL{#1\hskip.31in}\pctsign}\global\advance\refno by1\findarg}
\def\findarg#1#{\begingroup\obeylines\newlinechar=`\^^M\pass@rg}
{\obeylines\gdef\pass@rg#1{\writ@line\relax #1^^M\hbox{}^^M}%
\gdef\writ@line#1^^M{\expandafter\toks0\expandafter{\striprel@x #1}%
\edef\next{\the\toks0}\ifx\next\em@rk\let\next=\endgroup\else\ifx\next\empty%
\else\immediate\write\wfile{\the\toks0}\fi\let\next=\writ@line\fi\next\relax}}
\def\striprel@x#1{} \def\em@rk{\hbox{}}
\def\lref{\begingroup\obeylines\lr@f}
\def\lr@f#1#2{\DefWarn#1\gdef#1{\let#1=\UNd@FiNeD\ref#1{#2}}\endgroup\unskip}

\def\addref#1{\immediate\write\rfile{\noexpand\item{}#1}} 
\def\listrefs{\footatend\vfill\supereject\immediate\closeout\rfile\writestoppt
\baselineskip=\footskip\centerline{{\bf References}}\bigskip{\parindent=20pt%
\frenchspacing\escapechar=` \input \jobname.refs\vfill\eject}\nonfrenchspacing}
\def\startrefs#1{\immediate\openout\rfile=\jobname.refs\refno=#1}
\def\xref{\expandafter\xr@f}\def\xr@f[#1]{#1}
\def\refs#1{\count255=1[\r@fs #1{\hbox{}}]}
\def\r@fs#1{\ifx\UNd@FiNeD#1\message{reflabel \string#1 is undefined.}%
\nref#1{need to supply reference \string#1.}\fi%
\vphantom{\hphantom{#1}}{\let\hyperref=\relax\xdef\next{#1}}%
\ifx\next\em@rk\def\next{}%
\else\ifx\next#1\ifodd\count255\relax\xref#1\count255=0\fi%
\else#1\count255=1\fi\let\next=\r@fs\fi\next}
%

%
\newwrite\ffile\global\newcount\figno \global\figno=1
\def\fig{fig.~\hyperref{}{figure}{\the\figno}{\the\figno}\nfig}
\def\nfig#1{\DefWarn#1%
\xdef#1{fig.~\noexpand\hyperref{}{figure}{\the\figno}{\the\figno}}%
\writedef{#1\leftbracket fig.\noexpand~\xfig#1}%
\ifnum\figno=1\immediate\openout\ffile=\jobname.figs\fi\chardef\wfile=\ffile%
{\let\hyperref=\relax
\immediate\write\ffile{\noexpand\medskip\noexpand\item{Fig.\ %
\noexpand\hyperdef\noexpand\hypernoname{figure}{\the\figno}{\the\figno}. }
\reflabeL{#1\hskip.55in}\pctsign}}\global\advance\figno by1\findarg}
\def\listfigs{\vfill\eject\immediate\closeout\ffile{\parindent40pt
\baselineskip14pt\centerline{{\bf Figure Captions}}\nobreak\medskip
\escapechar=` \input \jobname.figs\vfill\eject}}
\def\xfig{\expandafter\xf@g}\def\xf@g fig.\penalty\@M\ {}
\def\figs#1{figs.~\f@gs #1{\hbox{}}}
\def\f@gs#1{{\let\hyperref=\relax\xdef\next{#1}}\ifx\next\em@rk\def\next{}\else
\ifx\next#1\xfig #1\else#1\fi\let\next=\f@gs\fi\next}
\def\figin{\epsfcheck\figin}\def\figins{\epsfcheck\figins}
\def\epsfcheck{\ifx\epsfbox\UNd@FiNeD
\message{(NO epsf.tex, FIGURES WILL BE IGNORED)}
\gdef\figin##1{\vskip2in}\gdef\figins##1{\hskip.5in}
\else\message{(FIGURES WILL BE INCLUDED)}%
\gdef\figin##1{##1}\gdef\figins##1{##1}\fi}
\def\DefWarn#1{}
\def\figinsert{\goodbreak\midinsert}
\def\ifig#1#2#3{\DefWarn#1\xdef#1{fig.~\noexpand\hyperref{}{figure}%
{\the\figno}{\the\figno}}\writedef{#1\leftbracket fig.\noexpand~\xfig#1}%
\figinsert\figin{\centerline{#3}}\medskip\centerline{\vbox{\baselineskip12pt
\advance\hsize by -1truein\noindent\wrlabeL{#1=#1}\footnotefont%
{\bf Fig.~\hyperdef\hypernoname{figure}{\the\figno}{\the\figno}:} #2}}
\bigskip\endinsert\global\advance\figno by1}
\newwrite\lfile
{\escapechar-1\xdef\pctsign{\string\%}\xdef\leftbracket{\string\{}
\xdef\rightbracket{\string\}}\xdef\numbersign{\string\#}}
\def\writedefs{\immediate\openout\lfile=\jobname.defs \def\writedef##1{%
{\let\hyperref=\relax\let\hyperdef=\relax\let\hypernoname=\relax
 \immediate\write\lfile{\string\def\string##1\rightbracket}}}}%
\def\writestop{\def\writestoppt{\immediate\write\lfile{\string\pageno
 \the\pageno\string\startrefs\leftbracket\the\refno\rightbracket
 \string\def\string\secsym\leftbracket\secsym\rightbracket
 \string\secno\the\secno\string\meqno\the\meqno}\immediate\closeout\lfile}}
\def\writestoppt{}\def\writedef#1{}
\def\seclab#1{\DefWarn#1%
\xdef #1{\noexpand\hyperref{}{section}{\the\secno}{\the\secno}}%
\writedef{#1\leftbracket#1}\wrlabeL{#1=#1}}
\def\subseclab#1{\DefWarn#1%
\xdef #1{\noexpand\hyperref{}{subsection}{\secn@m.\the\subsecno}%
{\secn@m.\the\subsecno}}\writedef{#1\leftbracket#1}\wrlabeL{#1=#1}}
\def\applab#1{\DefWarn#1%
\xdef #1{\noexpand\hyperref{}{appendix}{\secn@m}{\secn@m}}%
\writedef{#1\leftbracket#1}\wrlabeL{#1=#1}}
\newwrite\tfile \def\writetoca#1{}
\def\leaderfill{\leaders\hbox to 1em{\hss.\hss}\hfill}
\def\writetoc{\immediate\openout\tfile=\jobname.toc
   \def\writetoca##1{{\edef\next{\write\tfile{\noindent ##1
   \string\leaderfill {\string\hyperref{}{page}{\noexpand\number\pageno}%
                       {\noexpand\number\pageno}} \par}}\next}}}
\newread\ch@ckfile
\def\listtoc{\immediate\closeout\tfile\immediate\openin\ch@ckfile=\jobname.toc
\ifeof\ch@ckfile\message{no file \jobname.toc, no table of contents this pass}%
\else\closein\ch@ckfile\centerline{\bf Contents}\nobreak\medskip%
{\baselineskip=12pt\footnotefont\parskip=0pt\catcode`\@=11\input\jobname.toc
\catcode`\@=12\bigbreak\bigskip}\fi}
\catcode`\@=12 
%
\edef\tfontsize{\ifx\answ\bigans scaled\magstep3\else scaled\magstep4\fi}
\font\titlerm=cmr10 \tfontsize \font\titlerms=cmr7 \tfontsize
\font\titlermss=cmr5 \tfontsize \font\titlei=cmmi10 \tfontsize
\font\titleis=cmmi7 \tfontsize \font\titleiss=cmmi5 \tfontsize
\font\titlesy=cmsy10 \tfontsize \font\titlesys=cmsy7 \tfontsize
\font\titlesyss=cmsy5 \tfontsize \font\titleit=cmti10 \tfontsize
\skewchar\titlei='177 \skewchar\titleis='177 \skewchar\titleiss='177
\skewchar\titlesy='60 \skewchar\titlesys='60 \skewchar\titlesyss='60
\def\titlefont{\def\rm{\fam0\titlerm}
\textfont0=\titlerm \scriptfont0=\titlerms \scriptscriptfont0=\titlermss
\textfont1=\titlei \scriptfont1=\titleis \scriptscriptfont1=\titleiss
\textfont2=\titlesy \scriptfont2=\titlesys \scriptscriptfont2=\titlesyss
\textfont\itfam=\titleit \def\it{\fam\itfam\titleit}\rm}
 \ifx\answ\bigans\else scaled\magstep1\fi
\ifx\answ\bigans\def\abstractfont{\tenpoint}\else
\font\absit=cmti10 scaled \magstep1
\font\abssl=cmsl10 scaled \magstep1
\font\absrm=cmr10 scaled\magstep1 \font\absrms=cmr7 scaled\magstep1
\font\absrmss=cmr5 scaled\magstep1 \font\absi=cmmi10 scaled\magstep1
\font\absis=cmmi7 scaled\magstep1 \font\absiss=cmmi5 scaled\magstep1
\font\abssy=cmsy10 scaled\magstep1 \font\abssys=cmsy7 scaled\magstep1
\font\abssyss=cmsy5 scaled\magstep1 \font\absbf=cmbx10 scaled\magstep1
\skewchar\absi='177 \skewchar\absis='177 \skewchar\absiss='177
\skewchar\abssy='60 \skewchar\abssys='60 \skewchar\abssyss='60
\def\abstractfont{\def\rm{\fam0\absrm}
\textfont0=\absrm \scriptfont0=\absrms \scriptscriptfont0=\absrmss
\textfont1=\absi \scriptfont1=\absis \scriptscriptfont1=\absiss
\textfont2=\abssy \scriptfont2=\abssys \scriptscriptfont2=\abssyss
\textfont\itfam=\absit \def\it{\fam\itfam\absit}\def\footnotefont{\tenpoint}%
\textfont\slfam=\abssl \def\sl{\fam\slfam\abssl}%
\textfont\bffam=\absbf \def\bf{\fam\bffam\absbf}\rm}\fi
\def\tenpoint{\def\rm{\fam0\tenrm}
\textfont0=\tenrm \scriptfont0=\sevenrm \scriptscriptfont0=\fiverm
\textfont1=\teni  \scriptfont1=\seveni  \scriptscriptfont1=\fivei
\textfont2=\tensy \scriptfont2=\sevensy \scriptscriptfont2=\fivesy
\textfont\itfam=\tenit \def\it{\fam\itfam\tenit}\def\footnotefont{\ninepoint}%
\textfont\bffam=\tenbf \def\bf{\fam\bffam\tenbf}\def\sl{\fam\slfam\tensl}\rm}
\font\ninerm=cmr9 \font\sixrm=cmr6 \font\ninei=cmmi9 \font\sixi=cmmi6
\font\ninesy=cmsy9 \font\sixsy=cmsy6 \font\ninebf=cmbx9
\font\nineit=cmti9 \font\ninesl=cmsl9 \skewchar\ninei='177
\skewchar\sixi='177 \skewchar\ninesy='60 \skewchar\sixsy='60
\def\ninepoint{\def\rm{\fam0\ninerm}
\textfont0=\ninerm \scriptfont0=\sixrm \scriptscriptfont0=\fiverm
\textfont1=\ninei \scriptfont1=\sixi \scriptscriptfont1=\fivei
\textfont2=\ninesy \scriptfont2=\sixsy \scriptscriptfont2=\fivesy
\textfont\itfam=\ninei \def\it{\fam\itfam\nineit}\def\sl{\fam\slfam\ninesl}%
\textfont\bffam=\ninebf \def\bf{\fam\bffam\ninebf}\rm}
%
%

\hyphenation{anom-aly anom-alies coun-ter-term coun-ter-terms}
\def\inv{^{\raise.15ex\hbox{${\scriptscriptstyle -}$}\kern-.05em 1}}

\def\Dsl{\,\raise.15ex\hbox{/}\mkern-13.5mu D} 
\def\dsl{\raise.15ex\hbox{/}\kern-.57em\partial}

 \def\Tr{{\rm Tr}}
\def\lspace{\ifx\answ\bigans{}\else\qquad\fi}
\def\lbspace{\ifx\answ\bigans{}\else\hskip-.2in\fi} 
\def\boxeqn#1{\vcenter{\vbox{\hrule\hbox{\vrule\kern3pt\vbox{\kern3pt
	\hbox{${\displaystyle #1}$}\kern3pt}\kern3pt\vrule}\hrule}}}
\def\mbox#1#2{\vcenter{\hrule \hbox{\vrule height#2in
		\kern#1in \vrule} \hrule}}  
%

\def\darr#1{\raise1.5ex\hbox{$\leftrightarrow$}\mkern-16.5mu #1}

\def\roughly#1{\raise.3ex\hbox{$#1$\kern-.75em\lower1ex\hbox{$\sim$}}}

\def\bb{
\font\tenmsb=msbm10
\font\sevenmsb=msbm7
\font\fivemsb=msbm5
\textfont1=\tenmsb
\scriptfont1=\sevenmsb
\scriptscriptfont1=\fivemsb
}

\input amssym

\input epsf

\def\IZ{\relax\ifmmode\mathchoice
{\hbox{\cmss Z\kern-.4em Z}}{\hbox{\cmss Z\kern-.4em Z}} {\lower.9pt\hbox{\cmsss Z\kern-.4em Z}}
{\lower1.2pt\hbox{\cmsss Z\kern-.4em Z}}\else{\cmss Z\kern-.4em Z}\fi}

\newif\ifdraft\draftfalse
\newif\ifinter\interfalse
\ifdraft\draftmode\else\interfalse\fi
\def\journal#1&#2(#3){\unskip, \sl #1\ \bf #2 \rm(19#3) }
\def\andjournal#1&#2(#3){\sl #1~\bf #2 \rm (19#3) }

\def\ie{{\it i.e.}}
\def\eg{{\it e.g.}}

\def\frac#1#2{{#1\over#2}}

\def\inbar{\,\vrule height1.5ex width.4pt depth0pt}
\def\IC{\relax\hbox{$\inbar\kern-.3em{\rm C}$}}
\def\IR{\relax{\rm I\kern-.18em R}}
\def\IP{\relax{\rm I\kern-.18em P}}
\def\Z{{\bf Z}}

%
%


%
\catcode`\@=11
\def\slash#1{\mathord{\mathpalette\c@ncel{#1}}}
\overfullrule=0pt

\def\underrel#1\over#2{\mathrel{\mathop{\kern\z@#1}\limits_{#2}}}

\catcode`\@=12


%

\def\det{{\rm det}}

\def \sinh{{\rm sinh}}

\def\det{{\rm det}}
\def\exp{{\rm exp}}

\def\sra{ \;\; \;^{\longrightarrow}_{\Sigma \rightarrow \infty} \;\;\; }
\def\tra{ \;\; \;^{\longrightarrow}_{\tau \rightarrow 0} \;\;\; }

\def\rra{ \;\; \;^{\longrightarrow}_{r \rightarrow 0} \;\;\; }


\def\[{[}
\def\]{]}
\def\bq{{\bf {q}}}
\def\bz{{\bf {z}}}

\def\bw{{\bf {w}}}
\def\bs{{\bf s}}
\def\bn{{\bf n}}
\def\bm{{\bf m}}
\def\bu{{\bf u}}

\def\comment#1{ }

%
\def\draftnote#1{\ifdraft{\baselineskip2ex
                 \vbox{\kern1em\hrule\hbox{\vrule\kern1em\vbox{\kern1ex
                 \noindent \underbar{NOTE}: #1
             \vskip1ex}\kern1em\vrule}\hrule}}\fi}
\def\internote#1{\ifinter{\baselineskip2ex
                 \vbox{\kern1em\hrule\hbox{\vrule\kern1em\vbox{\kern1ex
                 \noindent \underbar{Internal Note}: #1
             \vskip1ex}\kern1em\vrule}\hrule}}\fi}

%
%



%
%
%
%

%

\def\inv{^{-1}}

\def\nlb{\item{$\bullet$}}


\def\Tr{{\rm Tr}}

\def\cN{{\cal N}}

\def\1{{\ds 1}}
\def\R{\hbox{$\bb R$}}

\def\Z{\hbox{$\bb Z$}}

\def\T{\hbox{$\bb T$}}

\def\S{\hbox{$\bb S$}}

\newfam\frakfam
\font\teneufm=eufm10
\font\seveneufm=eufm7
\font\fiveeufm=eufm5
\textfont\frakfam=\teneufm
\scriptfont\frakfam=\seveneufm
\scriptscriptfont\frakfam=\fiveeufm

\def\cW{{\cal W}}
\def\cZ{{\cal Z}}

\lref\NiarchosAH{
  V.~Niarchos,
  ``Seiberg dualities and the 3d/4d connection,''
JHEP {\bf 1207}, 075 (2012).
[arXiv:1205.2086 [hep-th]].
}

\lref\AharonyGP{
  O.~Aharony,
  ``IR duality in d = 3 N=2 supersymmetric USp(2N(c)) and U(N(c)) gauge theories,''
Phys.\ Lett.\ B {\bf 404}, 71 (1997).
[hep-th/9703215].
}

\lref\AffleckAS{
  I.~Affleck, J.~A.~Harvey and E.~Witten,
  ``Instantons and (Super)Symmetry Breaking in (2+1)-Dimensions,''
Nucl.\ Phys.\ B {\bf 206}, 413 (1982)..
}

\lref\IntriligatorID{
  K.~A.~Intriligator and N.~Seiberg,
  ``Duality, monopoles, dyons, confinement and oblique confinement in supersymmetric SO(N(c)) gauge theories,''
Nucl.\ Phys.\ B {\bf 444}, 125 (1995).
[hep-th/9503179].
}

\lref\PasquettiFJ{
  S.~Pasquetti,
  ``Factorisation of N = 2 Theories on the Squashed 3-Sphere,''
JHEP {\bf 1204}, 120 (2012).
[arXiv:1111.6905 [hep-th]].
}

\lref\BeemMB{
  C.~Beem, T.~Dimofte and S.~Pasquetti,
  ``Holomorphic Blocks in Three Dimensions,''
JHEP {\bf 1412}, 177 (2014).
[arXiv:1211.1986 [hep-th]].
}

\lref\SeibergPQ{
  N.~Seiberg,
  ``Electric - magnetic duality in supersymmetric nonAbelian gauge theories,''
Nucl.\ Phys.\ B {\bf 435}, 129 (1995).
[hep-th/9411149].
}

\lref\AharonyBX{
  O.~Aharony, A.~Hanany, K.~A.~Intriligator, N.~Seiberg and M.~J.~Strassler,
  ``Aspects of N=2 supersymmetric gauge theories in three-dimensions,''
Nucl.\ Phys.\ B {\bf 499}, 67 (1997).
[hep-th/9703110].
}

\lref\AharonyUYA{
  O.~Aharony and D.~Fleischer,
  ``IR Dualities in General 3d Supersymmetric SU(N) QCD Theories,''
JHEP {\bf 1502}, 162 (2015).
[arXiv:1411.5475 [hep-th]].
}

\lref\IntriligatorNE{
  K.~A.~Intriligator and P.~Pouliot,
  ``Exact superpotentials, quantum vacua and duality in supersymmetric SP(N(c)) gauge theories,''
Phys.\ Lett.\ B {\bf 353}, 471 (1995).
[hep-th/9505006].
}

\lref\KarchUX{
  A.~Karch,
  ``Seiberg duality in three-dimensions,''
Phys.\ Lett.\ B {\bf 405}, 79 (1997).
[hep-th/9703172].
}

\lref\SafdiRE{
  B.~R.~Safdi, I.~R.~Klebanov and J.~Lee,
  ``A Crack in the Conformal Window,''
[arXiv:1212.4502 [hep-th]].
}

\lref\AharonyTH{
  O.~Aharony, M.~Berkooz, S.~Kachru, N.~Seiberg and E.~Silverstein,
  ``Matrix description of interacting theories in six-dimensions,''
Adv.\ Theor.\ Math.\ Phys.\  {\bf 1}, 148 (1998).
[hep-th/9707079].
}

\lref\SchweigertTG{
  C.~Schweigert,
  ``On moduli spaces of flat connections with nonsimply connected structure group,''
Nucl.\ Phys.\ B {\bf 492}, 743 (1997).
[hep-th/9611092].
}

\lref\WittenYU{
  E.~Witten,
  ``On the conformal field theory of the Higgs branch,''
JHEP {\bf 9707}, 003 (1997).
[hep-th/9707093].
}

\lref\GiveonZN{
  A.~Giveon and D.~Kutasov,
  ``Seiberg Duality in Chern-Simons Theory,''
Nucl.\ Phys.\ B {\bf 812}, 1 (2009).
[arXiv:0808.0360 [hep-th]].
}

\lref\GaiottoBE{
  D.~Gaiotto, G.~W.~Moore and A.~Neitzke,
  ``Framed BPS States,''
[arXiv:1006.0146 [hep-th]].
}

\lref\SethiPA{
  S.~Sethi and M.~Stern,
  ``D-brane bound states redux,''
Commun.\ Math.\ Phys.\  {\bf 194}, 675 (1998).
[hep-th/9705046].
}

\lref\AharonyDW{
  O.~Aharony and M.~Berkooz,
  ``IR dynamics of D = 2, N=(4,4) gauge theories and DLCQ of 'little string theories',''
JHEP {\bf 9910}, 030 (1999).
[hep-th/9909101].
}

\lref\AldayRS{
  L.~F.~Alday, M.~Bullimore and M.~Fluder,
  ``On S-duality of the Superconformal Index on Lens Spaces and 2d TQFT,''
JHEP {\bf 1305}, 122 (2013).
[arXiv:1301.7486 [hep-th]].
}

\lref\RazamatJXA{
  S.~S.~Razamat and M.~Yamazaki,
  ``S-duality and the N=2 Lens Space Index,''
[arXiv:1306.1543 [hep-th]].
}

\lref\HoriDK{
  K.~Hori and D.~Tong,
  ``Aspects of Non-Abelian Gauge Dynamics in Two-Dimensional N=(2,2) Theories,''
JHEP {\bf 0705}, 079 (2007).
[hep-th/0609032].
}

\lref\NiarchosAH{
  V.~Niarchos,
  ``Seiberg dualities and the 3d/4d connection,''
JHEP {\bf 1207}, 075 (2012).
[arXiv:1205.2086 [hep-th]].
}

\lref\BeniniCZ{
  F.~Benini and N.~Bobev,
  ``Exact two-dimensional superconformal R-symmetry and c-extremization,''
Phys.\ Rev.\ Lett.\  {\bf 110}, no. 6, 061601 (2013).
[arXiv:1211.4030 [hep-th]].
}

\lref\almost{
  A.~Borel, R.~Friedman, J.~W.~Morgan,
  ``Almost commuting elements in compact Lie groups,''
arXiv:math/9907007.
}

\lref\HarveyNHA{
  J.~A.~Harvey, S.~Lee and S.~Murthy,
  ``Elliptic genera of ALE and ALF manifolds from gauged linear sigma models,''
JHEP {\bf 1502}, 110 (2015).
[arXiv:1406.6342 [hep-th]].
}
\lref\KapustinJM{
  A.~Kapustin and B.~Willett,
  ``Generalized Superconformal Index for Three Dimensional Field Theories,''
[arXiv:1106.2484 [hep-th]].
}

\lref\AharonyGP{
  O.~Aharony,
  ``IR duality in d = 3 N=2 supersymmetric USp(2N(c)) and U(N(c)) gauge theories,''
Phys.\ Lett.\ B {\bf 404}, 71 (1997).
[hep-th/9703215].
}

\lref\AharonyJKI{
  O.~Aharony, S.~S.~Razamat, N.~Seiberg and B.~Willett,
  ``The long flow to freedom,''
JHEP {\bf 1702}, 056 (2017).
[arXiv:1611.02763 [hep-th]].
}

\lref\FestucciaWS{
  G.~Festuccia and N.~Seiberg,
  ``Rigid Supersymmetric Theories in Curved Superspace,''
JHEP {\bf 1106}, 114 (2011).
[arXiv:1105.0689 [hep-th]].
}

\lref\RomelsbergerEG{
  C.~Romelsberger,
  ``Counting chiral primaries in N = 1, d=4 superconformal field theories,''
Nucl.\ Phys.\ B {\bf 747}, 329 (2006).
[hep-th/0510060].
}

\lref\KapustinKZ{
  A.~Kapustin, B.~Willett and I.~Yaakov,
  ``Exact Results for Wilson Loops in Superconformal Chern-Simons Theories with Matter,''
JHEP {\bf 1003}, 089 (2010).
[arXiv:0909.4559 [hep-th]].
}

\lref\DolanQI{
  F.~A.~Dolan and H.~Osborn,
  ``Applications of the Superconformal Index for Protected Operators and q-Hypergeometric Identities to N=1 Dual Theories,''
Nucl.\ Phys.\ B {\bf 818}, 137 (2009).
[arXiv:0801.4947 [hep-th]].
}

\lref\GaddeIA{
  A.~Gadde and W.~Yan,
  ``Reducing the 4d Index to the $S^3$ Partition Function,''
JHEP {\bf 1212}, 003 (2012).
[arXiv:1104.2592 [hep-th]].
}

\lref\DolanRP{
  F.~A.~H.~Dolan, V.~P.~Spiridonov and G.~S.~Vartanov,
  ``From 4d superconformal indices to 3d partition functions,''
Phys.\ Lett.\ B {\bf 704}, 234 (2011).
[arXiv:1104.1787 [hep-th]].
}

\lref\ImamuraUW{
  Y.~Imamura,
 ``Relation between the 4d superconformal index and the $S^3$ partition function,''
JHEP {\bf 1109}, 133 (2011).
[arXiv:1104.4482 [hep-th]].
}

\lref\HamaEA{
  N.~Hama, K.~Hosomichi and S.~Lee,
  ``SUSY Gauge Theories on Squashed Three-Spheres,''
JHEP {\bf 1105}, 014 (2011).
[arXiv:1102.4716 [hep-th]].
}

\lref\GaddeEN{
  A.~Gadde, L.~Rastelli, S.~S.~Razamat and W.~Yan,
  ``On the Superconformal Index of N=1 IR Fixed Points: A Holographic Check,''
JHEP {\bf 1103}, 041 (2011).
[arXiv:1011.5278 [hep-th]].
}

\lref\EagerHX{
  R.~Eager, J.~Schmude and Y.~Tachikawa,
  ``Superconformal Indices, Sasaki-Einstein Manifolds, and Cyclic Homologies,''
[arXiv:1207.0573 [hep-th]].
}

\lref\AffleckAS{
  I.~Affleck, J.~A.~Harvey and E.~Witten,
  ``Instantons and (Super)Symmetry Breaking in (2+1)-Dimensions,''
Nucl.\ Phys.\ B {\bf 206}, 413 (1982).
}

\lref\SeibergPQ{
  N.~Seiberg,
  ``Electric - magnetic duality in supersymmetric nonAbelian gauge theories,''
Nucl.\ Phys.\ B {\bf 435}, 129 (1995).
[hep-th/9411149].
}

\lref\AlvarezGaumeNF{
  L.~Alvarez-Gaume, S.~Della Pietra and G.~W.~Moore,
  ``Anomalies and Odd Dimensions,''
Annals Phys.\  {\bf 163}, 288 (1985)..
}

\lref\SeibergRSG{
  N.~Seiberg and E.~Witten,
  ``Gapped Boundary Phases of Topological Insulators via Weak Coupling,''
[arXiv:1602.04251 [cond-mat.str-el]].
}

\lref\debult{
  F.~van~de~Bult,
  ``Hyperbolic Hypergeometric Functions,''
University of Amsterdam Ph.D. thesis
}

\lref\Shamirthesis{
  I.~Shamir,
  ``Aspects of three dimensional Seiberg duality,''
  M. Sc. thesis submitted to the Weizmann Institute of Science, April 2010.
  }

\lref\slthreeZ{
  J.~Felder, A.~Varchenko,
  ``The elliptic gamma function and $SL(3,Z) \times Z^3$,'' $\;\;$
[arXiv:math/0001184].
}

\lref\BeniniHJO{
  F.~Benini and A.~Zaffaroni,
  ``Supersymmetric partition functions on Riemann surfaces,''
[arXiv:1605.06120 [hep-th]].
}

\lref\BeniniNC{
  F.~Benini, T.~Nishioka and M.~Yamazaki,
  ``4d Index to 3d Index and 2d TQFT,''
Phys.\ Rev.\ D {\bf 86}, 065015 (2012).
[arXiv:1109.0283 [hep-th]].
}

\lref\GaddeDDA{
  A.~Gadde and S.~Gukov,
  ``2d Index and Surface operators,''
JHEP {\bf 1403}, 080 (2014).
[arXiv:1305.0266 [hep-th]].
}

\lref\GaiottoWE{
  D.~Gaiotto,
  ``N=2 dualities,''
  JHEP {\bf 1208}, 034 (2012).
  [arXiv:0904.2715 [hep-th]].
}

\lref\WittenZH{
  E.~Witten,
  ``Some comments on string dynamics,''
In *Los Angeles 1995, Future perspectives in string theory* 501-523.
[hep-th/9507121].
}

\lref\SpiridonovZA{
  V.~P.~Spiridonov and G.~S.~Vartanov,
  ``Elliptic Hypergeometry of Supersymmetric Dualities,''
Commun.\ Math.\ Phys.\  {\bf 304}, 797 (2011).
[arXiv:0910.5944 [hep-th]].
}

\lref\BeniniMF{
  F.~Benini, C.~Closset and S.~Cremonesi,
  ``Comments on 3d Seiberg-like dualities,''
JHEP {\bf 1110}, 075 (2011).
[arXiv:1108.5373 [hep-th]].
}

\lref\ClossetVP{
  C.~Closset, T.~T.~Dumitrescu, G.~Festuccia, Z.~Komargodski and N.~Seiberg,
  ``Comments on Chern-Simons Contact Terms in Three Dimensions,''
JHEP {\bf 1209}, 091 (2012).
[arXiv:1206.5218 [hep-th]].
}

\lref\SpiridonovHF{
  V.~P.~Spiridonov and G.~S.~Vartanov,
  ``Elliptic hypergeometry of supersymmetric dualities II. Orthogonal groups, knots, and vortices,''
[arXiv:1107.5788 [hep-th]].
}

\lref\SpiridonovWW{
  V.~P.~Spiridonov and G.~S.~Vartanov,
  ``Elliptic hypergeometric integrals and 't Hooft anomaly matching conditions,''
JHEP {\bf 1206}, 016 (2012).
[arXiv:1203.5677 [hep-th]].
}

\lref\DimoftePY{
  T.~Dimofte, D.~Gaiotto and S.~Gukov,
  ``3-Manifolds and 3d Indices,''
[arXiv:1112.5179 [hep-th]].
}

\lref\GukovKMK{
  S.~Gukov, D.~Pei, P.~Putrov and C.~Vafa,
[arXiv:1701.06567 [hep-th]].
}

\lref\GukovGKN{
  S.~Gukov, P.~Putrov and C.~Vafa,
JHEP {\bf 1707}, 071 (2017).
[arXiv:1602.05302 [hep-th]].
}

\lref\GukovSNA{
  S.~Gukov and D.~Pei,
Commun.\ Math.\ Phys.\  {\bf 355}, no. 1, 1 (2017).
[arXiv:1501.01310 [hep-th]].
}

\lref\GaddeWQ{
  A.~Gadde, S.~Gukov and P.~Putrov,
JHEP {\bf 1405}, 047 (2014).
[arXiv:1302.0015 [hep-th]].
}

\lref\KapustinHPK{
  A.~Kapustin and B.~Willett,
[arXiv:1302.2164 [hep-th]].
}

\lref\ClossetZGF{
  C.~Closset, H.~Kim and B.~Willett,
  ``Supersymmetric partition functions and the three-dimensional A-twist,''
JHEP {\bf 1703}, 074 (2017).
[arXiv:1701.03171 [hep-th]].
}

\lref\ClossetARN{
  C.~Closset and H.~Kim,
  ``Comments on twisted indices in 3d supersymmetric gauge theories,''
JHEP {\bf 1608}, 059 (2016).
[arXiv:1605.06531 [hep-th]].
}

\lref\KimWB{
  S.~Kim,
  ``The Complete superconformal index for N=6 Chern-Simons theory,''
Nucl.\ Phys.\ B {\bf 821}, 241 (2009), [Erratum-ibid.\ B {\bf 864}, 884 (2012)].
[arXiv:0903.4172 [hep-th]].
}

\lref\WillettGP{
  B.~Willett and I.~Yaakov,
  ``N=2 Dualities and Z Extremization in Three Dimensions,''
[arXiv:1104.0487 [hep-th]].
}

\lref\ImamuraSU{
  Y.~Imamura and S.~Yokoyama,
  ``Index for three dimensional superconformal field theories with general R-charge assignments,''
JHEP {\bf 1104}, 007 (2011).
[arXiv:1101.0557 [hep-th]].
}

\lref\FreedYA{
  D.~S.~Freed, G.~W.~Moore and G.~Segal,
  ``The Uncertainty of Fluxes,''
Commun.\ Math.\ Phys.\  {\bf 271}, 247 (2007).
[hep-th/0605198].
}

\lref\HwangQT{
  C.~Hwang, H.~Kim, K.~-J.~Park and J.~Park,
  ``Index computation for 3d Chern-Simons matter theory: test of Seiberg-like duality,''
JHEP {\bf 1109}, 037 (2011).
[arXiv:1107.4942 [hep-th]].
}

\lref\ParkWTA{
  J.~Park and K.~J.~Park,
  ``Seiberg-like Dualities for 3d N=2 Theories with SU(N) gauge group,''
JHEP {\bf 1310}, 198 (2013).
[arXiv:1305.6280 [hep-th]].
}

\lref\JafferisNS{
  D.~Jafferis and X.~Yin,
  ``A Duality Appetizer,''
[arXiv:1103.5700 [hep-th]].
}

\lref\GreenDA{
  D.~Green, Z.~Komargodski, N.~Seiberg, Y.~Tachikawa and B.~Wecht,
  ``Exactly Marginal Deformations and Global Symmetries,''
JHEP {\bf 1006}, 106 (2010).
[arXiv:1005.3546 [hep-th]].
}

\lref\GaiottoXA{
  D.~Gaiotto, L.~Rastelli and S.~S.~Razamat,
  ``Bootstrapping the superconformal index with surface defects,''
[arXiv:1207.3577 [hep-th]].
}

\lref\TroostUD{
  J.~Troost,
 ``The non-compact elliptic genus: mock or modular,''
JHEP {\bf 1006}, 104 (2010).
[arXiv:1004.3649 [hep-th]].
}

\lref\BhattacharyaZY{
  J.~Bhattacharya, S.~Bhattacharyya, S.~Minwalla and S.~Raju,
  ``Indices for Superconformal Field Theories in 3,5 and 6 Dimensions,''
JHEP {\bf 0802}, 064 (2008).
[arXiv:0801.1435 [hep-th]].
}

\lref\IntriligatorID{
  K.~A.~Intriligator and N.~Seiberg,
  ``Duality, monopoles, dyons, confinement and oblique confinement in supersymmetric SO(N(c)) gauge theories,''
Nucl.\ Phys.\ B {\bf 444}, 125 (1995).
[hep-th/9503179].
}

\lref\SeibergNZ{
  N.~Seiberg and E.~Witten,
  ``Gauge dynamics and compactification to three-dimensions,''
In *Saclay 1996, The mathematical beauty of physics* 333-366.
[hep-th/9607163].
}

\lref\KinneyEJ{
  J.~Kinney, J.~M.~Maldacena, S.~Minwalla and S.~Raju,
  ``An Index for 4 dimensional super conformal theories,''
  Commun.\ Math.\ Phys.\  {\bf 275}, 209 (2007).
  [hep-th/0510251].
}

\lref\NakayamaUR{
  Y.~Nakayama,
  ``Index for supergravity on AdS(5) x T**1,1 and conifold gauge theory,''
Nucl.\ Phys.\ B {\bf 755}, 295 (2006).
[hep-th/0602284].
}

\lref\DiaconescuGU{
  D.~E.~Diaconescu and N.~Seiberg,
  ``The Coulomb branch of (4,4) supersymmetric field theories in two-dimensions,''
JHEP {\bf 9707}, 001 (1997).
[hep-th/9707158].
}

\lref\GaddeKB{
  A.~Gadde, E.~Pomoni, L.~Rastelli and S.~S.~Razamat,
  ``S-duality and 2d Topological QFT,''
JHEP {\bf 1003}, 032 (2010).
[arXiv:0910.2225 [hep-th]].
}

\lref\GaddeTE{
  A.~Gadde, L.~Rastelli, S.~S.~Razamat and W.~Yan,
  ``The Superconformal Index of the $E_6$ SCFT,''
JHEP {\bf 1008}, 107 (2010).
[arXiv:1003.4244 [hep-th]].
}

\lref\AharonyCI{
  O.~Aharony and I.~Shamir,
  ``On $O(N_c)$ d=3 N=2 supersymmetric QCD Theories,''
JHEP {\bf 1112}, 043 (2011).
[arXiv:1109.5081 [hep-th]].
}

\lref\ClossetVG{
  C.~Closset, T.~T.~Dumitrescu, G.~Festuccia, Z.~Komargodski and N.~Seiberg,
  ``Contact Terms, Unitarity, and F-Maximization in Three-Dimensional Superconformal Theories,''
JHEP {\bf 1210}, 053 (2012).
[arXiv:1205.4142 [hep-th]].
}

\lref\ChenPHA{
  H.~Y.~Chen, H.~Y.~Chen and J.~K.~Ho,
  ``Connecting Mirror Symmetry in 3d and 2d via Localization,''
[arXiv:1312.2361 [hep-th]].
}

\lref\GiveonSR{
  A.~Giveon and D.~Kutasov,
  ``Brane dynamics and gauge theory,''
Rev.\ Mod.\ Phys.\  {\bf 71}, 983 (1999).
[hep-th/9802067].
}

\lref\SpiridonovQV{
  V.~P.~Spiridonov and G.~S.~Vartanov,
  ``Superconformal indices of ${\cal N}=4$ SYM field theories,''
Lett.\ Math.\ Phys.\  {\bf 100}, 97 (2012).
[arXiv:1005.4196 [hep-th]].
}
\lref\GaddeUV{
  A.~Gadde, L.~Rastelli, S.~S.~Razamat and W.~Yan,
  ``Gauge Theories and Macdonald Polynomials,''
Commun.\ Math.\ Phys.\  {\bf 319}, 147 (2013).
[arXiv:1110.3740 [hep-th]].
}
\lref\KapustinGH{
  A.~Kapustin,
  ``Seiberg-like duality in three dimensions for orthogonal gauge groups,''
[arXiv:1104.0466 [hep-th]].
}

\lref\orthogpaper{O. Aharony, S. S. Razamat, N.~Seiberg and B.~Willett, 
``3d dualities from 4d dualities for orthogonal groups,''
JHEP {\bf 1308}, 099 (2013).
[arXiv:1307.0511 [hep-th]].
}

\lref\KapustinHA{
  A.~Kapustin and M.~J.~Strassler,
  ``On mirror symmetry in three-dimensional Abelian gauge theories,''
JHEP {\bf 9904}, 021 (1999).
[hep-th/9902033].
}

\lref\IntriligatorEX{
  K.~A.~Intriligator and N.~Seiberg,
  ``Mirror symmetry in three-dimensional gauge theories,''
Phys.\ Lett.\ B {\bf 387}, 513 (1996).
[hep-th/9607207].
}

\lref\deBoerMP{
  J.~de Boer, K.~Hori, H.~Ooguri and Y.~Oz,
  ``Mirror symmetry in three-dimensional gauge theories, quivers and D-branes,''
Nucl.\ Phys.\ B {\bf 493}, 101 (1997).
[hep-th/9611063].
}

\lref\readinglines{
  O.~Aharony, N.~Seiberg and Y.~Tachikawa,
  ``Reading between the lines of four-dimensional gauge theories,''
JHEP {\bf 1308}, 115 (2013).
[arXiv:1305.0318 [hep-th]].
}

\lref\AharonyKMA{
  O.~Aharony, S.~S.~Razamat, N.~Seiberg and B.~Willett,
  ``3$d$ dualities from 4$d$ dualities for orthogonal groups,''
JHEP {\bf 1308}, 099 (2013).
[arXiv:1307.0511, arXiv:1307.0511 [hep-th]].
}

\lref\RazamatOPA{
  S.~S.~Razamat and B.~Willett,
  ``Global Properties of Supersymmetric Theories and the Lens Space,''
Commun.\ Math.\ Phys.\  {\bf 334}, no. 2, 661 (2015).
[arXiv:1307.4381].
}

\lref\WittenNV{
  E.~Witten,
  ``Supersymmetric index in four-dimensional gauge theories,''
Adv.\ Theor.\ Math.\ Phys.\  {\bf 5}, 841 (2002).
[hep-th/0006010].
}

\lref\BeniniNC{
  F.~Benini, T.~Nishioka and M.~Yamazaki,
  ``4d Index to 3d Index and 2d TQFT,''
Phys.\ Rev.\ D {\bf 86}, 065015 (2012).
[arXiv:1109.0283 [hep-th]].
}

\lref\GaddeUV{
  A.~Gadde, L.~Rastelli, S.~S.~Razamat and W.~Yan,
  ``Gauge Theories and Macdonald Polynomials,''
Commun.\ Math.\ Phys.\  {\bf 319}, 147 (2013).
[arXiv:1110.3740 [hep-th]].
}

\lref\GaddeIK{
  A.~Gadde, L.~Rastelli, S.~S.~Razamat and W.~Yan,
  ``The 4d Superconformal Index from q-deformed 2d Yang-Mills,''
Phys.\ Rev.\ Lett.\  {\bf 106}, 241602 (2011).
[arXiv:1104.3850 [hep-th]].
}

\lref\GaiottoUQ{
  D.~Gaiotto and S.~S.~Razamat,
  ``Exceptional Indices,''
JHEP {\bf 1205}, 145 (2012).
[arXiv:1203.5517 [hep-th]].
}

\lref\JafferisUN{
  D.~L.~Jafferis,
  ``The Exact Superconformal R-Symmetry Extremizes Z,''
JHEP {\bf 1205}, 159 (2012).
[arXiv:1012.3210 [hep-th]].
}

\lref\RazamatUV{
  S.~S.~Razamat,
  ``On a modular property of N=2 superconformal theories in four dimensions,''
JHEP {\bf 1210}, 191 (2012).
[arXiv:1208.5056 [hep-th]].
}

\lref\noumi{
  Y.~Komori, M.~Noumi, J.~Shiraishi,
  ``Kernel Functions for Difference Operators of Ruijsenaars Type and Their Applications,''
SIGMA 5 (2009), 054.
[arXiv:0812.0279 [math.QA]].
}

\lref\RazamatJXA{
  S.~S.~Razamat and M.~Yamazaki,
  ``S-duality and the N=2 Lens Space Index,''
[arXiv:1306.1543 [hep-th]].
}

\lref\RazamatOPA{
  S.~S.~Razamat and B.~Willett,
  ``Global Properties of Supersymmetric Theories and the Lens Space,''
Commun.\ Math.\ Phys.\  {\bf 334}, no. 2, 661 (2015).
[arXiv:1307.4381 [hep-th]].
}

\lref\GaddeTE{
  A.~Gadde, L.~Rastelli, S.~S.~Razamat and W.~Yan,
  ``The Superconformal Index of the $E_6$ SCFT,''
JHEP {\bf 1008}, 107 (2010).
[arXiv:1003.4244 [hep-th]].
}

\lref\deBult{
  F.~J.~van~de~Bult,
  ``An elliptic hypergeometric integral with $W(F_4)$ symmetry,''
The Ramanujan Journal, Volume 25, Issue 1 (2011)
[arXiv:0909.4793[math.CA]].
}

\lref\GaddeKB{
  A.~Gadde, E.~Pomoni, L.~Rastelli and S.~S.~Razamat,
  ``S-duality and 2d Topological QFT,''
JHEP {\bf 1003}, 032 (2010).
[arXiv:0910.2225 [hep-th]].
}

\lref\ArgyresCN{
  P.~C.~Argyres and N.~Seiberg,
  ``S-duality in N=2 supersymmetric gauge theories,''
JHEP {\bf 0712}, 088 (2007).
[arXiv:0711.0054 [hep-th]].
}

\lref\SpirWarnaar{
  V.~P.~Spiridonov and S.~O.~Warnaar,
  ``Inversions of integral operators and elliptic beta integrals on root systems,''
Adv. Math. 207 (2006), 91-132
[arXiv:math/0411044].
}

\lref\SethiPA{
  S.~Sethi and M.~Stern,
 ``D-brane bound states redux,''
Commun.\ Math.\ Phys.\  {\bf 194}, 675 (1998).
[hep-th/9705046].
}

\lref\GerchkovitzGTA{
  E.~Gerchkovitz, J.~Gomis and Z.~Komargodski,
 ``Sphere Partition Functions and the Zamolodchikov Metric,''
[arXiv:1405.7271 [hep-th]].
}

\lref\GaiottoHG{
  D.~Gaiotto, G.~W.~Moore and A.~Neitzke,
  ``Wall-crossing, Hitchin Systems, and the WKB Approximation,''
[arXiv:0907.3987 [hep-th]].
}

\lref\RuijsenaarsVQ{
  S.~N.~M.~Ruijsenaars and H.~Schneider,
  ``A New Class Of Integrable Systems And Its Relation To Solitons,''
Annals Phys.\  {\bf 170}, 370 (1986).
}

\lref\GaiottoAK{
  D.~Gaiotto and E.~Witten,
  ``S-Duality of Boundary Conditions In N=4 Super Yang-Mills Theory,''
Adv.\ Theor.\ Math.\ Phys.\  {\bf 13}, 721 (2009).
[arXiv:0807.3720 [hep-th]].
}

\lref\RuijsenaarsPP{
  S.~N.~M.~Ruijsenaars,
  ``Complete Integrability Of Relativistic Calogero-moser Systems And Elliptic Function Identities,''
Commun.\ Math.\ Phys.\  {\bf 110}, 191 (1987).
}

\lref\HallnasNB{
  M.~Hallnas and S.~Ruijsenaars,
  ``Kernel functions and Baecklund transformations for relativistic Calogero-Moser and Toda systems,''
J.\ Math.\ Phys.\  {\bf 53}, 123512 (2012).
}

\lref\kernelA{
S.~Ruijsenaars,
  ``Elliptic integrable systems of Calogero-Moser type: Some new results on joint eigenfunctions'', in Proceedings of the 2004 Kyoto Workshop on "Elliptic integrable systems", (M. Noumi, K. Takasaki, Eds.), Rokko Lectures in Math., no. 18, Dept. of Math., Kobe Univ.
}

\lref\ellRSreview{
Y.~Komori and S.~Ruijsenaars,
  ``Elliptic integrable systems of Calogero-Moser type: A survey'', in Proceedings of the 2004 Kyoto Workshop on "Elliptic integrable systems", (M. Noumi, K. Takasaki, Eds.), Rokko Lectures in Math., no. 18, Dept. of Math., Kobe Univ.
}

\lref\langmann{
E.~Langmann,
  ``An explicit solution of the (quantum) elliptic Calogero-Sutherland model'', [arXiv:math-ph/0407050].
}

\lref\TachikawaWI{
  Y.~Tachikawa,
  ``4d partition function on $S^1 \times S^3$ and 2d Yang-Mills with nonzero area,''
PTEP {\bf 2013}, 013B01 (2013).
[arXiv:1207.3497 [hep-th]].
}

\lref\MinahanFG{
  J.~A.~Minahan and D.~Nemeschansky,
  ``An N=2 superconformal fixed point with E(6) global symmetry,''
Nucl.\ Phys.\ B {\bf 482}, 142 (1996).
[hep-th/9608047].
}

\lref\AldayKDA{
  L.~F.~Alday, M.~Bullimore, M.~Fluder and L.~Hollands,
  ``Surface defects, the superconformal index and q-deformed Yang-Mills,''
[arXiv:1303.4460 [hep-th]].
}

\lref\FukudaJR{
  Y.~Fukuda, T.~Kawano and N.~Matsumiya,
  ``5D SYM and 2D q-Deformed YM,''
Nucl.\ Phys.\ B {\bf 869}, 493 (2013).
[arXiv:1210.2855 [hep-th]].
}

\lref\HellermanZS{
  S.~Hellerman, A.~Henriques, T.~Pantev, E.~Sharpe and M.~Ando,
  ``Cluster decomposition, T-duality, and gerby CFT's,''
Adv.\ Theor.\ Math.\ Phys.\  {\bf 11}, no. 5, 751 (2007).
[hep-th/0606034].
}

\lref\XieHS{
  D.~Xie,
  ``General Argyres-Douglas Theory,''
JHEP {\bf 1301}, 100 (2013).
[arXiv:1204.2270 [hep-th]].
}

\lref\DrukkerSR{
  N.~Drukker, T.~Okuda and F.~Passerini,
  ``Exact results for vortex loop operators in 3d supersymmetric theories,''
[arXiv:1211.3409 [hep-th]].
}

\lref\qinteg{
  M.~Rahman, A.~Verma,
  ``A q-integral representation of Rogers' q-ultraspherical polynomials and some applications,''
Constructive Approximation
1986, Volume 2, Issue 1.
}

\lref\qintegOK{
  A.~Okounkov,
  ``(Shifted) Macdonald Polynomials: q-Integral Representation and Combinatorial Formula,''
Compositio Mathematica
June 1998, Volume 112, Issue 2. 
[arXiv:q-alg/9605013].
}

\lref\macNest{
 H.~Awata, S.~Odake, J.~Shiraishi,
  ``Integral Representations of the Macdonald Symmetric Functions,''
Commun. Math. Phys. 179 (1996) 647.
[arXiv:q-alg/9506006].
}

\lref\BeemYN{
  C.~Beem and A.~Gadde,
  ``The superconformal index of N=1 class S fixed points,''
[arXiv:1212.1467 [hep-th]].
}

\lref\GaddeFMA{
  A.~Gadde, K.~Maruyoshi, Y.~Tachikawa and W.~Yan,
  ``New N=1 Dualities,''
JHEP {\bf 1306}, 056 (2013).
[arXiv:1303.0836 [hep-th]].
}

\lref\BeniniNDA{
  F.~Benini, R.~Eager, K.~Hori and Y.~Tachikawa,
  ``Elliptic genera of two-dimensional N=2 gauge theories with rank-one gauge groups,''
Lett.\ Math.\ Phys.\  {\bf 104}, 465 (2014).
[arXiv:1305.0533 [hep-th]].
}

\lref\DiFrancescoTY{
  P.~Di Francesco, O.~Aharony and S.~Yankielowicz,
  ``Elliptic genera and the Landau-Ginzburg approach to N=2 orbifolds,''
Nucl.\ Phys.\ B {\bf 411}, 584 (1994).
[hep-th/9306157].
}

\lref\NekrasovUH{
  N.~A.~Nekrasov and S.~L.~Shatashvili,
  ``Supersymmetric vacua and Bethe ansatz,''
Nucl.\ Phys.\ Proc.\ Suppl.\  {\bf 192-193}, 91 (2009).
[arXiv:0901.4744 [hep-th]].
}

\lref\GorskyTN{
  A.~Gorsky,
  ``Dualities in integrable systems and N=2 SUSY theories,''
J.\ Phys.\ A {\bf 34}, 2389 (2001).
[hep-th/9911037].
}

\lref\NekrasovXAA{
  N.~A.~Nekrasov and S.~L.~Shatashvili,
  ``Bethe/Gauge correspondence on curved spaces,''
JHEP {\bf 1501}, 100 (2015).
[arXiv:1405.6046 [hep-th]].
}

\lref\IntriligatorLCA{
  K.~Intriligator and N.~Seiberg,
  ``Aspects of 3d N=2 Chern-Simons-Matter Theories,''
JHEP {\bf 1307}, 079 (2013).
[arXiv:1305.1633 [hep-th]].
}

\lref\FockAE{
  V.~Fock, A.~Gorsky, N.~Nekrasov and V.~Rubtsov,
  ``Duality in integrable systems and gauge theories,''
JHEP {\bf 0007}, 028 (2000).
[hep-th/9906235].
}

\lref\CsakiCU{
  C.~Csaki, M.~Schmaltz, W.~Skiba and J.~Terning,
  ``Selfdual N=1 SUSY gauge theories,''
Phys.\ Rev.\ D {\bf 56}, 1228 (1997).
[hep-th/9701191].
}

\lref\BeniniUI{
  F.~Benini and S.~Cremonesi,
  ``Partition functions of $N=(2,2)$ gauge theories on $S^2$ and vortices,''
Commun.\ Math.\ Phys.\  {\bf 334}, no. 3, 1483 (2015).
[arXiv:1206.2356 [hep-th]].
}

\lref\DoroudXW{
  N.~Doroud, J.~Gomis, B.~Le Floch and S.~Lee,
  ``Exact Results in D=2 Supersymmetric Gauge Theories,''
JHEP {\bf 1305}, 093 (2013).
[arXiv:1206.2606 [hep-th]].
}

\lref\GomisWY{
  J.~Gomis and S.~Lee,
  ``Exact Kahler Potential from Gauge Theory and Mirror Symmetry,''
JHEP {\bf 1304}, 019 (2013).
[arXiv:1210.6022 [hep-th]].
}

\lref\GomisYAA{
  J.~Gomis, P.~S.~Hsin, Z.~Komargodski, A.~Schwimmer, N.~Seiberg and S.~Theisen,
  ``Anomalies, Conformal Manifolds, and Spheres,''
JHEP {\bf 1603}, 022 (2016).
[arXiv:1509.08511 [hep-th]].
}

\lref\HwangNOP{
  C.~Hwang and P.~Yi,
  ``Twisted Partition Functions and $H$-Saddles,''
JHEP {\bf 1706}, 045 (2017).
[arXiv:1704.08285 [hep-th]].
}

\lref\AganagicUW{
  M.~Aganagic, K.~Hori, A.~Karch and D.~Tong,
  ``Mirror symmetry in (2+1)-dimensions and (1+1)-dimensions,''
JHEP {\bf 0107}, 022 (2001).
[hep-th/0105075].
}

\lref\HoriKT{
  K.~Hori and C.~Vafa,
  ``Mirror symmetry,''
[hep-th/0002222].
}

\lref\SeibergBD{
  N.~Seiberg,
  ``Five-dimensional SUSY field theories, nontrivial fixed points and string dynamics,''
Phys.\ Lett.\ B {\bf 388}, 753 (1996).
[hep-th/9608111].
}

\lref\AharonyDHA{
  O.~Aharony, S.~S.~Razamat, N.~Seiberg and B.~Willett,
  ``3d dualities from 4d dualities,''
JHEP {\bf 1307}, 149 (2013).
[arXiv:1305.3924 [hep-th]].
}

\lref\HoriPD{
  K.~Hori,
  ``Duality In Two-Dimensional (2,2) Supersymmetric Non-Abelian Gauge Theories,''
JHEP {\bf 1310}, 121 (2013).
[arXiv:1104.2853 [hep-th]].
}

\lref\DoroudPKA{
  N.~Doroud and J.~Gomis,
  ``Gauge Theory Dynamics and Kahler Potential for Calabi-Yau Complex Moduli,''
JHEP {\bf 1312}, 099 (2013).
[arXiv:1309.2305 [hep-th]].
}

\lref\WittenYC{
  E.~Witten,
  ``Phases of N=2 theories in two-dimensions,''
Nucl.\ Phys.\ B {\bf 403}, 159 (1993).
[hep-th/9301042].
}

\lref\GerchkovitzGTA{
  E.~Gerchkovitz, J.~Gomis and Z.~Komargodski,
  ``Sphere Partition Functions and the Zamolodchikov Metric,''
JHEP {\bf 1411}, 001 (2014).
[arXiv:1405.7271 [hep-th]].
}

\lref\ImamuraSU{
  Y.~Imamura and S.~Yokoyama,
  ``Index for three dimensional superconformal field theories with general R-charge assignments,''
JHEP {\bf 1104}, 007 (2011).
[arXiv:1101.0557 [hep-th]].
}

\lref\DimofteJU{
  T.~Dimofte, D.~Gaiotto and S.~Gukov,
  ``Gauge Theories Labelled by Three-Manifolds,''
Commun.\ Math.\ Phys.\  {\bf 325}, 367 (2014).
[arXiv:1108.4389 [hep-th]].
}

\lref\qformsym{
D.~Gaiotto, A.~Kapustin, N.~Seiberg, B.~Willett, 
``Generalized global symmetries,''
JHEP {\bf 1502}, 172 (2015).
[arXiv:1412.5148 [hep-th]].
}

\lref\BeniniMIA{
  F.~Benini, D.~S.~Park and P.~Zhao,
  ``Cluster algebras from dualities of 2d N=(2,2) quiver gauge theories,''
Commun.\ Math.\ Phys.\  {\bf 340}, 47 (2015).
[arXiv:1406.2699 [hep-th]].
}

\lref\DoreyRB{
  N.~Dorey and D.~Tong,
  ``Mirror symmetry and toric geometry in three-dimensional gauge theories,''
JHEP {\bf 0005}, 018 (2000).
[hep-th/9911094].
}

\lref\DiPietroBCA{
  L.~Di Pietro and Z.~Komargodski,
  ``Cardy Formulae for SUSY Theories in d=4 and d=6,''
JHEP {\bf 1412}, 031 (2014).
[arXiv:1407.6061 [hep-th]].
}

\lref\DijkgraafVV{
  R.~Dijkgraaf, E.~P.~Verlinde and H.~L.~Verlinde,
  ``Matrix string theory,''
Nucl.\ Phys.\ B {\bf 500}, 43 (1997).
[hep-th/9703030].
}

\lref\HoriAX{
  K.~Hori and A.~Kapustin,
  ``Duality of the fermionic 2-D black hole and N=2 liouville theory as mirror symmetry,''
JHEP {\bf 0108}, 045 (2001).
[hep-th/0104202].
}

\lref\GiveonPX{
  A.~Giveon and D.~Kutasov,
  ``Little string theory in a double scaling limit,''
JHEP {\bf 9910}, 034 (1999).
[hep-th/9909110].
}

\lref\HoriEWA{
  K.~Hori, C.~Y.~Park and Y.~Tachikawa,
  ``2d SCFTs from M2-branes,''
JHEP {\bf 1311}, 147 (2013).
[arXiv:1309.3036 [hep-th]].
}

\lref\ClossetVVL{
  C.~Closset, N.~Mekareeya and D.~S.~Park,
  ``A-twisted correlators and Hori dualities,''
JHEP {\bf 1708}, 101 (2017).
[arXiv:1705.04137 [hep-th]].
}

\Title{\vbox{\baselineskip12pt
}}
{\vbox{\centerline{From $3d$ duality to $2d$ duality}
\vskip7pt
\centerline{}
}
}

\centerline{Ofer Aharony,$^a$ Shlomo S. Razamat,$^b$ and Brian Willett$^{c}$}
\bigskip
\centerline{$^a$ {\it Department of Particle Physics and Astrophysics, Weizmann Institute of Science, Rehovot 76100, Israel}}
\centerline{$^b$ {\it Department of Physics, Technion, Haifa,  32000, Israel}}
\centerline{$^c$ {\it KITP, Santa Barbara, CA, USA}}

\vskip.1in \vskip.2in \centerline{\bf Abstract}

In this paper we discuss $3d$ ${\cal N}=2$ supersymmetric gauge theories and their IR dualities when they are compactified  on a circle of radius $r$, and when we take the $2d$ limit in which $r\to 0$. The $2d$ limit depends on how the mass parameters are scaled as $r\to 0$, and often vacua become infinitely distant in the $2d$ limit, leading to a direct sum of different $2d$ theories. For generic mass parameters, when we take the same limit on both sides of a duality, we obtain $2d$ dualities (between gauge theories and/or Landau-Ginzburg theories) that pass all the usual tests. However, when there are non-compact branches the discussion is subtle because the metric on the moduli space, which is not controlled by supersymmetry, plays an important role in the low-energy dynamics after compactification. Generally speaking, for IR dualities of gauge theories, we conjecture that dualities involving non-compact Higgs branches survive. On the other hand when there is a non-compact Coulomb branch on at least one side of the duality, the duality fails already when the $3d$ theories are compactified on a circle. Using the valid reductions we reproduce many known $2d$ IR dualities, giving further evidence for their validity, and we also find new $2d$ dualities.

\vskip.2in

\noindent

\vfill

\Date{October 2017}

\newsec{Introduction and summary}

Often different high energy theories are equivalent at low energies. This universality has been observed to happen in many examples of supersymmetric gauge theories in various dimensions. The mechanism responsible for these dualities is still poorly understood in spite of the plethora of known examples. To try to sieve through the large amount of cases to extract the essential properties, it is useful to understand what are the minimal sets of dualities from which all the rest can be derived. A useful guiding principle in the search for such a minimal set is to start with instances with the largest number of degrees of freedom, and then derive other examples by getting rid of some of the degrees of freedom. 

In this paper we continue our program of assuming the validity of dualities between $d$-dimensional quantum field theories with four supercharges and reducing them to lower dimensions. The goal is to understand the fate of the dualities in this reduction and try to derive known and new dualities in lower dimensions. In \AharonyDHA\ this strategy was applied to reductions on a circle from four to three dimensions. Indeed, it was possible to derive all the known (non-mirror) IR equivalences in three dimensions starting from four, and we also found new dualities. Here we will discuss the next step in the program, namely further reduction of three dimensional theories on a circle.  As we will explain here, two dimensional conformal theories present us with new challenges which significantly complicate the answers to the posed question. In the rest of the introduction we will detail the new issues one encounters in two dimensions and in three dimensional theories on a circle, and briefly summarize our results.

\subsec{New issues in two dimensional theories}

There are several new issues arising in two dimensions compared to higher
dimensions, which complicate the analysis of the IR behavior of $2d$ theories
and of $3d$ theories on a circle.

Supersymmetric theories with four supercharges often have a classical moduli  space of vacua (at least for some value of their parameters), and $2d$ theories
are no exception. This space can be modified or lifted by quantum corrections.
In higher dimensions, there is a vacuum for each point on the quantum moduli
space, which is in a different superselection sector from other points.  Moreover, the low-energy theory at generic points on the moduli space is
free, while interacting theories arise at low energies only at singular points on the moduli space. On the
other hand, in $2d$, because of quantum fluctuations, the ground state (and
all other states) explore all of the `moduli space' (which is called a target space).
Furthermore, the metric is classically marginal in $2d$ (unlike in higher dimensions, where it is always irrelevant), and
generally acquires a beta function quantum mechanically, meaning that it
cannot be ignored in an analysis of the IR behavior. 
Thus, in $2d$, knowledge of the full target space, and in
particular of the metric on this target space, is required for understanding
the IR behavior.  
This is particularly significant when the target space is non-compact, or when there is a runaway behavior such that the only supersymmetric
configurations are infinitely far away.
In such cases there is generally no normalizable ground state, but
there can be a continuum of states going down to zero energy
(depending on the K\"ahler potential at infinity). Such a
continuum can contribute to supersymmetric indices in ways that differ
from discrete vacua; for instance it can lead to fractional contributions
to the Witten index \BeniniNDA.

An interesting possibility that can occur in two dimensions and not in
higher dimensions is that a given high-energy supersymmetric
theory can flow to more than one superconformal theory (SCFT) at low
energies (namely, it flows to a direct sum of decoupled superconformal
theories, in which the Hilbert space is the sum of the two Hilbert spaces and not their product) \refs{\AharonyTH,\WittenYU,\HellermanZS}. Note that this is stronger than the decoupling between
different points (`vacua') on a connected moduli space in higher dimensions,
where one can construct large regions of one vacuum inside the other
at arbitrarily low energies. In general the decoupled SCFTs
may have different central charges. When they arise from
different regions in the classical moduli space, there are often non-compact semi-infinite `throats'
in the target spaces of each of the disconnected SCFTs, located where
the target spaces were classically connected.

\subsec{New issues in three dimensional theories on a circle}

When we compactify a $3d$ ${\cal N}=2$ supersymmetric (SUSY) theory on a circle
of radius $r$ and go to low energies compared to $1/r$, we can
describe the theory by a two dimensional effective action, involving
the two dimensional light fields.  Many of the characteristic features of two dimensional field theories discussed above then become important when describing the system.

The theory on a circle can have various dimensionless parameters, such as $g^2 r$ for $3d$ gauge couplings $g$, or $mr$ and $\zeta r$ for $3d$ masses $m$ and Fayet-Iliopoulos (FI)
terms $\zeta$. Naively one expects that the low-energy
theory should be independent of $r$, and thus of these dimensionless
parameters.  In reductions from $4d$ to $3d$ this is essentially true, and
in many cases one obtains at low energies (or as $r\to 0$)
the IR limit of the $3d$ theory that one obtains by dimensionally
reducing the $4d$ theory. For instance, in a $4d$ gauge theory
one can take $r\to 0$ keeping the $3d$ gauge coupling fixed;
generally $3d$ theories do not have any symmetry-preserving marginal deformations,
so any other way of taking $r\to 0$ must give the same answer.
However, one has to be aware of the following fact.
When reducing a $4d$
gauge field to $3d$ one generates two new scalar fields, from
the holonomy of the gauge field and from the dual of the $3d$
gauge field. So the moduli spaces of $3d$ theories are larger
than those of their $3d$ parents (classically, and often also
quantum mechanically), and even if one starts from
a specific point on the $4d$ moduli space, one has to specify
precisely which point one is analyzing on the $3d$ moduli space;
the low-energy effective theory can be different at different
points.

When going down from $3d$ to $2d$, naively the situation
is simpler since no new scalar degrees of freedom arise.
However, the situation is actually more complicated. One
reason is that
the K\"ahler potential cannot be ignored, and it often leads
to $2d$ marginal deformations, implying that different ways of
taking $r\to 0$ can lead to different CFTs, differing by
marginal deformations. As a simple example consider a
free $3d$ vector multiplet with coupling constant $e$ 
on a circle. In $3d$ one can
dualize the vector field to a compact free scalar field.
Upon reduction on a circle, this scalar field remains
compact, and it is the T-dual of the scalar field coming
from the holonomy of the gauge field on a circle. Canonically
normalizing this scalar field (and ignoring numerical constants), the radius of its target space is
$\sqrt{e^2 r}$ (or, equivalently, $1 / \sqrt{e^2 r}$). Thus,
already in this free case, the low-energy theories that one
gets for different dimensionless parameters $e^2 r$ are not
the same. There is a specific $r\to 0$ limit where one keeps
fixed the $2d$ gauge coupling $e^2 / r$, and this limit
has the same low-energy limit as that of the corresponding
$2d$ gauge theory, but other $r\to 0$ limits give different results.  Similar statements apply to more general gauge theories, and, in addition, the parameters $r m$ and $r \zeta$ can lead to marginal twisted chiral deformations in the $2d$ description.  

Thus, when we take an $r\to 0$ limit, it is important to specify
precisely how all the parameters -- masses, FI terms and gauge
couplings -- scale as a function of $r$. Different scalings
can give different theories; these can then have marginal parameters
arising from further changes in the parameters (that do not
change their scaling with $r$). For gauge theories, one natural `scaling limit' is the one
related to a $2d$ gauge theory. In this limit it is natural
to keep masses of charged fields fixed as $r\to 0$, and to scale $3d$ FI terms
as $1/r$ (with possible additional logarithmic dependence
on $r$ due to the running of the $2d$ FI term). However,
other limits are also interesting. In particular, $3d$ mirror
symmetry exchanges masses with FI terms, so a natural
limit for some $3d$ gauge theory maps to an `unnatural'
limit for its mirror theory.

In addition, regions of the moduli space which are separated
by distances of order $1/r$ can become separated and
decouple in the IR as $r\to 0$ (in addition to the issue mentioned above,
where single $2d$ gauge theories lead at low energies to decoupled branches). So we often get a direct sum of several $2d$ theories arising from the reduction of a single $3d$ theory.
In particular, note that the scalar field coming from the
holonomy of a $3d$ gauge field has periodicity proportional to $1/r$
(when normalized in the same way as $2d$ vector multiplets),
so this distance scale naturally appears in the reduction of
gauge theories.  This decoupling can be seen even in the mass-deformed version of the theory; there one finds various sets of discrete vacua, where the difference of the (twisted) superpotential between different sets diverges as $r \to 0$, implying that the BPS solitons interpolating between these vacua become infinitely massive, and they decouple.

It is interesting to ask whether the $2d$ low-energy theories that we get as
$r\to 0$ can also be described as low-energy limits of
well-defined $2d$ theories, which we can call their
`UV completions'. For the `natural scaling limit' described
above one expects such a UV completion to be given by
the corresponding $2d$ gauge theory. However, when
there are marginal K\"ahler deformations, 
one has to check whether the K\"ahler potential arising from the
$2d$ UV completion is the same as that of the $3d$
theory on a circle. In particular, this may be true for
some specific scaling of $e^2$ with $r$, but not for other
scalings. In other scaling limits the $2d$ gauge theory does
not provide a UV completion, but we will see that there are
often other UV completions, that can be Landau-Ginzburg
(LG) type theories of twisted chiral superfields with some
twisted superpotential.  A useful method for building such UV completions is to study properties of the theory which are protected by supersymmetry, such as the (twisted) chiral ring and supersymmetric partition functions.   Often one is able to construct a $2d$ theory with the same protected data as the $3d$ theory on a circle, which gives an indication they have the same low energy description.  However, this only gives partial evidence; for example, as mentioned above, one may also need to match
the K\"ahler potentials between the $2d$ UV completion
and the $3d$ theory on a circle.

\subsec{New issues in reduction of dualities}

In \AharonyDHA\ it was found that the reduction of a $4d$ low-energy duality on
a circle is subtle, and while it can be related to $3d$ dualities, their
form is often not the same as that of the $4d$ duality. One
reason for this was mentioned above -- the $3d$ moduli space is
larger than the $4d$ moduli space, so one has to be careful about
how to map different points on the $3d$ moduli space across the
duality.  Another issue is that the $r\to 0$ limit does not commute with the
low-energy limit in $4d$, in which the two theories are equivalent.
Dualities between asymptotically free $4d$ gauge theories are
valid below their dynamical scales $\Lambda$, ${\tilde \Lambda}$.
However, if one attempts to take $r\to 0$ while staying below
both of these scales, one finds that one cannot keep both $3d$
gauge couplings fixed, so one does not obtain the IR limit of the
corresponding $3d$ gauge theories. As discussed in \AharonyDHA, in order
to preserve the duality one must then add additional terms to the
action on both sides.

As mentioned above, when reducing $3d$ dualities to $2d$, 
the dimension of the moduli space does not increase, though we still need to worry about precisely where in the moduli space we end up as $r\to 0$.
However, the second issue becomes much
worse. When two $3d$ gauge theories with gauge couplings $e^2$,
${\tilde e}^2$ are dual, this typically means that their low-energy behavior
at the origin of the $3d$ moduli space is the same. Far out on the
moduli space, the metric on the moduli space is usually different --
for instance, scalar fields coming from dual photons generally
have a periodicity proportional to $e$ --  but the low-energy flat space
field theory in such regions is free, so this has no effect on the low-energy
equivalence of dual theories.  However, upon reduction on a circle, the metric on the moduli space becomes
a classically marginal deformation, and the theory explores all parts
of the moduli space, so the fact that the metrics far on the moduli
space differ becomes important. Generally, when there are
non-compact target spaces (or runaways), this effect destroys
the low-energy equivalence of dual theories for any radius $r$.
Indeed, we saw above that even theories of a free $3d$ vector multiplet
with different gauge couplings on a circle are not equivalent at low energies,
let alone any duals that they may have.

For generic values of the parameters the vacua are discrete
and this problem does not arise. Since we know the mapping between
the $3d$ parameters across dualities, we can match their discrete
vacua for any radius, and even as $r\to 0$, if we perform the same
scaling of the parameters on both sides. Generally the low-energy
theories at such discrete vacua are trivial, but we can sometimes tune
parameters so that they become non-trivial SCFTs like LG models.
But problems arise when we take limits in which the vacua
become continuous, since then we have to make sure that we
obtain the same K\"ahler potential on both sides of a duality.
In particular, even when we can provide a $2d$ completion for
both sides of a duality between $3d$ theories on a circle, it is
not always the case that the two $2d$ completions lead to the same
K\"ahler potential at low energies, and, when they do not, we will
not obtain a $2d$ duality between these completions. Note that, as discussed in \AharonyJKI, even when the classical K\"ahler potentials are different on two sides of the duality, and even when they differ asymptotically, their renormalization group flows may still lead to the same K\"ahler potential at low energies (at any finite position on the target space).

\subsec{Outline of the paper and summary of the results}

Let us now outline the structure of the paper and detail briefly the new results. We start the discussion in Section $2$ by addressing in detail general properties of reductions of three dimensional theories on a circle in the relatively simpler case where the moduli spaces are lifted to discrete vacua by mass deformations. In particular, we stress the existence of different scaling limits of parameters when the radius of the circle is shrunk. In this section we also introduce the technology of computing the twisted superpotential for theories on a circle, which is our main quantitative tool. We discuss here several examples of reductions of theories on a circle, and claim that typically such reductions should be thought of as giving rise to a direct sum of theories, possibly with different central charges, in two dimensions.   In Section $3$ we discuss the fate of three dimensional dualities  with discrete vacua (no non-compact branches) when these are considered on a circle.  In particular we derive several known dualities in two dimensions starting from three dimensions, as well as one instance of a new duality, between $SU(N_c)$ gauge theories that have both fundamental and anti-fundamental flavors. 
In Section $4$ we discuss  dualities with a continuous spectrum (e.g. with non-compact branches), and new issues that arise there.
 Some appendices complement the text with background material, details of computations, and additional examples of the statements presented in the bulk. 

The main results of the paper include derivations of known and new dualities in two dimensions starting from $3d$. There are also many questions for which we do not provide conclusive verdicts.   
When the moduli space is compact so that the spectrum is discrete, one can derive a consistent set of arguments in favor of $2d$ dualities arising from the compactification. When there is a non-compact Higgs branch, we provide considerable evidence that $2d$ dualities still arise, but it is not conclusive. On the other hand, when at least one side of the duality has a non-compact Coulomb branch, the $3d$ duality does not reduce to a $2d$ duality in general, even though some protected objects do still match, suggesting that it may be possible to fix the reduction in some way. We discuss various examples, including cases where both non-compact Higgs and non-compact Coulomb branches exist.
At the technical level we discuss how the subtlety in the reductions can be traced into issues with commuting the limit of performing the computation of partition functions, which involves integrals and infinite sums, and the   limit of taking the radius to zero.\foot{ For a related recent discussion see \HwangNOP.}  

It would be interesting to understand these issues better, and to study many additional examples. It would also be interesting to generalize our results to theories with larger or smaller amounts of supersymmetry.

\newsec{Massive $3d$ $\cN=2$ theories on a circle}

In this section we discuss the properties of $3d$ $\cN=2$ theories upon reduction on a circle, in the case where any continuous moduli space is lifted by real mass parameters.  The presence of a continuous moduli space leads to additional complications which we discuss in Section $4$. Already in the case where the moduli space is lifted there are various subtleties in the reduction, both in the case of a single theory, considered in the present section, and for a pair of dual theories, which we discuss in the following section.  

Our main tool for describing the reduction of the mass-deformed theory will be the effective twisted superpotential of the theory compactified on a circle of radius $r$, which we introduce next.  Related observables are the supersymmetric partition functions on $\Sigma_g \times S^1$, including the supersymmetric index (the $\S^2 \times \S^1$ partition function).  These observables, which we will describe in more detail below, turn out to be sensitive only to the information contained in the mass-deformed theory, and we will need more refined observables to probe the theories with continuous moduli spaces.

\subsec {Effective twisted superpotential for a $3d$ theory on a circle}

As reviewed in Appendix A, the twisted superpotential of a $2d$ $\cN=(2,2)$ theory depends only on the twisted chiral multipets in the theory, and is protected under renormalization group flow.  Given a $3d$ $\cN=2$ gauge theory, we may study the effective twisted superpotential of the compactified theory on $\R^2 \times \S^1_r$, which can be thought of as a $2d$ $\cN=(2,2)$ theory with an infinite number of fields.  This provides information about the vacua of the theory in flat space, and is closely related to certain supersymmetric partition functions on compact manifolds \refs{\NekrasovUH,\NekrasovXAA,\ClossetARN,\BeniniHJO,\ClossetZGF,\GukovGKN}.\foot{The twisted superpotential has also played an important role in understanding the algebra of loop operators in $3d$ \refs{\KapustinHPK,\GaddeWQ,\ClossetARN}, and in the context of the $3d-3d$ correspondence \refs{\GukovSNA,\GukovGKN,\GukovKMK}.}  In the present context, we will use it to understand the reduction of $3d$ $\cN=2$ theories on $\S^1_r$ by studying the limit of this function as $r \rightarrow 0$.  A similar approach was taken in \AganagicUW, and we recover some of their results below. 

Given a $3d$ $\cN=2$ gauge theory compactified on $\S^1_r$, we may compute the effective twisted superpotential by summing over the contributions from the massive fields in the Kaluza-Klein (KK) towers.  As discussed in Appendix A, this is a function of twisted chiral field strength multiplets for the gauge and flavor symmetries, $G$ and $H$, which we denote by $\Sigma_i$, $i=1,...,r_G$, and $m_a$, $a=1,...,r_H$, respectively, where we work in some basis of the Cartan subalgebras of the respective groups.  The former multiplets are dynamical, and the latter may be viewed as fixed background fields. The bottom component of these twisted chiral multiplets is a complex scalar field which is built out of the $3d$ fields as:

\eqn\sdef{ \Sigma_i = \sigma^R_i + i {A_3}_i, \;\;\;\; m_a = m^R_a + i {A^{BG}_3}_a  }
where $\sigma^R_i$ is the real scalar in $3d$, ${A_3}_i$ is the component of the gauge field along $\S^1$, and similarly for the background parameters, $m^R_a$ is a $3d$ `real mass' parameter and ${A^{BG}}_a$ the background gauge field coupling to the corresponding flavor symmetry.  These fields are periodically identified due to large gauge transformations around the $\S^1$:
\eqn\largegauge{
\Sigma_i \sim \Sigma_i + \frac{i}{r}, \;\;\; m_a \sim m_a + \frac{i}{r} \,.}
We will interchangeably use these variables to refer to either the twisted chiral multiplet or its bottom scalar component.  It is also sometimes useful to work in terms of the gauge-invariant (up to Weyl transformations) fields $x_i \equiv e^{2 \pi r \Sigma_i}$ and the parameters $\nu_a \equiv e^{2 \pi r m_a}$.

For a general $3d$ $\cN=2$ gauge theory on a circle, the twisted superpotential is given by:
\eqn\threedwdef{
\eqalign{
\cW&(\Sigma_i,m_a)  = \sum_\alpha \cW^{(3d)}_\chi(Q^i_\alpha\Sigma_i + S^a_\alpha m_a)  \cr
& + \pi r  \sum_{i,j} k_{GG}^{ij} (\Sigma_i \Sigma_j + \delta_{i,j} \frac{i}{r} \Sigma_i) +\pi r \sum_{a,b} k_{FF}^{ab} (m_a m_b + \delta_{a,b} \frac{i}{r} m_a) + 2 \pi r \sum_{a,i} k_{FG}^{ai} m_a \Sigma_i\,.
}}
Here the first term is the contribution of the chiral multiplets, where the $\alpha$'th chiral has gauge charges $Q^i_\alpha$ and flavor charges $S^a_\alpha$, and $\cW_\chi^{(3d)}$ is the contribution of a single $3d$ chiral multiplet. This can be computed by summing the contributions from the KK modes:
\eqn\chiralcont{
\cW_\chi^{(3d)}(\Sigma) = \sum_{n \in \Z} W_{\chi}^{(2d)}\left(\Sigma + \frac{in}{r}\right) = {1 \over 2 \pi r} {\rm Li}_2(e^{-2 \pi r \Sigma}) \,,
}
where $W_{\chi}^{(2d)}(\Sigma)=\Sigma (\log(\Sigma/\mu)-1)$, as discussed in Appendix A.  When performing this infinite sum, one must regularize it in a way which sacrifices either parity or gauge-invariance; we have chosen to preserve the latter, and in our convention the chiral multiplet contribution \chiralcont\ includes an effective ``level $-\frac{1}{2}$ Chern-Simons term'' for the gauge field coupled to it, which breaks parity.\foot{This name is a short-hand notation for regularizing the theory with the chiral multiplet in a way that leads to pure Chern-Simons theories with levels $0$ or $(-1)$ upon giving a mass to the chiral multiplet.}  The remaining terms in \threedwdef\ are the contributions of the bare Chern-Simons (CS) terms, with  levels $k_{GG}^{ij} ,k_{FF}^{ab} ,k_{FG}^{ai}$ for the gauge-gauge, flavor-flavor, and gauge-flavor CS terms, respectively.  With our convention for the normalization of the twisted chiral multiplets, these levels must all be integers.

The effective twisted superpotential determines the supersymmetric vacua of the theory via the equations:
\eqn\vaceqn{
1 = \exp \bigg( {{\partial \cW}\over{\partial \Sigma_i }}\bigg), \;\;\;\; i =1,\cdots,r_G\,. }
Using:
\eqn\wchiderivative{
 {{d \cW^{(3d)}_\chi(\Sigma)} \over {d \Sigma}} = \log (1- e^{-2\pi r \Sigma})\,, 
}
one finds that the right-hand side of \vaceqn\ is a rational function of $x_i =e^{2 \pi r \Sigma_i}$ and $\nu_a =e^{2 \pi r m_a}$.  Thus the vacua are determined by solving a system of polynomial equations.  In the non-Abelian case, we discard any solutions which are not acted on freely by the Weyl symmetry, as these naive vacua, which have enhanced gauge symmetry, break supersymmetry \refs{\HoriDK,\AharonyJKI} (the effective action \threedwdef\ is not valid there).

In the following, we will study the reduction of $3d$ theories by considering the $r \rightarrow 0$ limit of the effective twisted superpotential.  To take a simple example, consider a free chiral multiplet, which we may couple to a background gauge field with charge $1$, giving it an effective real mass, $m$.  This has the effective twisted superpotential $\cW_\chi^{(3d)}(m)$ in \chiralcont.  If we take the $r \rightarrow 0$ limit while holding $m$ fixed we find (up to a constant):

\eqn\chirallimit{
\cW_\chi^{(3d)}(m)  \rightarrow \cW_\chi^{(2d)}(m) + m \log(2 \pi r \mu) \;\;\;\;{\rm as} \; r \rightarrow 0.}
This reflects the fact that as we take $r \rightarrow 0$, the non-zero KK modes become very massive, and can be integrated out, leaving only the zeroth KK mode, which represents a $2d$ chiral multiplet of mass $m$.  The non-zero KK modes generate a divergent linear term in the effective twisted superpotential, which can be interpreted as a shift of the effective (running) FI parameter for this background $U(1)$ symmetry. 

\subsec{General features of the reduction}

Let us now describe some general features of the reduction of massive $3d$ $\cN=2$ theories.  We start with a UV description of a $3d$ $\cN=2$ gauge theory.  This typically has several relevant parameters, which can be classified into gauge couplings, $g_i^2$, real mass parameters, $m_a$, including FI terms for $U(1)$ factors of the gauge group, and superpotential deformations, such as complex masses.  When we consider the system on a finite circle, $\S^1_r$, it is natural to form the dimensionless combinations:
\eqn\parameters{
\gamma_i = g_i^2 r, \;\;\;\; \mu_a = m_a r\,. }

At low energies compared to $r^{-1}$, the system is effectively two dimensional, and the $2d$ physics we obtain will depend on these parameters.  It is also typically strongly coupled, which complicates the analysis of this dependence.  To make progress it will be important to study properties of these models which are protected by supersymmetry.  In theories with extended supersymmetry ($\cN \geq 3$), the metric on the moduli space is not renormalized, which allows us to deduce many aspects of the low energy behavior using the UV description.  For $\cN=2$ theories, however, this metric is generically renormalized, and depends on the $\gamma_i$, which appear in the $D$-terms of the action and so are not protected.  We will return to these issues in Section $4$.

For now, we will consider the theory at generic, non-zero values of the mass parameters, in which case the moduli space is lifted, and the $\gamma_i$ are not relevant for describing the low energy theory.  As we have seen above, the parameters $\mu_a$ arising from the real masses live in protected short ``twisted chiral'' multiplets in $2d$, and control the twisted chiral ring of the theory.  We may study the protected data in the theory using the effective twisted superpotential, introduced above.  This will allow us to deduce many features of the $2d$ low energy theory coming from the $3d$ theory on a circle, and, in many cases, to guess a $2d$ UV description which flows to the same low energy theory.

Given a $3d$ Lagrangian, there are in general many distinct $2d$ theories one may obtain, depending on how one scales the mass parameters with $r$.  Let us write, schematically:
\eqn\miscal{ m_a = m_a(t_a,r) }
where $t_a$ is a finite parameter that controls the specific scaling with $r$, which will become a twisted chiral parameter in the $2d$ low energy description.  
Typical examples are $m_a=t_a$ or $m_a=t_a/r$. Then we may consider the effective twisted superpotential as a function of $r$,
\eqn\wscal{ \cW_{3d} (\Sigma_i,m_a;r) =\cW_{3d}( \Sigma_i,m_a(t_a,r) ;r).}

As we take the limit where $r$ becomes small, different supersymmetric vacua of $\cW_{3d}(r)$  can scale in different ways as a function of $r$. In this section we assume that (for any value of $r$) these vacua are isolated and gapped. In each of these vacua we can (for an appropriate choice of \miscal) obtain a finite effective twisted superpotential near that vacuum as $r\to 0$, after some
redefinition of the gauge variables $\Sigma_i$,
\eqn\uiscal{ \Sigma_i = \Sigma_i(X_i,r), }
where the $X_i$ will become dynamical twisted chiral fields in $2d$.  
Then we find:
\eqn\wlim{\cW_{3d}( \Sigma_i(X_i,r),m_a(t_a,r);r) \rra \cW_{2d}(X_i,t_a) + \cdots }
where the dots denote subleading terms, and/or divergent terms depending only on background fields, which can be removed by the addition of appropriate counterterms.  
In general the redefinition \uiscal\ will be different for different vacua when they become infinitely separated as $r\to 0$; the flow from $3d$ then leads to a direct sum of different theories, one for each set of separated vacua. For instance, some vacua may be at finite values of $\Sigma$, and others at values scaling as $1/r$, while additional vacua may have some $\Sigma_i$'s remaining fixed and others scaling as $1/r$, requiring different redefinitions for different $\Sigma_i$'s. 

We assume here that the K\"ahler potential remains finite as $r\to 0$ in the $2d$ variables used in \wlim; in some cases this will be true for the standard UV K\"ahler potential of the $3d$ gauge theory, while in other cases we may need to also rescale the UV K\"ahler potential for this. In some cases the mass gap around some vacuum may not be small compared to the KK scale $1/r$, and then there is no $2d$ description.

In many cases one may interpret $\cW_{2d}(X_i, t_a)$ as the effective twisted superpotential of some known $2d$ theory.  In some cases one finds that $\cW_{2d}$ is simply quadratic, leading to an uninteresting massive theory, while in other cases $\cW_{2d}$ is the twisted superpotential of a $2d$ gauge theory or interacting Landau-Ginzburg model, and our $r\to 0$ effective theory (when we choose a non-singular K\"ahler potential for it) will be equivalent to the IR limit of that theory. If we start from a $3d$ gauge theory there is always a limit of the parameters in which we keep the $2d$ gauge coupling $g_i^2/r$ and the $2d$ gauge theory parameters fixed (up to possible logarithmic runnings), and if there is a corresponding vacuum that remains at a finite position in the $2d$ gauge theory variables $\Sigma_i$ as $r\to 0$, then this limit gives a $2d$ gauge theory. In general we may obtain a direct sum of this with other theories.

\subsec{Abelian examples}

Having described the general properties of the reduction of massive theories, we now turn to discuss several examples which illustrate these features concretely.  We begin with $3d$ $\cN=2$ gauge theories with Abelian gauge groups.

\

\noindent {a.} {\it $U(1)_{-1/2}$ with one charged chiral superfield}

\

First we consider the $3d$ $U(1)$ theory with a single charged chiral multiplet.\foot{The reduction of this theory using the twisted superpotential was first studied in \AganagicUW; here we review it to illustrate some features of the more general case.}  In order to preserve gauge invariance we must include an appropriate ``half-integer-level Chern-Simons term,'' and we choose $k=-\frac12$.  The theory has one real mass parameter, the Fayet-Iliopolous (FI) parameter $\zeta$.  When $\zeta$ vanishes the moduli space is a semi-infinite cigar labeled by the monopole operator $V_+$, while otherwise there is a unique vacuum (the Witten index is $1$).  The moduli space is smooth such that at any point the low-energy theory is dual to a single free chiral superfield; we will discuss the consequences of this duality in the next section.  When this theory is compactified on a circle, we can parameterize the moduli space by the twisted chiral superfield $\Sigma$ instead of the chiral superfield $V_+$; the two descriptions are related asymptotically by T-duality along the compact direction of the cigar.  

We will consider this theory at a non-zero value of the FI parameter $\zeta$, in which case this moduli space is lifted.  The effective twisted superpotential of this theory is then
\eqn\uonehalfb{
\cW = \cW^{(3d)}_\chi(\Sigma) + 2\pi r \zeta \Sigma
\qquad \Rightarrow \qquad { {\partial \cW}\over{\partial \Sigma}} = \log(1-e^{-2 \pi r \Sigma})+ 2 \pi r \zeta\,.
}
The vacuum equation leads to a single vacuum:
\eqn\uonevac{
\exp\left(\frac{\partial \cW}{\partial \Sigma}\right) = (1-e^{-2 \pi r \Sigma}) e^{2 \pi r \zeta} = 1 \;\;\;\; \Rightarrow \;\;\; \Sigma= -\frac{1}{2 \pi r} \log(1-e^{-2 \pi r \zeta})\,. }

There are multiple $2d$ limits that one can consider. First, we can keep 
$\zeta$ fixed as $r \to 0$.  Then from \uonevac, one finds that $\Sigma$
goes to infinity as $\Sigma \sim -\log(2\pi r \zeta) / 2\pi r$. Thus in this limit it is natural to
define a new variable $X \equiv 2 \pi r \Sigma + \log(2 \pi r \mu)$ that remains
finite as $r\to 0$. In terms of this, we obtain in this $2d$ limit the effective superpotential
\eqn\uonetwod{\cW = \zeta X + \mu  e^{-X} - \zeta \log(2 \pi r \mu) +  O(r)\,.}
Here the third term is a divergent function of the background fields, related to the running FI term, which is generated by integrating out massive KK modes, and can be removed by an appropriate counterterm.  For generic $\zeta$ this model has a single discrete vacuum, and at $\zeta=0$ we find the superpotential of the Liouville theory.

We can also discuss another limit, in which $\zeta$ diverges and we keep fixed the
effective $2d$ FI term, $t = 2 \pi r \zeta + \log(2 \pi r \mu)$, fixed.  This limit gives the $2d$ $U(1)$ theory with a single charged chiral multiplet, which has a single massive vacuum. In this limit the supersymmetric vacuum remains finite as $r\to 0$, and in particular $r\Sigma \ll 1$, 
so that there is some range of energies where we have the
$2d$ running of the FI term, and we then find a finite $2d$ limit for the twisted superpotential:
\eqn\twoduonelimit{ \cW = \cW^{(2d)}_\chi(\Sigma) + t \Sigma + O(r)\,,}
which agrees with the effective twisted superpotential of the $2d$ $U(1)$ theory.  This has a vacuum at $\Sigma =\mu e^{-t}$. Since
the cutoff $\mu$ is of order $1/r$, this is a good description for large $t$.

\

\noindent{b.} {\it $U(1)$ with one flavor}

\

Next consider a $3d$ $U(1)$ theory with a single flavor, \ie, a pair of chiral multiplets of charges $\pm 1$.  This theory has a $U(1)_A \times U(1)_J$ flavor symmetry, where $U(1)_A$ acts on the chirals with the same charge.  Correspondingly, there are two real mass parameters, the axial mass, $m$, and the FI parameter, $\zeta$.

For the theory on a circle, we can again compute the effective superpotential,\foot{Here we add a bare Chern-Simons term at level $1$ to cancel the implicit ``level $-\frac{1}{2}$ CS terms'' coming from the two chiral multiplets in our convention.}
\eqn\uoneone{ \cW = \cW_\chi^{(3d)}(\Sigma+m) + \cW_\chi^{(3d)}(-\Sigma+m) + \pi r \Sigma(\Sigma+\frac{i}{r}) + 2 \pi r \zeta \Sigma.}
We may write the vacuum equation conveniently in terms of the exponential variable $x=e^{2\pi r \Sigma}$, and the parameters $\nu=e^{2 \pi r m}$ and  $z=e^{2 \pi r \zeta}$, as:
\eqn\uoneoneb{1 =  \exp\bigg( {{\partial \cW}\over {\partial \Sigma}}\bigg) =  z {{x \nu -1}\over {x-\nu } }\,.}
This has a single solution at:
\eqn\uoneonevac{  \Sigma= \frac{1}{2 \pi r} \log (x) =\frac{1}{2 \pi r} \log\left( \frac{z-\nu}{z \nu - 1} \right)\,. }

We can discuss several possible $2d$ limits.  First, suppose we keep $m$ fixed and scale $\zeta \propto 1/r$.  This
is naturally related to the $2d$ SQED theory with a finite FI parameter $t = 2 \pi r \zeta + \pi i$, a finite mass $m$, and with a finite value for $\Sigma$ (that goes to infinity as $t\to \pi i$)\foot{That is, this vacuum goes to infinity when the real FI parameter, $\hat{r}$, is set to zero and the theta angle, $\theta$, is set to $\pi$, so that $t=\hat{r}+i \theta$ is as above.  In the massless theory, this corresponds to the choice of parameters where a Coulomb branch appears.}.  Indeed, one finds that the twisted superpotential \uoneone\ behaves as:
\eqn\uoneonestanlim{ \cW \rightarrow \cW_\chi^{(2d)}(\Sigma+m) + \cW_\chi^{(2d)}(-\Sigma+m) + t \Sigma +2 m \log (2 \pi r \mu),}
agreeing with that of the $2d$ gauge theory, up to a counterterm. 

Next note that if we redefine $\Sigma \rightarrow \Sigma-m$ in \uoneone, we obtain:
\eqn\uoneonec{ {{\partial \cW}\over {\partial \Sigma}} =  2\pi r (\zeta -m) +  \log(1-e^{2 \pi r \Sigma}) - \log(1-e^{2 \pi r (\Sigma-2m)})\,. }
If we now take $\zeta,m \rightarrow \infty$ while keeping their difference, $\zeta-m$, finite, we see the third term is negligible, and the first two have the same form as \uonehalfb.  Thus we find the Liouville theory at low energies, with $\zeta-m$ now playing the role of the mass term.  In this limit we are preserving one half of the Coulomb branch, and the Liouville theory we find is dual to the cigar theory describing this branch.  Similarly, we may keep $\zeta+m$ finite to find a Liouville theory describing the other half of the Coulomb branch.

Finally, if we keep both $\zeta$ and $m$ finite as $r\to 0$, then the vacuum goes to infinity as $1/r$. We can then define a new variable $X = 2 \pi r \Sigma$, and obtain for this variable a finite effective twisted superpotential (up to counterterms)
\eqn\neww{W =  \zeta X + 2 m \log\left(2\,\sinh({X\over 2})\right) + \cdots }
This has Liouville like behavior at $X\to \pm \infty$ just as above, controlled by $\zeta+m$ and $\zeta-m$. As $m$ becomes small we have a vacuum at $X \propto m / \zeta$, connecting to where the Higgs branch used to be. This is also the low-energy behavior of the $3d$ theory on a circle of finite radius $r$.

We can conjecture that the same low-energy theory may arise from a $2d$ Landau-Ginzburg  theory with the superpotential \neww. The protected information of this theory is the same as that of the $3d$ theory on a circle, but
it is not clear if the K\"ahler potential is
the same, for some scaling of $g^2$ as $r\to 0$, and for some choice of the UV K\"ahler potential for $X$.  We will return to the fate of the moduli space in Section $4$.

\

\noindent{c.} {\it  $U(1)_k$ with $N_f$ flavors}

\

Next let us consider a more general $U(1)$ theory, with $N_f$ flavors and a Chern-Simons term at level $k$.  In this example we encounter for the first time a common feature of the reduction of $3d$ theories, mentioned above, which is that the resulting $2d$ description is described by a direct sum of decoupled sectors.

We start by writing the effective twisted superpotential:
\eqn\dwdsuonek{  \cW= \sum_{a=1}^{N_f} \bigg(\cW^{(3d)}_\chi(\Sigma+m_a) + \cW^{(3d)}_\chi(-\Sigma+\tilde{m}_a) \bigg)  + 2 \pi r \zeta \Sigma + \pi r (k+N_f) \Sigma(\Sigma+\frac{i}{r})\,. }
We may write the vacuum equation, $\exp\big(\frac{\partial \cW}{\partial \Sigma}\big)=1$, in terms of exponential variables as above, giving a polynomial equation:
\eqn\uonenfkvaceqn{ z (-x)^k \prod_{a=1}^{N_f} (\nu_a^{-1}-x)  = \prod_{a=1}^{N_f} (1-x {\tilde{\nu}_a}^{-1})\,. }
This has $N_f+k$ solutions, and so the theory has $N_f+k$ vacua.  Let us analyze the fate of these vacua as we take various $2d$ limits.

First we take the limit which gives the $2d$ gauge theory, holding $t=2 \pi r \zeta+\pi i (k+N_f)$ and the masses fixed as $r \rightarrow 0$.  If we focus on the region with $r \Sigma \ll 1$, then the Chern-Simons contribution to $\cW$ of order $r \Sigma^2$ is suppressed. By a similar argument to the one leading to \uoneonestanlim, we find the twisted superpotential of a $2d$ $U(1)$ theory with $N_f$ chiral multiplets:
\eqn\uonenfkssl{ \cW \;\;  \rightarrow_{r\to 0, r\Sigma \ll 1} \;\;\;\sum_{a=1}^{N_f} \big(\cW_\chi^{(2d)}(\Sigma+m_a) + \cW_\chi^{(2d)}(-\Sigma+\tilde{m}_a) \big) + t \Sigma + ...
}
up to counterterms depending on the background fields.  Thus we expect the theory in this region of field space to be described at low energies by this $2d$ $U(1)$ theory with $N_f$ flavors. 

One can check that this $2d$ gauge theory has $N_f$ vacua for a generic FI parameter.  Thus we have accounted for $N_f$ of the vacua of the $3d$ theory, but there are still $k$ missing, which lie outside of the region $r \Sigma \ll 1$ we restricted to above.  To find these, let us take $X = 2\pi r \Sigma$ and look for solutions at finite $X$.  The effective twisted superpotential for $X$ can be expanded for small $r$ as:
\eqn\uonenfkx{
\cW =  {{1}\over{2 \pi r}}\left({1\over2} k X^2 + (t + \pi i N_f) X\right) + O(1).
}
Then one finds that the $k$ additional vacua lie at:
\eqn\uonenfsn{
X_n = -{{t_n}\over k}+ O(r) , \;\;\; t_n=t+2 \pi i (n + \frac{N_f}{2}), \;\;\;n=1,\cdots,k\,,
}
where we may restrict to $n=1,\cdots,k$ because $X \sim X+2 \pi i$.  
The fluctuations around each vacuum are described by a single massive twisted chiral multiplet, up to a constant term in the superpotential $\cW = -t_n^2  / 4 \pi k r + O(1)$, depending only on the background parameters.   Note that the vacua \uonenfsn\ are all separated by distances of order $1\over{r}$ in $\Sigma$ space.  This suggests that the different vacua decouple into separate theories at low energies, but this argument depends on the precise K\"ahler potential. A more precise way to see this decoupling is to note that the values of the twisted superpotential in distinct vacua differ by a factor which diverges as $r \to 0$.  This implies that solitons which connect the different vacua become infinitely massive as $r \to 0$, such that these vacua are completely decoupled in the $2d$ limit.  To summarize, we find, in this limit, that the theory is described at low energies by a direct sum of a $2d$ $U(1)$ theory with $N_f$ flavors, and $k$ copies of the theory of a massive twisted chiral multiplet.  

We can also consider a limit where we hold the masses finite as $r \rightarrow 0$ and take $\zeta$ close to $-\frac{i k}{2 r}$.  This corresponds to taking $t \rightarrow \pi i N_f$ in the $2d$ gauge theory, which is the choice that leaves an unlifted Coulomb branch in $2d$.  Then one finds that only $(N_f-1)$ of the small solutions ($r \Sigma \ll 1$) remain at finite $\Sigma$, while one runs off to infinity.  In addition, out of the $k$ large solutions \uonenfsn, $(k-1)$ stay large while one gets small.  What are the fates of these two runaway vacua, the small solution that is getting large, and the large solution that is getting small?

In order to study these it is convenient to define a different scaling of the chiral superfield,
\eqn\halfscal{ \Sigma = (2 \pi r \mu)^{-1/2} Y\,. }
Then one finds for this variable as $r\to 0$ the LG model:
\eqn\halfscalsuperpot{ \cW = {{k}\over{2 \mu }} Y^2 + m_A \log (\frac{Y}{\mu})\,,  }
where $m_A = \sum_{a=1}^{N_f} (m_a+\tilde{m}_a)$.
This has two vacua, and we claim it describes the two runaway vacua from above.  Note that $\zeta$ does not appear here.  We can regain the dependence on $\zeta$ by modifying its $r\to 0$ scaling to include a subleading term:
\eqn\halfscalzeta{ \zeta = -\frac{i k}{2r} + (2 \pi r \mu)^{-1/2} Z. }
Here $t=2 \pi r \zeta+\pi i (k+N_f)$ is still going to $\pi i N_f$, so the behavior of the gauge theory and of the massive vacua is unaffected, but the LG model is modified to:
\eqn\halfscalsuperpottwo{ \cW = {{k}\over{2\mu}} Y^2 + \frac{Z Y}{\mu} + m_A \log  (\frac{Y}{\mu}) \,, }
with explicit dependence on the FI parameter.  To summarize, in this scaling \halfscalzeta\ of the parameters $(N_f-1)$ of the vacua stay finite as $r\to 0$, $(k-1)$ scale as $\Sigma \propto 1/r$, and two scale as $\Sigma \propto 1/\sqrt{r}$. As $r\to 0$ we get a direct sum of the $2d$ gauge theory, with $(k-1)$ massive vacua, and with the LG model in \halfscalsuperpottwo.

\

\noindent{d.} {\it Coulomb and Higgs limits}

\

Now we take the lessons we learned from the examples above and apply them to a general Abelian gauge theory.   Consider a theory with gauge group $U(1)^N$, and $M$ chiral multiplets of charges $Q_{a,i}$, $a=1,\cdots,M$, $i=1,\cdots,N$, and a matrix of CS levels $k_{ij}$.  There is a real mass parameter $m_a$ for each chiral multiplet and an FI parameter $\zeta_i$ for each gauge factor.  Of these $N+M$ parameters, $N$ can be absorbed into shifts of the gauge scalars $\Sigma_i$, so only $M$ are physical, but it will be convenient to keep this redundancy in the description.  

First, we have seen above that if we keep the real mass parameters finite, and appropriately scale the FI parameter(s), we may obtain a $2d$ $U(1)$ gauge theory.  In general, we define the ``Higgs limit'' as follows: 
\eqn\higgslimit{ m_a \; {\rm finite}, \;\;\; \zeta_i = {{1}\over{2 \pi r}}( t_i - \log(2 \pi r \mu) \sum_a Q_{a,i} )\,, }
with $t_i$ fixed as $r\to 0$.
Then by an argument similar to the ones in examples above, in the region $r \Sigma \ll 1$ the twisted superpotential is well approximated by that of the $2d$ $U(1)^N$ gauge theory with the same charges $Q_{a,i}$, twisted masses $m_a$, and FI parameters $t_i$.\foot{More precisely, the definition of $t_i$ used here may differ from the natural definition in $2d$ by a finite shift due to a contribution from the Chern-Simons terms.} In addition, there will typically be other vacua beyond those accounted for here, and the $2d$ description will be as a direct sum of this $2d$ gauge theory and contributions from these other vacua.\foot{These other vacua may be massive, described by non-trivial LG models, or may be described by gauge theories of lower rank.  We will see explicit examples  of the latter in the next subsection.}  We call this the Higgs limit because when we take the real masses to vanish and the FI terms non-zero in $3d$, we find that the Higgs branch is preserved while the Coulomb branch is lifted, and indeed the $2d$ $U(1)^N$ theory that we obtain has, in the massless limit, the same Higgs branch as the $3d$ theory.  

Next we define the ``Coulomb limit.''  By analogy to the previous case, this will correspond to a limit of real mass parameters which preserves a $3d$ Coulomb branch.  Recall that a continuous $3d$ Coulomb branch may arise in flat space when there exists some choice of $\epsilon_a \in \{ \pm 1\}$, $a=1,\cdots,M$, such that the following conditions are satisfied \IntriligatorLCA{:
\foot{Here $\sigma^R$ and $m^R$ are the $3d$ real scalar and real mass, respectively, which we distinguish from their complexifications, $\Sigma$ and $m$, which arise when we compactify the theory on a circle.}
\eqn\coulconda{ \epsilon_a (Q_{a,i} \sigma^R_i + m^R_a) > 0\,, }
\eqn\coulcondb{ k_{ij} = {1\over2} \sum_{a=1}^M \epsilon_a Q_{a,i} Q_{a,j} \,, }
\eqn\coulcondc{ \zeta_i = {1\over2} \sum_{a=1}^M \epsilon_a Q_{a,i} m^R_a \,.}
The first condition defines a (possibly non-compact) polyhedral region in the $N$-dimensional $\sigma^R$-plane, and the Coulomb branch is a $\T^N$ fibration over this region, where circles in the fibration degenerate along the boundaries.  The second and third conditions impose that the effective CS and FI terms are zero in this region.

When the second condition is satisfied, it is possible to rewrite the effective twisted superpotential for the theory on a circle as:
\eqn\coulimsp{ 
{{\partial \cW}\over{\partial \Sigma_i}} = \sum_{a=1}^M Q_{a,i} \log(1-e^{2 \pi r \epsilon_a( \sum_j Q_{a,j} \Sigma_j + m_a)}) + 2 \pi r \rho_i \,,}
where we have defined:
\eqn\rhodef{
\rho_i = \zeta_i - {1\over2} \sum_{a=1}^M \epsilon_a Q_{a,i} m_a + {{i}\over{2 r}} \sum_{a | \epsilon_a = 1} Q_{a,i}\,. }
Then we claim that the limit of the parameters which preserves the Coulomb branch is\foot{Recall that $3d$ mirror symmetry \IntriligatorEX\ exchanges Higgs and Coulomb branches, and exchanges $m_a$ with $\zeta_i$, so this can be viewed as the mirror of the ``Higgs limit''.}:
\eqn\coullimit{ m_a = {{1}\over{2 \pi r}}(s_a + \epsilon_a \log(2 \pi r \mu) ) ,\;\;\; \rho_i = \; {\rm finite,\ \ as} \ r\ \to 0\,. }
Then one can expand the effective twisted superpotential to leading order in $r$ to find, defining $X_i = 2 \pi r \Sigma_i$:
\eqn\coulimspexp{ 
{{\partial \cW}\over{\partial \Sigma_i}} = 2 \pi r \bigg( -\sum_{a=1}^M \mu Q_{a,i} e^{\epsilon_a( \sum_j Q_{a,j} X_j + s_a)}+ \rho_i \bigg) + \cdots \,.
}
But this means ${{\partial \cW}\over{\partial X_i}} ={{1}\over{2 \pi r}} {{\partial \cW}\over{\partial \Sigma_i}}$ is finite as $r \rightarrow 0$, and integrating it we find:
\eqn\coulimspint{ 
\cW(X_i)= -\sum_{a=1}^M \epsilon_a \mu e^{\epsilon_a( \sum_j Q_{a,j} X_j + s_a)} + \rho_i X_i + \cdots\,.
}
This is the twisted superpotential of a Liouville theory, so we expect this theory to give a good description of this region of the field space.  As usual, there may also be other vacua in distant regions of field space, which appear in a direct sum with this theory in the complete $2d$ description.  We emphasize that for a given theory, there may be many inequivalent $r\to 0$ Coulomb limits covering different regions of the $3d$ moduli space, leading to different Liouville theories.  

In addition to the Higgs and Coulomb limits, which exist for general Abelian gauge theories, there may be other limits which exist in special cases, such as the limits leading to the LG models in \neww\ and \halfscalsuperpottwo.

\subsec{Non-Abelian examples}

In this section we consider the reduction of some non-Abelian gauge theories with matter, to illustrate some new features that arise here.

\

\noindent 
{a.}  {\it $U(N_c)$ with $N_f$ flavors and CS level $k$}

\

Consider a $U(N_c)$ gauge theory with $N_f$ flavors and Chern-Simons level $k$.  The effective twisted superpotential is given by:
\eqn\dwdsunk{ \cW = \sum_{j=1}^{N_c} \bigg( \sum_{a=1}^{N_f} \bigg( \cW_\chi^{(3d)}(\Sigma_j + m_a) + \cW_\chi^{(3d)}(-\Sigma_j+\tilde{m}_a) \bigg)   + \pi r (k+N_f) \Sigma_j(\Sigma_j + \frac{i}{r}) + 2 \pi r \zeta \Sigma_j \bigg)\,. } 
This is just $N_c$ copies of \dwdsuonek.
Defining $x_j=e^{2 \pi r \Sigma_j}, \nu_a=e^{2 \pi r m_a}, \tilde{\nu}_a=e^{2 \pi r \tilde{m}_a}, z=e^{2 \pi r \zeta}$, we may write the vacuum equation $\exp\big({{{\partial \cW}\over{\partial \Sigma_j}} }\big) = 1$ as:
\eqn\unvaceqn{ z (-x_j)^k \prod_{a=1}^{N_f} (x_j -\nu_a^{-1})  = \prod_{a=1}^{N_f} (1-x_j {\tilde{\nu}_a}^{-1}), \;\;\;\; j=1,\cdots,N_c\,. }
This is the same as \uonenfkvaceqn, and has $N_f+k$ solutions for each $\Sigma_j$.  A vacuum of the $U(N_c)$ theory is given by a simultaneous solution to these equations for all $j$.  In order that these are acted on freely by the Weyl symmetry, we must also impose that none of the $\Sigma_j$ are equal.  Since the equation for each $j$ is the same, due to the Weyl symmetry, this is equivalent to picking $N_c$ distinct solutions to this equation, and so this theory has $\pmatrix{N_f+k \cr N_c}$ vacua.  In particular, if $N_c>N_f+k$, the theory is SUSY breaking and has no supersymmetric vacua.

Now let us consider the ``Higgs'' $2d$ limit, where we hold $t=2 \pi r \zeta$ and the other real masses fixed as $r \rightarrow 0$.  As in the Abelian case, if we constrain a given $\Sigma_j$ to have $r \Sigma_j \ll 1$, then the dependence of the effective twisted superpotential on $\Sigma_j$ is given by \uonenfkssl, which has $N_f$ vacuum solutions, while the remaining $k$ solutions lie at finite values of $r \Sigma_j$, and are given given by \uonenfsn.  To construct a vacuum of the $U(N_c)$ theory, we must pick the $N_c$ eigenvalues to be distributed amongst these $N_f+k$ solutions in some way.  Suppose we take $\ell$ of them to have $r \Sigma_j\ll 1$, and the remaining $N_c-\ell$ to be large.  Since the eigenvalues must be distinct, this requires $\ell \leq N_f$ and $N_c-\ell \leq k$ (in addition, of course, to $0 \leq \ell \leq N_c$), so that:
\eqn\lconst{
{\rm max}(0,N_c-k) \leq \ell \leq {\rm min}(N_c,N_f)\,. }
For such a choice of $\ell$, the $\ell$ small eigenvalues are governed by the same effective twisted superpotential as the eigenvalues of a $2d$ $U(\ell)$ theory with $N_f$ flavors.  The large eigenvalues, on the other hand, behave as massive twisted chirals, and their dynamics is trivial.  Thus we conjecture that the theory is described in this region of field space by a $2d$ $U(\ell)$ gauge theory with $N_f$ flavors.  The full low energy theory will be a direct sum of such low energy gauge theory descriptions corresponding to each way of distributing the large and small eigenvalues.  Since, for a given $\ell$, there are $\pmatrix{k \cr N_c-\ell}$ ways to arrange the large eigenvalues, the $U(\ell)$ theory appears with this multiplicity in the direct sum.  Thus, to summarize, we find the low-energy description (ignoring decoupled massive twisted chiral multiplets):
\eqn\unfkle{
\bigoplus_{\ell={\rm max}(0,N_c-k)}^{{\rm min}(N_c,N_f)} \pmatrix{k \cr N_c-\ell} \times U(\ell)_{N_f \; {\rm flavors}} }
Here when $\ell=0$ we have simply a theory of massive twisted chiral multiplets. For $k=0$ we obtain in this limit just the $2d$ $U(N_c)_{N_f}$ theory.

\

\noindent{b.} {\it  $USp(2N_c)$ theories }

\

The reduction of $USp(2N_c)$ theories is very similar to that of $U(N_c)$ theories, with a few extra subtleties.  We will just state the results here, and refer to Appendix B for the details.

Consider an $USp(2N_c)$ theory with $2N_f$ fundamental chiral multiplets, and Chern-Simons level $k$.  Here $k$ and $N_f$ may be half-integer, but cancellation of the global anomaly imposes $N_f+k \in \Z$, and we assume $k \geq 0$.  Let us again take the $2d$ Higgs limit, where the masses of the chiral multiplets are held finite (note there is no FI term in this case).  When $2N_f$ is odd one finds, much as in the $U(N_c)$ case, that the resulting $2d$ theory is a direct sum of $2d$ $USp$ theories:
\eqn\spodddecompmt{ USp(2N_c)_k , \; 2N_f \; \rm{odd}\;\;\; \rightarrow \;\;\; \bigoplus_{\ell={\rm max} (0,N_c-k+\frac{1}{2})}^{{\rm min}(N_f-\frac{1}{2},N_c)} \pmatrix{{k-\frac{1}{2}}\cr{N_c-\ell}} USp(2 \ell)_{N_f} }

When $2N_f$ is even, the behavior is more subtle.  For the case $k=0$, the gauge theory reduces straightforwardly to the $2d$ $USp(2N_c)$ theory.  However, for $k>0$, one finds that in addition to a direct sum of gauge theories as we found above, there are also vacua where one of the eigenvalues is governed by the twisted superpotential of a non-trivial LG model,
\eqn\wlgsp{ \cW_{LG} = e^{{\hat Y}} + {\hat Y} \sum_{a=1}^{2N_f} m_a. }
The full $2d$ theory we obtain in the reduction is then:
\eqn\spevendecompmtapp{ \eqalign{USp(2N_c)_k , \; 2N_f \; \rm{even}\;\;\; \rightarrow \;\;\; \bigg(\bigoplus_{\ell={\rm max} (0,N_c-k+1)}^{{\rm min}(N_f-1,N_c)} \pmatrix{k-1\cr N_c-\ell} USp(2 \ell)_{N_f} \bigg) \cr
\bigoplus \bigg(\bigoplus_{\ell={\rm max} (0,N_c-k)}^{{\rm min}(N_f-1,N_c)} \pmatrix{k-1\cr N_c-\ell-1}  USp(2\ell)_{N_f} + \; {\rm LG \; model \ \wlgsp}\bigg)\,.  }} 
It is not obvious that the K\"ahler potential of the LG model \wlgsp\ obtained from the $3d$ reduction agrees with the natural K\"ahler potential for this $2d$ LG model. We will return to this issue when we discuss dualities below.

\

\noindent{c.} {\it $SU(N_c)$ theories }

\

Finally, we consider the reduction of $SU(N_c)$ theories with (anti-)fundamental matter.  A useful way to obtain these theories is to start from a $U(N_c)$ theory with the same matter content and promote the $U(1)_J$ symmetry to a gauge symmetry. For many purposes this has the effect of ``ungauging'' the $U(1)$ baryon number symmetry, leaving an $SU(N_c)$ theory.  

Let us consider the $SU(N_c)$ theory with $N_f$ fundamental chiral multiplets and $N_a$ anti-fundamentals, and CS level $k$.  The effective twisted superpotential is given by:
\eqn\wsunk{\eqalign{ &\cW =\cr &\,
 \sum_{j=1}^{N_c} \bigg( \sum_{a=1}^{N_f}  \cW_\chi^{(3d)}(\Sigma_j + m_a) +  \sum_{a=1}^{N_a} \cW_\chi^{(3d)}(-\Sigma_j+\tilde{m}_a) \bigg)   + \pi r (k+\frac{N_f+N_a}{2}) \Sigma_j(\Sigma_j + \frac{i}{r}) + 2 \pi r \lambda \Sigma_j \bigg)\,.} } 
This is identical to the twisted superpotential of the $U(N_c)$ theory with $N_f$ fundamentals and $N_a$ anti-fundamentals, except we have renamed $\zeta \rightarrow \lambda$ to emphasize that we now treat it as a dynamical variable.  It acts as a Lagrange multiplier, with its vacuum equation imposing the tracelessness condition, $\sum_j \Sigma_j=0$.  More precisely, the vacuum equations for this theory are:

\eqn\sunvac{ \prod_{a=1}^{N_f}(1- \nu_a x_j) = \Lambda x^{k+\frac{N_f-N_a}{2} }\prod_{a=1}^{N_a}(x_j-\tilde{\nu}_a) , \;\;\;\;\;\;\;\; \prod_{j=1}^{N_c} x_j = 1. }
The first condition gives a polynomial equation for each $x_j$, and to find a vacuum, one must find those values of $\Lambda\equiv e^{2 \pi i \lambda}$ for which it is possible to pick $N_c$ distinct solutions to this polynomial satisfying $\prod_{j=1}^{N_c} x_j = 1$.  

Let us consider the limit of this equation as $r \rightarrow 0$.  Then, if we focus on the region $r \Sigma_j \ll 1$, we find the effective twisted superpotential is approximated by:
\eqn\wsunktwod{ \cW = \sum_{j=1}^{N_c} \bigg( \sum_{a=1}^{N_f}  \cW_\chi^{(2d)}(\Sigma_j + m_a) +  \sum_{a=1}^{N_a} \cW_\chi^{(2d)}(-\Sigma_j+\tilde{m}_a) \bigg)   + 2 \pi r (\lambda + (N_f-N_a) \log (2 \pi r \mu)) \Sigma_j \bigg)\,. } 
After a redefinition of the Lagrange multiplier, $\lambda \rightarrow \lambda- (N_f-N_a) \log (2 \pi r \mu)$, this is precisely the effective twisted superpotential of the $2d$ $SU(N_c)$ theory with $N_f$ fundamental and $N_a$ anti-fundamental flavors.  The vacuum equations here are:
\eqn\sunvactwod{ \prod_{a=1}^{N_f}(\Sigma_j+m_a) = \Lambda' \prod_{a=1}^{N_a}(-\Sigma_j+\tilde{m}_a) , \;\;\;\; \sum_{j=1}^{N_c} \Sigma_j = 0. }

This implies that all of the solutions of the $2d$ vacuum equations arise as solutions to the $3d$ vacuum equations in the $r \to 0$ limit.  Thus we generically expect that in the $2d$ reduction of the $3d$ $SU(N_c)$ theory, we will find:
\eqn\sudecomp{ SU(N_c)_k , \; N_f,N_a \;\;\; \rightarrow \;\;\; SU(N_c), \; N_f,N_a\;\; \bigoplus \;\; ...  }
where there may or may not be extra terms appearing in the direct sum.  This question is essentially equivalent to the question of whether there are strictly more vacua in the $3d$ theory than in the $2d$ theory (the above argument implies there are at least as many in $3d$), and the answer will depend on the specific values of $N_c,N_f,N_a$, as well as on the choice of $k$.  The counting of vacua in $SU(N_c)$ theories is more difficult than in $U(N_c)$ and $USp(2N_c)$ theories, and in general there is not a simple closed formula. Generally one finds that there are strictly more vacua in $3d$ than in $2d$, so that a non-trivial direct sum does arise.  We will discuss the implications for duality in the next section.

\subsec{Reduction using supersymmetric partition functions}

Let us now take an alternative approach to studying the reduction of mass-deformed theories, which uses supersymmetric partition functions.  The definitions and some basic properties of the relevant partition functions are reviewed in Appendix A.  

We start with a convenient  $3d$ partition function, the $\S^2 \times \S^1$ partition function (also known as the supersymmetric index), and take a limit where the size of the $\S^1$ factor goes to zero, upon which we obtain the $\S^2$ partition function of its $2d$ reduction. 
The partition functions are especially useful since they very intuitively encode some of the essential details of a Lagrangian.  Reducing the index to the sphere partition function in this representation we can then obtain not just the value of the partition function, but sometimes also a natural guess for a Lagrangian in two dimensions that will give this partition function. However, there are various subtleties in taking this limit, as we will discuss below.  

If we start with a dual pair in $3d$, the duality implies the identity of their supersymmetric indices, and so upon taking this limit on both sides we derive the identity of the $\S^2$ partition functions of their respective $2d$ reductions, which provides non-trivial evidence for a $2d$ duality.
However, it is important to perform further tests of this $2d$ duality to verify that it is correct, and we find in some cases that, although the $\S^2$ partition functions of a putative dual pair agree, they fail other, more refined tests, and so the duality does not hold. For instance, the elliptic genus of a $2d$ theory does not have any $3d$ origin, so its matching does not follow from a $3d$ duality.

The study of the reduction of a $3d$ theory by taking a limit of its supersymmetric index should be equivalent to the analysis we performed above using the twisted superpotential, and we will see that this is indeed the case. However, the two approaches use a somewhat different point of view, and illuminate different aspects of the reduction.


As reviewed in Appendix A, the $\S^2$ partition function of a $2d$ $\cN=(2,2)$ gauge theory/LG model is given by the following integral formula\foot{More precisely, the $\Sigma_i$ are constrained to satisfy $\Sigma_i+\bar{\Sigma}_i\in \Z$ due to quantization of the flux on $\S^2$, so this is really an integral and an infinite sum, but the essential arguments below are not affected by this point.} (here we work in units where the radius of $\S^2$ is one):
\eqn\stwowpm{\eqalign{& \cZ_{\S^2}(m_a,X_\beta) =\cr &\qquad\, \frac{1}{|W|}\int \prod_i \frac{d\Sigma_i d\bar{\Sigma}_i}{\pi} \prod_\alpha\frac{dY_\alpha d\bar{Y}_\alpha}{\pi} \exp \bigg( \cW_{\S^2}^{(+)}(\Sigma_i,m_a,Y_\alpha,X_\beta) -\cW_{\S^2}^{(-)}(\bar{\Sigma}_i,\bar{m}_a,\bar{Y}_\alpha,\bar{X}_\beta) \bigg), }}
where $\Sigma_i$ ($m_a$) run over the dynamical (background) gauge multiplets in the theory, $Y_\alpha$ ($X_\beta$) over the dynamical (background) twisted chiral multiplets, $|W|$ is the order of the Weyl group, and we define a curved space analogue of the effective twisted superpotential:
\eqn\wpmtwod{ \cW_{\S^2}^{(\pm)}=  \cW_{UV,2d} \pm \Omega_{UV,2d} + \sum_{I \in {\rm chirals}} \cW^{(\pm)}_{\chi,\S^2}(Q_I^i \Sigma_i + q_I^a m_a). }
Here $\cW_{UV,2d}$ is the twisted superpotential appearing in the flat space UV action, while the contribution of a chiral multiplet of R-charge $\Delta$ is given by:
\eqn\twodstwochiral{ \cW^{(\pm)}_{\chi,\S^2}(\Sigma) =  \log \bigg(\Gamma (\frac{1}{2} + \Sigma \pm \frac{1-\Delta}{2} )\bigg).}
We note this approaches the flat space effective twisted superpotential of a chiral multiplet as $\Sigma \rightarrow \infty$ (recall that we took the radius of $\S^2$ to one, so this really means $\Sigma \cdot r_{\S^2} \to \infty$).  We also allow for a so-called ``effective dilaton'' $\Omega_{UV,2d}$, which controls the coupling of the theory to curved space \NekrasovXAA.

We also consider the $\S^2 \times \S^1_\tau$ partition function of a $3d$ $\cN=2$ theory, with $\tau$ equal to the radius of the $\S^1$ (namely, the ratio between the radii of the $\S^1$ and the $\S^2$).  This can be defined by a trace over states on $\S^2$:
\eqn\indextrace{ {\rm Tr}_{{{\cal H}}_{n_a}} [(-1)^{2J_3} \bq^{\frac{1}{2}(E+R)} \prod_a {{\bf \nu}_a}^{F_a}], }
where $\bq = e^{-\tau}$, $J_3$ is the angular momentum on $\S^2$, $F_a$ are the changes under a basis of the maximal torus of the flavor symmetry group of the theory, and we work in the space of states with a background flux ${\bf n}_a$ on $\S^2$ for the flavor symmetries.  This can be computed by writing the index of the free UV theory, depending also on fugacities $z_i$ for the gauge symmetry, and then projecting onto gauge-invariant states by integrating over $z_i$ and summing over gauge fluxes.  For our purpose, it is useful to rewrite this in terms of an integral formula similar to \stwowpm:
\eqn\stwosone{ \cZ_{\S^2 \times \S^1_\tau}(m_a) = \frac{1}{|W|} \int \prod_i \frac{d\Sigma_i d\bar{\Sigma}_i}{\pi} \exp \bigg( \cW_{\S^2 \times \S^1_\tau}^{(+)}(\Sigma_i,m_a)-\cW_{\S^2 \times \S^1_\tau}^{(-)}(\bar{\Sigma}_i,\bar{m}_a) \bigg)\,, }
where we have identified:
\eqn\indparrel{
e^{ \tau m_a} = (e^{\pi i}\bq)^{-\bn_a/2} {\bf \nu}_a, \;\;\;\;e^{\tau \bar{m}_a} =(e^{-\pi i} \bq)^{-\bn_a/2} {\bf \nu}_a^{-1}\,.
}
Here we have:
\eqn\wpmthreed{ \cW_{\S^2 \times \S^1_\tau}^{(\pm)}=  \cW_{CS} + \sum_{I \in {\rm chirals}} \cW^{(\pm)}_{\chi,\S^2 \times \S^1_\tau}(Q_I^i \Sigma_i + q_I^a m_a) }
with $\cW_{CS}$ equal to the contribution to the twisted superpotential of the Chern-Simons terms, and:
\eqn\threedindexchiral{ \cW^{(\pm)}_{\chi,\S^2 \times \S^1_\tau}(\Sigma) =  - \log \bigg((e^{\tau \Sigma} \bq^{1 \mp (1-\Delta)};\bq)\bigg), }
where $(\bz;\bq) \equiv \prod_{j=0}^\infty (1- \bz \bq^j)$.  As above, this approaches the effective twisted superpotential of a $3d$ chiral multiplet on $\R^2 \times \S^1_\tau$ as $\Sigma$ is taken large.

Now we claim that, in a suitable sense, the $\tau \rightarrow 0$ limit of the supersymmetric index of a theory gives the $\S^2$ partition function of its reduction.  Moreover, this limit precisely mirrors the procedure for the reduction of a $3d$ theory using the twisted superpotential, discussed above.

In more detail, we start, as in Section $2.2$, by choosing the scaling of the real mass parameters, $m_a$.  In the theory on $\S^2 \times \S^1_\tau$, we should scale the background multiplets $m_a$ in the same way as in our flat-space discussion (where now we replace $2 \pi r$ with $\tau$).  Next, recall that in order to focus on different vacua (or different regions of the moduli space) it was necessary to scale the dynamical gauge multiplets in a specific way.  In the context of the $\S^2 \times \S^1_\tau$ partition function, these dynamical multiplets are the integration variables, $\Sigma_i$, and so for any finite $\tau$ such a rescaling has no effect.  However, in the $\tau \rightarrow 0$ limit, we may need to perform such a rescaling in order to capture the dominant contributions to the integral.  A useful way to think about this rescaling is that the limit of reduction to two dimensions and the integrations/sums of the matrix model computing the index do not necessarily commute.   In general, finding the correct change of variables that allows us to make these limits commute is a difficult analysis problem. However, we conjecture that the choice dictated by the flat space analysis of the twisted superpotential always leads to the correct change of variables in the limit of the index; as discussed above, in some cases we expect different scalings of the dynamical fields to focus on different aspects of the low-energy physics, that are relevant near different supersymmetric vacua.  We can check this conjecture by computing the limit of the index numerically, or by verifying that the identities for supersymmetric indices implied by dualities in three dimensions reduce to identities when we take the limit of vanishing $\tau$ and focus on the relevant physical vacua.

Let us analyze how various contributions to the index behave as we take $\tau \rightarrow 0$.   First, if the mass, $\Sigma$, of a chiral multplet stays finite in this limit, we find:
\eqn\lightchcirallim{  \cW^{(\pm)}_{\chi,\S^2 \times \S^1_\tau}(\Sigma) \to \cW^{(\pm)}_{\chi,\S^2}(\Sigma) +  (-\Sigma \pm \frac{\Delta-1}{2})  \log (\tau) + \cdots\,.
} 
On the other hand, if we take a limit where the mass $\Sigma$ of a chiral multiplet becomes large as $\tau \rightarrow 0$, then we may approximate the chiral multiplet contribution to the index by the flat space effective twisted superpotential.  Similarly, the contribution of a Chern-Simons term is the same as that of the flat space effective twisted superpotential.  Thus we see that the integrand of the $\S^2 \times \S^1$ partition function approaches that of the $\S^2$ partition function of a certain theory, whose chiral multiplets come from the light chiral multiplets of the $3d$ theory, and whose twisted superpotential comes from the limit of the effective twisted superpotential of the $3d$ theory on a circle. We are then led to precisely the same theory as we were led to by the twisted superpotential analysis above.\foot{One should also consider the limit of the effective dilaton, which can be derived by a similar procedure as the effective twisted superpotential.  Studying its limit gives a slightly more refined observable probing the $3d$ reduction, \eg, determining how the R-charge assignments behave upon reduction, but we omit this analysis here.}

More precisely, we find that in the $\tau \rightarrow 0$ limit of the index, there may be a divergent overall factor.  This factor arises from the pieces of the effective twisted superpotential and effective dilaton depending only on background fields, which may diverge as $\tau \rightarrow 0$.  For a given scaling of the $\Sigma$'s, we generically find a saddle point for which the limit behaves as:
\eqn\indexlimitdiv{ \cZ_{\S^2 \times \S^1_\tau}(m_a(t_a,\tau);\tau) \tra f_{div}(t_a,\tau ) \cZ_{\S^2}(t_a) + \cdots \,,  }
where $m_a$ and $t_a$ are as in Section $2.2$, and $f_{div}$ is a function which diverges as $\tau \rightarrow 0$.  We expect that $f_{div}$ encodes anomalies or other protected information about the $2d$ theory, along the lines of \DiPietroBCA, but we do not explore this issue here.  

A pair of dual $3d$ theories has equal supersymmetric indices.  Then, provided we take the same limit of parameters on both sides, we find an expression of the form \indexlimitdiv\ for both theories.  The divergent pre-factors necessarily agree, so we may strip them off and infer the identity of the finite piece, which gives the identity of the $\S^2$ partition functions of the corresponding $2d$ reductions. This provides non-trivial direct evidence for their duality. However, a question still remains of what is the exact two dimensional theory which gives rise to these $\S^2$ partition functions.
We will stress in examples below that equality of $3d$ indices leads to equality of $\S^2$ partition functions of a certain pair of theories, but not necessarily to their duality.

One further complication that may arise is when the theory reduces to a non-trivial direct sum of $2d$ descriptions.  In that case, there will be several competing saddles in the $\tau \to 0$ limit of the integral \stwosone\ corresponding to these summands.  For a given range of parameters (\ie, the $m_a$ and the R-charges of the chiral multiplets), one of these saddles will typically be dominant over the others, and the $\S^2$ partition function we find in the limit will be that of the theory corresponding to this term in the direct sum.  As we vary parameters, we may find the theories corresponding to other terms in the direct sum.  We will see examples of this behavior in the next section when we discuss dualities between pairs of $3d$ theories.

Finally, we mention that, in addition to $\S^2 \times \S^1_\tau$ index discussed above, one can also study the reduction using other $3d$ partition functions.  For example, we may consider the $\S^3/\Z_p$ partition function, which reduces to the $\S^2$ partition function in the $p \rightarrow \infty$ limit \BeniniNC, or the $b \rightarrow 0$ limit of the $\S^3_b$ partition function, which is related to the twisted superpotential of the $2d$ reduction \BeemMB.  In addition, one may consider the $\Sigma_g \times \S^1_\tau$ partition function \refs{\NekrasovXAA,\ClossetARN,\BeniniHJO}, which we claim reduces to the $\Sigma_g$ partition function as the radius $\tau$ goes to zero.  Specifically, the former is computed in terms of the twisted superpotential and effective dilaton of the $3d$ theory on $\S^1_\tau$, namely:
\eqn\sigmagsone{
\cZ_{\Sigma_g \times \S^1_\tau}(m_a,s_a) = \sum_{\hat{\Sigma}_i \in {\cal S}_{vac}} {\Pi_a}(\hat{\Sigma}_i,m_a)^{s_a} {\cal H}(\hat{\Sigma}_i,m_a)^{g-1},
}
where $\hat{\Sigma}_i \in {\cal S}_{vac}$ runs over the solutions to the vacuum equations $\exp \big( \partial_{\Sigma_i} \cW \big)=1$, $s_a$ are fluxes for the background gauge fields coupled to flavor symmetries, and:
\eqn\pihdef{
\Pi_a= \exp \big( \partial_{m_a} \cW \big) , \;\;\;\; {\cal H} = \exp \big(\Omega \big) \det_{i,j} \frac{\partial^2 \cW}{\partial \Sigma_i \partial \Sigma_j}.
}
The argument that the twisted superpotential (and effective dilaton) of the $3d$ theory reduces to that of the $2d$ theory then directly implies that these ingredients reduce to those of the $2d$ reduction, and so the $\tau \to 0$ limit of the $\Sigma_g \times \S^1_\tau$ partition function is the $\Sigma_g$ partition function of the $2d$ theory.  This was recently checked to match for the $2d$ dualities of Hori-Tong and Hori in \ClossetVVL, and we expect that the identities found there can be obtained from the limit of identities of the $\Sigma_g \times \S^1_\tau$ partition function for appropriate $3d$ dualities.  More precisely, one generally finds (after appropriately rescaling the dynamical twisted chiral fields) that contributions from multiple terms in the direct sum will appear in the sum over vacua above, and one of these terms is typically dominant in the $\tau \to 0$ limit, as discussed for the $\S^2 \times \S^1_\tau$ index above.

\newsec{Reductions of dualities between discrete vacua}

Next, we discuss examples of reductions of dualities from three to two dimensions. We start in this section from cases where the parameters are such that there are only discrete vacua; the theory in these vacua may be gapped, or the low-energy theory may be a non-trivial conformal field theory with a discrete spectrum. In the next section we analyze the special cases in which the low-energy theory has a continuous spectrum, leading to additional complications.

There exist numerous IR dualities between gauge theories with ${\cal N}=2$ and ${\cal N}=4$ supersymmetry in three dimensions. For some of these dualities a derivation is known based on assuming analogous dualities in four dimensions and using reductions and deformations of those. For other dualities, for example
mirror dual pairs with ${\cal N}=4$  supersymmetry, no derivation of this type is known. 

When considering four dimensional gauge theories on a finite circle, the effective theory in three dimensions is a gauge theory with the same matter content as the one in four dimensions, but typically with additional  superpotentials which break explicitly symmetries that are broken by anomalies in four dimensions \AharonyDHA. Another complication is that the scalar fields descending from the components of the vector field along the circle are compact. One can typically discover $3d$ UV completions of the low-energy $3d$ effective theories as standard $3d$ gauge theories, albeit generally with a superpotential involving monopole operators which is needed to obtain a bona fide three dimensional IR duality. In some cases  the compact nature of the scalar fields introduces additional complications which affect the rank of the gauge groups \AharonyKMA. 

We saw in the previous section that considering a $3d$ gauge theory on a circle can lead to a wider variety of behaviors. In particular the UV completion that we find for the effective theory is not always a gauge theory but rather it can be an LG model. 
We will argue that when the vacua are discrete we can deduce dualities in two dimensions assuming dualities in three dimensions. Some of the dualities already appear in the literature but some are new. When there is a (compact) non-trivial low-energy CFT, the IR duality states that this CFT is the same on both sides. When there is a massive vacuum, the IR duality states that all observables that are protected by supersymmetry are the same on both sides -- in particular the twisted chiral ring and the masses of BPS states (such as domain walls between different vacua).

\subsec{Reduction of dualities: ${\cal N}=2$ mirror symmetry}

We begin with a discussion of $3d$ ${\cal N}=2$ Abelian mirror duals and the implications for dualities in two dimensions. The discussion of the reduction here is not new and was considered in  \AganagicUW. 
We repeat this here as it is the simplest example and it illustrates some of the general features that we have considered.

\

\item{a.}{\it Basic example of ${\cal N}=2$ mirror symmetry}

Mirror symmetry dualities with $3d$ ${\cal N}=2$ supersymmetry exchange monopole and Higgs branch operators. Let us consider the most basic example:

\nlb {\it Theory A}  - A $U(1)$ gauge theory with Chern-Simons term at level $-\frac{1}{2}$, and a single charge-one chiral.  This theory has a `topological' $U(1)$ symmetry for which we can add a real mass, which gives the FI parameter $\zeta$. We discussed the reduction of this theory in the previous section.

\nlb {\it Theory B}  - A free chiral multiplet.  This has a $U(1)$ flavor symmetry for which we can add a real mass $m$, which maps to the FI parameter $\zeta$ of theory $A$ via $\zeta=-m$.

\

We can obtain this duality by beginning with the duality \AharonyBX\ between the $3d$ $U(1)$ gauge theory with
chiral multiplets of charges $\pm 1$, and the theory of three chiral multiplets with $W = X Y Z$,
and deforming it by an appropriate combination of its two real masses that leaves one of the three
chiral multiplets in the second description massless. We then take this combination to infinity,
while leaving fixed the other combination (which becomes the real mass parameter mentioned above) \DoreyRB.

Let us consider placing these theories on $\S^1_r$ and see what $2d$ theories we obtain.  We take a limit where the real mass parameter of the chiral multiplet, $m$, is held finite as $r \rightarrow 0$.  On the free chiral side, we saw in \chirallimit\ that the effective twisted superpotential behaves as follows in the $r \rightarrow 0$ limit:
\eqn\chirallimitb{
\cW_\chi^{(3d)}(m)  \rightarrow \cW_\chi^{(2d)}(m) + m \log(2 \pi r \mu) \;\;\;\;{\rm as} \;\; r \rightarrow 0.}
Similarly, on the gauge theory side, as in \uonetwod\ we find that it is necessary to rescale the gauge field as $2 \pi r \Sigma = X - \log(2 \pi r)$, and the effective twisted superpotential behaves as:
\eqn\efw{\cW \rightarrow  -m X + \mu  e^{-X} + m \log(2 \pi r \mu) .}
Note that the leading divergent terms in \chirallimitb\ and \efw\ match.  This serves as a very simple test of the duality, and means one can consistently add a counterterm on both sides of the duality to subtract this piece and leave a finite twisted superpotential.

The $2d$ Landau-Ginzburg model defined by \efw\ was argued in \HoriKT\ to be dual to a free chiral superfield in $2d$ with a ``real mass'' $m$.
So, as first discussed in \AganagicUW, the reduction from $3d$ seems to imply that this $2d$ duality follows from this basic example of $3d$ mirror symmetry.  However, this analysis only holds in the presence of a mass deformation.  When we turn the mass off, a continuous moduli space arises, and the low energy description becomes sensitive to the $D$-terms in the action and, in particular, to the gauge coupling.  We will return to these issues in Section $4$.

\

\item{b.}{\it Partition function}

Let us now repeat these arguments at the level of the supersymmetric partition functions.  The $3d$ duality implies the identity of the $\S^2 \times \S^1$ partition functions (indices) of the two theories:
\eqn\mirrorB{
 {\cal I}_{A} \equiv \sum_{\bm \in \Z} \oint \frac{d\bz}{2 \pi i \bz} \bz^{\bn} \bw^{\bm} {\cal I}_\chi^{\Delta=\frac{1}{3}}(\bz,\bm) \,,
}
\eqn\mirrorA{
 {\cal I}_{B} \equiv \bw^\bn\;{\cal I}_\chi^{\Delta=\frac{1}{3}}(\bw,\bn) \,,
}
where $\bw$ and $\bn$ are the index parameters corresponding to the $U(1)$ global symmetry, and ${\cal I}_\chi^{\Delta}$ is the index of a chiral multiplet with a level $-\frac{1}{2}$ CS term regulator and R-charge $\Delta$, which we set to $\frac{1}{3}$ here for convenience.\foot{One can mix the R-charge with the $U(1)$ global symmetry by taking $\bw \rightarrow \bw \bq^c$.  This introduces additional prefactors in the index which are related to $U(1)_R-U(1)_{global}$ contact terms.  The choice $\Delta=\frac{1}{3}$ gives the simplest expressions.}

Above we kept the real mass parameter $m$ corresponding to the $U(1)$ symmetry fixed as $r\to 0$.  Then from \indparrel, we see that we should scale the index parameters as:
\eqn\uonelimtwo{
\bw= (-1)^\bn e^{i \tau W}\,, \qquad\quad\quad \bn = n\,,
}
where $m = \frac{n}{2} + i W$ is held finite.  On the free chiral side, using \lightchcirallim, we obtain the $\S^2$ partition function of a $2d$ free chiral (following from \twodstwochiral), up to a divergent prefactor:
\eqn\tetrAlimtwo{
{\cal I}_B(\bw,\bn) \rightarrow \tau^{-\frac{2}{3}-2i W} Z_\chi^{\Delta=\frac{1}{3}}(W,n)\,.
}

On the gauge theory side, we must determine how to scale the integration variables, $\bz$ and $\bm$, in order to focus on the dominant terms in the integral. In Section $2.3$ we saw it was appropriate to scale the gauge variables as:
\eqn\uogs{
 2 \pi r \Sigma = X - \log (2 \pi r).
}
Then from \indparrel, and replacing $2 \pi r \rightarrow \tau$, we see that we should rescale the integration variables in the index as:
\eqn\zmscale{
\bz = (-1)^{\bm} e^{i \beta}\,,\qquad\quad\quad \bm =\frac{2}{\tau}(\alpha-\log (\tau) ) \,,
}
where $X= \alpha+ i\beta \sim X+2 \pi i $.  The scaling of $\bm$ means we can approximate $\sum_{\bm \in \Z}$ as $\frac{2}{\tau} \int d\alpha$ as $\tau \rightarrow 0$, and we find
\eqn\tetraBrewrite{
{\cal I}_A(w,n) \rightarrow  \frac{1}{\pi \tau} \int d\alpha d\beta  \;\tau^{1/3-2 i W} \exp \bigg( 2 i \alpha W  + i \beta n  - \frac{\alpha}{3}  + e^{-\alpha + i \beta} - e^{-\alpha - i \beta} \bigg)\,.
}
We see that the divergent scaling agrees with \tetrAlimtwo, and so we can strip it off and we find that also the finite pieces agree.  In the first case, \tetrAlimtwo, we have the $\S^2$ partition function of a free chiral coupled to a background gauge multiplet $m =  \frac{n}{2} + i W$.  In the second, \tetraBrewrite, we have the $\S^2$ partition function of the LG model with dynamical twisted chiral multiplet $X$ with (curved space) twisted superpotential,
\eqn\uonemirrtwistpot{
\cW_{S^2}^{(\pm)}(X,m) = m \; X - e^{-X} \mp \frac{1}{6} X\,.
}
The first two terms correspond to the flat space twisted superpotential of this LG model \uonetwod, while the third gives the dilaton coupling corresponding to the chosen R-symmetry.  The equality of the finite parts of the limits of the $3d$ indices implies that the $\S^2$ partition function of these theories agree.  This follows from the $3d$ duality, and in $2d$ it agrees precisely with the prediction of the Hori-Vafa duality \refs{\HoriKT,\GomisWY}.

\
\item{c.}{\it General $\cN=2$ mirror symmetry}

The duality described above is the basic building block of a more general $2d$ duality between Abelian gauge theories \DoreyRB.  The $3d$  theories in question are:

\

{\it Theory A}: $U(1)^M$ gauge theory with $N$ chiral multiplets of charge $Q_{a,j}$, $a=1,\cdots,N$, $j=1,\cdots,M$, where $N \geq M$.  Here we take $Q_{a,j}$ to be an integer matrix of maximal rank, $M$.    We also include a Chern-Simons term at level $k_j = -\frac{1}{2} \sum_a {Q_{a,j}}^2$ for the $j$'th gauge group.  This theory has $M$ $U(1)_J$ symmetries and $(N-M)$ independent $U(1)$ flavor symmetries acting on the chirals.

{\it Theory B}: $U(1)^{N-M}$ gauge theory with $N$ chiral multiplets of charge $\hat{Q}_{\hat{j},a}$, where $\hat{j}=1,\cdots,N-M$, with Chern-Simons level $\hat{k}_{\hat j} = \frac{1}{2} \sum_a {\hat{Q}_{\hat{j},a}}^2$  for the ${\hat j}$'th gauge group, and where we gauge a certain discrete subgroup $\otimes_j \Z_{d_j}$ of the flavor symmetry.
The form of the dual charge matrix $\hat{Q}_{\hat{j}a}$ is given in \DoreyRB.  This theory has $(N-M)$ $U(1)_J$ symmetries and $M$ independent $U(1)$ flavor symmetries acting on the chirals.  The $U(1)_J$ and flavor symmetries of the two theories are exchanged under the duality.

\

To reduce this duality, in the langauge of Section $2.3$, we take a Higgs limit of theory A.  This leaves a $2d$ $U(1)^M$ gauge theory with chiral multiplets of charges $Q_{a,j}$.  On the dual side, one can check that this limit of the mass and FI parameters maps precisely to a Coulomb limit of Theory B, and one obtains an LG model, which is precisely the dual description of \HoriKT.  The reduction of this duality is discussed in detail in  \AganagicUW.

\subsec{Reduction of $U(1)$ $N_f=1$ $\leftrightarrow$ $XYZ$}

Another Abelian example, which falls outside the class considered above, is the duality between the $U(1)$ theory with one flavor and the $XYZ$ model.  The mapping of $U(1)$ global symmetries across the duality is shown in the table below.  The reduction of the former theory was considered in the previous section, and we saw that there were several $2d$ limits one can consider.  
\eqn\uonetab{
\vbox{\offinterlineskip\tabskip=0pt
\halign{\strut\vrule#
&~$#$~\hfil\vrule
&~$#$~\hfil\vrule
&~$#$~\hfil
&\vrule#
\cr
\noalign{\hrule}
& & U(1)_A & U(1)_J &\cr
\noalign{\hrule}
&  Q       &   \quad  1 &  \quad  0   &\cr
&  \tilde{Q}       &   \quad  1 &  \quad  0   &\cr
\noalign{\hrule}
&  X       &   \,  -1 &  \quad  1   &\cr
&  Y       &   \,     -1 &  \,     -1   &\cr
&  Z       &   \quad  2 &  \quad  0   &\cr
}
\hrule}}

First, taking the Higgs limit gives a $2d$ $U(1)$ gauge theory with one flavor.  This corresponds to taking the mass for the $U(1)_A$ symmetry finite, while scaling that for the $U(1)_J$ symmetry to infinity.  Then on the dual side, $Z$ stays light, while $X$ and $Y$ become heavy and can be integrated out, and so the superpotential becomes trivial.  The $2d$ $U(1)$ theory then has a dual description as a free chiral multiplet.  This duality is a special case of the class of dualities considered in \BeniniMIA, and we will discuss the more general case below.  
We may also consider a different limit, where we hold both masses finite.  Then all fields of the $XYZ$ theory stay light, and so we simply obtain the $2d$ XYZ theory.  In the gauge theory, we saw above that we obtain a certain LG model, with twisted chiral field $x \sim r \Sigma$, which is governed by the twisted superpotential \neww:
\eqn\newwds{W =  \zeta x + 2 m \log\left(2\,\sinh({x\over 2})\right).}
Then it is natural to conjecture that this LG model is dual to the $2d$ $XYZ$ theory.  Indeed, at the level of the mass-deformed theories, this duality seems to hold.  Namely, one can compute the $\S^2$ partition functions of the two theories, which are defined for non-zero mass parameters, and they agree (this follows from the general discussion in Section $2.5$).  We must be more careful when we take the masses to zero, and we will see below that this duality fails to hold at zero mass.
\subsec{Reductions of $U(N_c)$ dualities}

Here we consider dualities relating $3d$ ${\cal N}=2$ theories with $U(N_c)$ gauge group and (anti-)fundamental matter.  These dualities are reminscent of four dimensional Seiberg dualities \SeibergPQ, and in fact can be derived from them by dimensional reduction \AharonyDHA.  A new feature in the three dimensional dualities is the Chern-Simons term, which plays an important role in determining the dual description.  We will see that, upon reduction on a circle, we recover in all cases the $2d$ dualities of \BeniniMIA\ and \BeniniUI, although the details of how this happens are non-trivial, and depend on the $3d$ starting point.

\

\item{a.}{\it The $k=0$ duality}

We start with a duality \AharonyGP\ relating the following theories:

\

\nlb {\it Theory $A$} - A $3d$ $U(N_c)$ gauge theory with $N_f$ fundamental flavors $(Q_a,\tilde{Q}_b)$.

\nlb {\it Theory $B$} - A $3d$ $U(N_f-N_c)$ gauge theory with $N_f$ fundamental flavors $(q^a,\tilde{q}^b)$ and singlet chirals $M_{ab}$ and $V_\pm$ with superpotential:
\eqn\ahaw{ W = q^a M_{ab} \tilde{q}^b + V_+ \tilde{V}_- + V_- \tilde{V}_+ \,,}
where $\tilde{V}_\pm$ are monopole operators which parameterize the Coulomb branch of the theory.

\

Under the duality, the monopole operators of theory $A$ map to the singlets $V_\pm$ of theory $B$, and $Q_a \tilde{Q}_b$ maps to $M_{ab}$.  The global symmetry group of the two theories is $SU(N_f) \times SU(N_f) \times U(1)_A \times U(1)_J \times U(1)_R$, and the fields are charged as:
\eqn\ahacharges{
\matrix{
\quad & SU(N_f) & SU(N_f) & U(1)_A & U(1)_R & U(1)_J \cr
Q_a & {\bf N_f} & {\bf 1} & 1 & 0 & 0 \cr
\tilde{Q}_b & {\bf 1} & {\bf N_f} & 1 & 0 & 0 \cr
q^a & {\bf \bar{N_f}} & {\bf 1} & -1 & 1 & 0 \cr
\tilde{q}^b & {\bf 1} & {\bf \bar{N_f}} & -1 & 1 & 0 \cr
V_\pm  & {\bf 1} & {\bf 1 }& -N_f & N_f-N_c+1 & \pm 1 }
}

We will take the specific $r\to 0$ limit of this theory that leads in Theory $A$ to the $2d$ gauge theory, namely (as discussed in the previous section) we keep the real masses for flavor symmetries finite, while we set $2 \pi r \zeta =  t$ and hold $t$ finite. For theory $A$, a $2d$ description is then given by the $2d$ $U(N_c)$ theory with $N_f$ flavors and FI parameter $t$.  

For theory $B$, we must understand the effect of the extra singlets and the superpotential.  Ignoring the $V_\pm$ fields, we have a `Higgs limit' also in this theory, so we would obtain a $2d$ $U(N_f-N_c)$ gauge theory with $N_f$ fundamental flavors and singlets $M_{ab}$ with superpotential $W=q^a M_{ab} \tilde{q}^b$.  To see the effect of the $V_\pm$ fields, note that they are charged under the $U(1)_J$ symmetry, and so become very heavy in the limit above where $\zeta \to \infty$.  Thus they should be integrated out, and one computes their contribution to the effective twisted superpotential as:
\eqn\vpmw{\cW_{V_\pm} = \cW_{\chi}^{(3d)}(\frac{t}{2 \pi r} - N_f m_A) + \cW_{\chi}^{(3d)}(-\frac{t}{2 \pi r} - N_f m_A) + \cW_{CS}  \,, }
where $m_A$ is the real mass for the $U(1)_A$ symmetry, and $\cW_{CS}$ is the contribution of a level $1$ CS term introduced to cancel the background terms implicit in the regularization of the chiral multiplets.  This has the same structure as the limit we took in \neww, and one finds that in the $r \rightarrow 0$ limit this becomes:
\eqn\vpmlim{ \cW_{V_\pm} = -N_f m_A \log \left( 2\; \sinh (\frac{t}{2})\right) + O(r)\,. }

Thus to summarize, the $3d$ duality on a circle leads to the following $2d$ duality, which was first found in \BeniniMIA:

\nlb {\it Theory $a$} - A $2d$ $U(N_c)$ gauge theory with $N_f$ fundamental flavors $(Q_a,\tilde{Q}_b)$.

\nlb {\it Theory $b$} - A $2d$ $U(N_f-N_c)$ gauge theory with $N_f$ fundamental flavors $(q^a,\tilde{q}^b)$ and singlet chirals $M_{ab}$, with superpotential $W=q^a M_{ab} \tilde{q}^b$.  In addition, there is a background twisted superpotential for the background twisted chiral fields:
\eqn\wbgb{\cW_{BG} = -N_ f m_A \log \left(2\; \sinh (\frac{t}{2})\right)\,.}
Although this is only a function of background fields, it is important to include it to ensure the correct matching of certain observables, such as partition functions, and certain correlation functions which can be obtained from partition functions by differentiating with respect to these background fields.

\

\item{b.}{\it Giveon-Kutasov duality}

Next we consider the duality of Giveon and Kutasov \GiveonZN.  This relates the following theories:

\nlb {\it Theory $A$} - A $3d$ $U(N_c)$ gauge theory with $N_f$ fundamental flavors $(Q_a,\tilde{Q}_b)$ and Chern-Simons level $k$, which we may take to be positive.

\nlb {\it Theory $B$} - A $3d$ $U(k+N_f-N_c)$ gauge theory with $N_f$ fundamental flavors $(q^a,\tilde{q}^b)$, Chern-Simons level $(-k)$, and singlet chirals $M_{ab}$ with superpotential $W=q^a M_{ab} \tilde{q}^b$.

\

The Chern-Simons level lifts the Coulomb branch of both theories, and hence there are no longer chiral monopole operators, and correspondingly the $V_\pm$ singlet fields are not present.  The global symmetries act on the other fields as in the previous subsection.

The reduction of theory $A$ was considered in Section $2.4$.  We found that it gives rise to a direct sum of $2d$ gauge theories, which we call theory $a$:
\eqn\ghared{
\bigoplus_{\ell={\rm max}(0,N_c-k)}^{{\rm min}(N_c,N_f)} \pmatrix{k \cr N_c-\ell} \times U(\ell)_{N_f \; {\rm flavors}}\,.
}
For theory $B$, the reduction is similar, and theory $b$ is also a direct sum:
\eqn\ghared{
\bigoplus_{\hat{\ell}={\rm max}(0,N_f-N_c)}^{{\rm min}(k+N_f-N_c,N_f)} \pmatrix{k \cr k+N_f-N_c-\hat{\ell}} \times U(\hat{\ell})_{N_f \; {\rm flavors + mesons}}
}
We see that if we match the $\ell$'th term on side $a$ with the $\hat{\ell}$'th term on side $b$ for $\hat{\ell}=N_f-\ell$, the degeneracies match and we find the same duality as in the previous subsection.  If one carefully keeps track of the twisted superpotential as a function of the background fields, including the contribution of relative Chern-Simons contact terms across the duality, one recovers the $2d$ background term, $\cW_{BG}$, above.  There are additionally contributions to $\cW$ which diverge as $r \to 0$ in different ways for different terms in the direct sum. For each such term, these can be checked to match with the corresponding term in the direct sum of the dual theory, providing further evidence for the duality mapping.

\

\item{c.}{\it Partition function}

Turning now to the partition function, this duality implies the following identity of supersymmetric indices:
\eqn\unindexident{\eqalign{&
{\cal I}_{U(N_c)_{k,N_f}}({\bf \nu}_a,\tilde{{\bf \nu}}_a,\bw; \bs_a,\tilde{\bs}_a,\bn) =\cr&\;\, \qquad {\cal I}_{U(k+N_f-N_c)_{-k,N_f}}({{\bf \nu}_a}^{-1},{\tilde{{\bf \nu}}_a}^{-1},\bw; -\bs_a,-\tilde{\bs}_a,\bn) {\cal I}_{mesons}({\bf \nu}_a,\tilde{\bf\nu}_a;\bs_a,\tilde{\bs}_a)\,,}}
where ${\bf \nu}_a,{\bf \tilde{\nu}}_a$, $a=1,\cdots,N_f$, and $\bw$ are the fugacities for the $U(N_f) \times U(N_f) \times U(1)_J$ global symmetry, and $\bs_a,\tilde{\bs}_a$, and $\bn$ are the corresponding fluxes.  We take all the chiral multiplets in theory A to have a common R-charge, $\Delta$, and then the chirals in theory B have R-charge $1-\Delta$.

Again we take the Higgs limit for the gauge theories, where the mass parameters for the flavor symmetry are held finite and the FI parameters are scaled to infinity.  This corresponds to the following limit of the index parameters:
\eqn\unindscal{
\eqalign{
{\bf \nu}_a = (-1)^{{\bf s}_a} & e^{i \tau M_a}, \;\;\; {\bf \tilde{\nu}}_a = (-1)^{{\bf \tilde{s}}_a} e^{i \tau \tilde{M}_a}, \;\;\; \bw = (-1)^{\bn} e^{i \theta}, \cr
\bs_a &= s_a, \;\;\;\; \tilde{\bs}_a = \tilde{s}_a, \;\;\;\;\; \bn =  \frac{2}{\tau} \hat{r},
}
}
where $t=\hat{r}+i \theta$ is the $2d$ complex FI parameter.\foot{We denote the real FI parameter, $\hat{r}$, with a hat to distinguish it from the radius of $\S^1$.  Here $\bn$ is an integer, but $\hat{r}$ becomes valued in $\R$ as we take $\tau \to 0$.}

Next we must choose how to scale the integration variables in the index, which correspond to the gauge fields, in order to focus on saddle points as $\tau\to 0$.  Let us first focus on theory A; a similar analysis will hold for theory B.  As discussed in Section $2.4$, there are various ways to scale the gauge field strength multiplets, each of which lead to different terms in the direct sum of theories that describes the reduction of this theory.  Correspondingly, there are several ways to scale the integration parameters in the index corresponding to different saddle points of the integral in the $\tau \to 0$ limit.   We will see that these saddle point contributions come with different divergent prefactors and, depending on the choice of background fields, one of them will be dominant and describe the leading piece in the $\tau \to 0$ limit.

Let us first consider the naive limit, where we scale the gauge parameters as:
\eqn\ungaugescala{
\bz_j = (-1)^{\bm_j} e^{i \tau Z_j}, \;\;\; \bm_j = m_j, \;\;\;\;\; j=1,\cdots,N_c.
}
Then, using \lightchcirallim\ to take the limit of the constituent chiral multiplets in the index, it is straightforward to check that limit of the index behaves as 
\eqn\ungjlimone{
{\rm Limit \; 1:} \;\;\; {\cal I}_{U(N_c)_{k,N_f}} \tra \tau^{{N_c}^2 + 2 N_c  N_f ( \Delta-1 -2 i M)} {\cal Z}_{U(N_c)_{N_f}} \,,}
where $M=\frac{1}{N_f}\sum_a (M_a+\tilde{M}_a)$, and
the finite piece on the right-hand side is the $\S^2$ partition function of the $2d$ $U(N_c)$ theory with $N_f$ flavors.

However, as expected from our discussion in Section 2.4, there are also other saddle points of the integral in the $\tau \to 0$ limit, which we can isolate by redefining the integration variables before taking the limit.  To motivate this from the point of view of the index, note that the CS and FI terms contribute to the index as
\eqn\uncsfigk{
\prod_j {\bz_j}^{k \bm_j} {\bz_j}^\bn \bw^{\bm_j} = \prod_j \bw^{-\bn/k} ( \bz_j \bw^{1/k})^{k \bm_j + \bn} \,. }
Then we may expect dominant contributions to the integral when this phase is one, which can be arranged by shifting:
\eqn\uncsgklimtwovars{
\bz_j \rightarrow \bw^{-1/k} e^{2 \pi i a_j/k} \bz_j = e^{i (-\theta + 2 \pi a_j)/k} \bz_j, \;\;\;\; \bm_j \rightarrow \bm_j - \frac{1}{k} n = \bm_j - \frac{2}{k \tau} \hat{r} }
for some choice of $a_j \in \Z_k$, and then focusing on the region near $\bz_j=1, \bm_j=0$.  If we take this scaling for $\ell$ of the variables, the contribution from the matter and vector multiplet for these variables is\foot{Here we define $M=\frac{1}{2N_f} \sum_a (M_a+\tilde{M}_a)$ and $s=\frac{1}{2N_f} \sum_a (s_a+\tilde{s}_a)$.  The limit of the contribution of the chiral multiplets is derived by noting, as in the previous section, that $\cW^{(\pm)}_{\chi,\S^2 \times \S^1_\tau}(\Sigma)$ goes over to the flat space twisted superpotential for large $\Sigma$, and taking a limit of the latter similar to the one giving \vpmlim.}
\eqn\ungkscalgenvec{
\eqalign{&\;\;\;\;\; e^{-i 2\ell \theta \hat{r}/k} {\bz_j}^{k \bm_j} \prod_{1 \leq i \leq j \leq \ell} ( 1 - e^{2 \pi i (a_i-a_j)/k}) 
 \prod_{j=1}^\ell e^{ -\frac{\hat{r}}{k}( N_c-\ell + 2N_f (\Delta-1 -  i M)) - \frac{i}{k} (-\theta+2 \pi a_j) N_f s}
\times\cr & \qquad \qquad \qquad \qquad \prod_\pm ( 1- e^{\frac{1}{k}(-\hat{r} \pm(-i \theta + 2 \pi i a_j))})^{  N_c-\ell + 2N_f (\Delta-1) - 2 i N_f M \mp N_f s }\,. 
}}
From the middle factor on the first line we see that we should set all the $a_j$ distinct, and so $\ell \leq k$.  Let us assume $k \leq N_c$, and take the maximal choice, $\ell=k$.  Then, at the cost of a symmetry factor $k!$, we can set $a_j=j$, and this simplifies to
\eqn\ungkscalgenvecsimp{ k^k e^{-2i \theta \hat{r}} {\bz_j}^{k \bm_j} e^{ -\hat{r}( N_c-k + 2N_f (\Delta-1- i M) )+ i \theta N_f s} \prod_{\pm} ( 1- e^{-\hat{r} \mp i \theta })^{  N_c-k + 2N_f (\Delta-1) - 2 i N_f M \mp N_f s }\,.
}
Then the sum over $\bm_j$ gives $\frac{1}{k} \delta(\bz_j-1)$, cancelling the $k^k$ factor.  What remains is precisely $\cZ_{BG} \equiv e^{\cW_{BG} - \bar{\cW}_{BG}}$, with $\cW_{BG}$ defined in \wbgb.  For the remaining $N_c-k$ variables we take the usual gauge theory limit.  The $k!$ symmetry factor combines with a factor of $( \,^{N_c}_{\,k} )$, from the ways of choosing which $k$ variables to shift, and with the $\frac{1}{N_c!}$ Weyl symmetry factor, to produce the $\frac{1}{(N_c-k)!}$ Weyl symmetry factor for $U(N_c-k)$, and so one finds
\eqn\gklimtwo{
{\rm Limit \; 2:} \;\;\; {\cal I}_{U(N_c)_{k,N_f}} \tra e^{-2i \theta \hat{r}} \tau^{(N_c-k)^2 +2N_f (N_c-k) (\Delta-1 - i M)} {\cal Z}_{U(N_c-k)_{N_f}} {\cal Z}_{BG}\,.
}

One can similarly take limits where $0<\ell<k$, and the various choices correspond directly to the various terms in the direct sum \unfkle.  These will all appear with different divergent prefactors, which depend on the background fields and choice of R-symmetry, as in \ungjlimone\ and \gklimtwo.  One of these saddle points will be dominant, and will determine the $\tau \to 0$ limit of the index, but as we vary parameters the dominant saddle may change. 

One can perform analogous limits on theory B.   Here one must be careful when comparing the two sides, because there is a relative background CS contact term for the $U(1)_J$ symmetry.  If we choose to put this on theory B, it is at level $-1$, and contributes a divergent phase $e^{-2 i \hat{r} \theta/\tau}$ in the limit.  Working out the corresponding two limits on theory B we find
\eqn\gklimB{
\eqalign{
& {\rm Limit \; 1:} \;\;\; {\cal I}_{U(k+N_f-N_c)_{-k,N_f}+mesons} \tra\cr
&\qquad\qquad\qquad\quad e^{-2i \theta \hat{r}} \tau^{(N_c-k)^2 +2N_f (N_c-k) (\Delta-1 - i M)} {\cal Z}_{U(k+N_f-N_c)_{N_f} + mesons} \cr & \cr
&{\rm Limit \; 2:} \;\;\; {\cal I}_{U(k+N_f-N_c)_{-k,N_f}+mesons} \tra \tau^{{N_c}^2 +2N_f {N_c} (\Delta-1 - i M)} {\cal Z}_{U(N_f-N_c)_{N_f}+mesons} {\cal Z}_{BG}\,.
}}
Comparing the scalings, we find, as above, that the limit $1$ (or $2$) of side A scales in the same way as limit $2$ (or $1$) of side B.  Thus we conjecture that when limit $1$ is valid on side A, limit $2$ is valid on side B, so that we obtain the following identity in $2d$,

\eqn\gklimfin{
{\cal Z}_{U(N_c)_{N_f}} =  {\cal Z}_{U(N_f-N_c)_{N_f}+mesons} {\cal Z}_{BG}\,.
}
This is the statement of the matching of the $\S^2$ partition functions for the duality of \refs{\BeniniMIA,\BeniniUI}.

Note that for a given choice of parameters,  we only obtain as $\tau \to 0$ the duality for  the term in the direct sum corresponding to the dominant saddle.  In general this will be the dominant saddle for some open subset of parameter space, and then we may argue for the duality in general by analytic continuation.  For other regions in parameter space, we may obtain the duality for other terms in the direct sum.

\

To summarize, we started in three dimensions with dualities depending on the level of the Chern-Simons term. Upon reduction the information of the level is encoded in the summands appearing in the direct sum of theories. The consistency of the reduction of the duality web can be viewed as a non trivial check of the dualites in both two and three dimensions.
One can also consider the reduction of the most general $U(N_c)$ dualities involving fundamental and anti-fundamental matter \BeniniMF, and one similarly finds they lead to the $2d$ $U(N_c)$ dualities discussed in \BeniniMIA.

\subsec{More dualities}

In this section we discuss reductions of dualities with gauge groups $USp(2N_c)$ and $SU(N_c)$.  The arguments here are similar to the $U(N_c)$ cases above and so we will be brief.  Some further details are discussed in Appendix B.

\

\item{a.}{\it $USp(2N_c)$ dualities: half integer CS level $\to$ $USp(2N_c)$ Hori duality}

The duality we consider \WillettGP\ is between theory A, a $USp(2N_c)$ theory with CS level $k$ and $2N_f$ fundamental chiral multiplets $Q_a$, and theory B, a $USp(2(N_f+k-N_c-1))$ theory with CS level $-k$, $2N_f$ fundamental chirals $q_a$, and $N_f(2N_f-1)$ singlet chirals $M^{ab}$, with a superpotential $\sum_{1\leq a<b\leq 2N_f} q_a q_b M^{ab}$.  Here we may take $k$ and $N_f$ half-integer provided their sum is integer, which ensures cancellation of the global anomaly.  

We first consider the case when $k>0$ is half-integer. The number of fundamental chiral fields, $2N_f$, is then odd. In this case, as we discussed in Section $2.4$, the effective theory on the circle splits to a direct sum
\eqn\spodddecompA{(A)\;\;\;\;\;  USp(2N_c)_k , \; 2N_f \;\;\; \rightarrow \;\;\; \bigoplus_{\ell={\rm max} (0,N_c-k+\frac{1}{2})}^{{\rm min}(N_f-\frac{1}{2},N_c)} \pmatrix{{k-\frac{1}{2}}\cr{N_c-\ell}} USp(2 \ell) , \;N_f}

The dual theory splits to 
\eqn\spodddecompB{(B)\;\;  USp(2(N_f+k-N_c-1))_{-k} , 2N_f \;\; \rightarrow \;\; \bigoplus_{{\hat \ell}={\rm max} (0,N_f-N_c-\frac{1}{2})}^{{\rm min}(N_f-\frac{1}{2},N_f+k-N_c-1)} \pmatrix{{k-\frac{1}{2}}\cr{N_f+k-N_c-1-{\hat \ell}}} USp(2 {\hat \ell}) , \;N_f}
Here we also have the singlet fields $M_{ab}$ which reduce to two dimensions. Then we claim that the term with $\ell$ in the sum appearing in the reduction of $(A)$ is dual to the term with $(N_f-\frac{1}{2}-{\hat \ell})$ in the sum for $(B)$. All these dualities are of the same form, mapping $USp(2N_c)$ with $2N_f$ fundamental chirals to a dual $USp(2(N_f-N_c-\frac{1}{2}))$ with $2N_f$ fundamental chirals and singlet mesons. This is the $2d$ duality discovered by Hori \HoriPD.

\

\item{b.}{\it $USp(2N_c)$ dualities: integer CS level}

We can also consider the reduction of dualities when $k$ is integer.  Let us start with the simplest case of vanishing $k$ \AharonyGP.   Here theory $A$ is a $USp(2N_c)$ gauge theory with $2N_f$  (even) fundamental chiral multiplets $Q_a$.  Theory $B$ is a $USp(2(N_f-N_c-1))$ gauge theory with $2N_f$ fundamental chiral multiplets $q_a$, $N_f(2N_f-1)$ chirals $M^{ab}$, $a,b=1,\cdots,N_f$, and a singlet chiral $Y$, with a superpotential
\eqn\spahapot{
W = \sum_{a<b} M^{ab} q_a q_b + Y \widetilde{Y}\,,
}
where $\widetilde{Y}$ is the monopole operator, which parameterizes the Coulomb branch of the massless theory.  We saw in Section $2.4$ that these theories reduce in a straightforward way to the corresponding $2d$ gauge theories.  Thus we are led to the
following two $2d$ theories being dual :

Theory A : $USp(2N_c)$ with $2N_f$ fundamental chirals, 

Theory B : $USp(2(N_f-N_c-1))$ with $2N_f$ fundamental chirals with singlet mesons and $Y$.  

Since the monopole operator $\widetilde{Y}$ becomes very heavy and decouples in the $2d$ limit we are considering, we expect the superpotential \spahapot\ for $Y$ to become trivial in $2d$, such that $Y$ becomes a free field in theory B. This leads to a prediction that the dual theory A has an additional accidental $U(1)$ symmetry at low energies, acting on a free field.  It would be interesting to gather additional evidence for this conjecture. Theories A and B have exactly matching $\S^2$ partition functions, as long as we fix the global charge of $Y$ to be the one imposed by the monopole superpotential \spahapot\ in $3d$. In principle the R-symmetry could mix with the new accidental symmetry, but we do not know how to take this into account in theory A.

One can similarly reduce the duality at integer $k>0$.  Here one finds that the terms in the first line of \spevendecompmtapp\ on one side of the duality map to the terms on the second line on the dual side, and vice-versa.  This means that on one side of the duality there is an extra twisted LG model, which may be dualized by the Hori-Vafa duality into a chiral field $Y$ charged under the $U(1)_A$ global symmetry.  Assuming this we are led to precisely the same duality as above.

\

\item{c.}{\it $SU(N_c)$ dualities $\to$ $SU(N_c)$ Hori-Tong duality and new dualities}

Here we briefly comment on the reduction of $SU(N_c)$ dualities.  Our starting point is the duality of \refs{\AharonyDHA,\ParkWTA,\AharonyUYA}, which relates:

\

\noindent{\it Theory A} - $SU(N_c)$ with Chern-Simons level $k$, $N_f$ fundamental flavors, and $N_a$ anti-fundamental flavors.  Here we can use charge conjugation and parity to take $k\geq 0$ and $N_f \geq N_a$, and we also impose:
\eqn\sundualitycondition{ 0 \leq  k  < \frac{N_f-N_a}{2} }

\noindent{\it Theory B} - Whenever \sundualitycondition\ is satisfied, this has a dual $3d$ description with gauge group $SU(N_f-N_c)$, Chern-Simons level $-k$, and $N_f$ fundamental flavors and $N_a$ anti-fundamental flavors, plus $N_f N_a$ uncharged mesons, $M_{ab}$, which couple via a superpotential:
\eqn\sundualitysuperpot{ W = M_{ab} q^a \tilde{q}^b. }

One can also find a dual description when \sundualitycondition\ is not satisfied, where the dual theory has a $U(1) \times SU(\hat{N}_c)$ gauge group, but we will not consider this case here.

The reduction of these theories was discussed in the previous section.  We found that the reduction of the $3d$ $SU(N_c)$ theory with a given matter content is, in general, the direct sum of the $2d$ $SU(N_c)$ theory with the same content, plus possible additional contributions.  Although we were not able to characterize these additional contributions in complete generality, we conjecture that, provided we impose \sundualitycondition\ as above, the summand corresponding to the $2d$ $SU(N_c)$ theory maps under the duality to the summand corresponding to the $2d$ $SU(N_f-N_c)$ theory.  This would imply the following $2d$ duality:

\

\noindent{\it Theory a} - $SU(N_c)$ with $N_f$ fundamental flavors and $N_a<N_f-1$ anti-fundamental flavors.\foot{The condition $N_a<N_f-1$ arises because we impose $0 \leq  2k  < N_f-N_a$, which excludes $N_f-N_a=0,1$. Note that we must have $k+(N_f-N_a)/2 \in \Z$.}

\noindent{\it Theory b} - $SU(N_f-N_c)$ with $N_f$ fundamental flavors, $N_a$ anti-fundamental flavors, and $N_f N_a$ mesons, with the same superpotential as in \sundualitysuperpot.

\

For $N_a=0$, this reproduces a known duality, some superpotential deformations of which were discussed in \HoriDK.  For $0<N_a<N_f$, we believe this is a new duality.  However, we will see in Section $4$ that further restrictions must be imposed on the parameters in order for this duality to hold in the massless theory.

\

\item{d.}{\it Duality appetizer $\to$ $SO(1)/O(1)$ Hori duality}

Finally, we consider the $3d$ duality of ~\JafferisNS:

\nlb\noindent{\it Theory $A$} - $SU(2)$ gauge theory with a level $1$ Chern-Simons term and a single adjoint field, $Z$.

\nlb\noindent{\it Theory $B$} - A free chiral field, $X$.\foot{In addition there is conjectured to be a decoupled topological theory, which is important to obtain a precise matching of vacua across the duality, as discussed in Appendix B.}

\

Under this duality the operator $X$ is identified with $\Tr(Z^2)$. 
 Both descriptions have a $U(1)$ global flavor symmetry and a $U(1)$ R-symmetry.  To be consistent with the duality we assign to $X$ twice the charges of the field $Z$.  Evidence for this duality can be obtained, for example, by showing that various partition functions for this pair of theories agree.

As before, the reduction on a circle of the free side of the duality is straightforward;
we take the usual limit for the global symmetry and obtain
the zero momentum $KK$ mode as the $2d$ chiral field. On the other side of the duality the reduction
is trickier due to the presence of the CS term. The CS term is crucial for the duality to hold in three
dimensions and it has a non-trivial effect on the reduction.  

In Appendix B we discuss the reduction of this side of the duality in more detail.  As described there, 
we can obtain from this reduction the following duality of \refs{\HoriPD,\DiFrancescoTY}:

\nlb\noindent{\it Theory $a$ - } The $\Z_2$ orbifold of a free chiral field $X$.  Because of the orbifold, the basic gauge-invariant operator is $X^2$, and there is a quantum $\Z_2$ global symmetry acting on twisted sector states.

\nlb\noindent{\it Theory $b$ - } An LG model with chiral fields $Y$ and $Z$ and a superpotential $Z Y^2$.  Here the field $Z$ maps to $X^2$ on the dual side, while $Y$ maps to a twisted sector state.  The quantum $\Z_2$ symmetry of theory $a$ maps to the symmetry taking $Y \rightarrow -Y$.

\

We describe further evidence for this duality using supersymmetric partition functions in Appendix B.

\newsec{The reduction of theories with non-compact moduli spaces}

In three space-time dimensions or more, the matching of the effective superpotential and all chiral observables, discussed in the previous sections, provides very convincing evidence for IR dualities. In two space-time dimensions, this is not always the case in the presence of spaces of classical supersymmetric vacua (``moduli spaces''), because the metric on the moduli space, which is not protected by supersymmetry, is a classically marginal operator which cannot be ignored in the IR dynamics. In addition, wavefunctions spread out over the whole moduli
space (which is called the ``target space''). 

For compact target spaces these are not serious problems, as there is typically a finite number of supersymmetric vacua (given by the Witten index). However, for non-compact target spaces there may be some normalizable supersymmetric vacuum near the origin, but there is always a continuum of states describing scattering states coming in from infinity and going back out. The conformal field theory describing the low-energy dynamics then has a continuous spectrum of operators.

This leads to several issues in formulating and testing IR dualities for $2d$ supersymmetric theories. Given two theories that are conjectured to be dual to each other, the UV metrics (K\"ahler potentials) on the moduli space are generally not the same. They are marginal operators that flow in a way that is not controlled by supersymmetry, and it is hard to say whether they flow to the same fixed point or not; this is necessary for an IR duality, in addition to the matching of all protected objects. Moreover, the metrics of putative dual theories are often different even in the asymptotic regions; this is not an issue in higher dimensions where the metric is an irrelevant operator, but it is a serious problem in two dimensions where the theory explores the full moduli space and is sensitive to its metric. In addition, the $\S^2$ partition function often diverges in these cases, and the elliptic genus generically gives non-integer values that are difficult to interpret, making it difficult to test dualities by the usual tools.

Note that having a non-compact moduli space is not distinguishable from having supersymmetric vacua at infinity, since these also lead to a continuum. A famous example is the cigar/Liouville duality. So when we refer to non-compact moduli spaces in this paper, we will refer to both cases.

Note also that these problems arise already for $3d$ theories compactified on a very large circle, since their low-energy theory is already two dimensional. So even two theories that are IR-dual in three dimensions will often not be dual when they are compactified on a finite circle, because of their different metrics.

We begin in Section $4.1$ by describing how the partition functions behave in non-compact CFTs. In Section $4.2$ we discuss IR dualities between $2d$ non-compact CFTs. In Section $4.3$ we discuss the reduction on a circle of $3d$ theories with non-compact moduli spaces, and in Section $4.4$ we discuss some examples of reductions of dual theories. Finally, in Section $4.5$ we discuss the behavior of theories with several non-compact branches of their moduli space.

\subsec{Partition functions of non-compact CFTs}

Some useful tools for studying $\cN=(2,2)$ theories are the supersymmetric partition functions on $\S^2$ and $\T^2$, reviewed in Appendix A.  Here we mention some issues that arise in computing these partition functions for theories with a non-compact target space.

First consider the $\S^2$ partition function.  As discussed in Appendix A, this depends on background twisted chiral multiplets.\foot{More precisely, this is true when we put the theory on $\S^2$ in a way which preserves the $U(1)_V$ R-symmetry; there is another compactification preserving the $U(1)_A$ R-symmetry, which is related by a $\Z_2$ automorphism of the supersymmetry algebra, but we will not consider it in this paper.}  For a compact theory, the $\S^2$ partition function computes the K\"ahler potential $K_{t.c.}$ on the portion of the conformal manifold involving exactly marginal deformations by twisted chiral multiplets $\lambda$ \GomisYAA:
\eqn\stwok{
{\cal Z}_{\S^2} = (l \Lambda_{UV} )^{c/3} e^{-K_{t.c.}(\lambda,\bar{\lambda})}\,,
}
where $l$ is the radius of the $\S^2$.
On the other hand, for a non-compact theory, the $\S^2$ partition function  does not converge unless one also turns on background values of vector multiplets coupled to flavor symmetries, which lift the non-compact moduli space.  The resulting object is a function of these background values, with poles along loci in parameter space where non-compact branches open up.  Because the non-compact directions on the target space are lifted whenever the $\S^2$ partition function is well-defined, it is insensitive to the asymptotic K\"ahler potential in these directions.

Next consider the $\T^2$ partition function, or elliptic genus.  For a compact $\cN=(2,2)$ theory, which has a discrete spectrum of states on $\S^1$, this has the interpretation of a trace over states in the RR sector:
\eqn\ellgentrace{ {\cal I}(y) = {\rm Tr} \left[(-1)^F y^{J_L} \right]\,,}
where $J_L$ is the left-moving $R$-charge.  If we take the limit $y \rightarrow 1$, we obtain the Witten index, which is an integer.

As with the $\S^2$ partition function, for a non-compact theory the flat directions lead to a divergence in the naive $\T^2$ partition function above, and we typically need to turn on background values for vector multiplets coupled to flavor symmetries, to lift at least parts of the non-compact moduli space (``Higgs branches'').  Here these are flat background connections (Wilson lines on the cycles $\gamma_{1,2}$ of the torus) for the $i$'th component in the flavor symmetry group.  If we define:
\eqn\flavorparameters{ \nu_a = e^{2 \pi i (\int_{\gamma_1} A^{(a)} + \tau \int_{\gamma_2} A^{(a)})}\,, }
then the elliptic genus is in general a meromorphic function of $y$ and the $\nu_a$.  We may again interpret this as a trace:
\eqn\ellgentraceflavor{ {\cal I}(y,\nu_a) = {\rm Tr}_{{\cal H}_{\nu_a}} \left[(-1)^F y^{J_L} {\nu_a}^{F_a}\right]}
where $F_a$ are the flavor symmetry generators.   The trace is now taken in a ``twisted sector'' of the Hilbert space, ${\cal H}_{\nu_a}$, where the fields are given twisted boundary conditions on the spatial circle due to the flat connection for the flavor symmetry group.

The limit $y \rightarrow 1$ is more subtle in the non-compact case. In general we may obtain a fractional answer after regulating the contributions from infinity.  For example, the elliptic genus of the $SU(2)$ gauge theory with $N_f$ flavors satisfies:
\eqn\sutwowi{ \lim_{y \rightarrow 1} {\cal I}(y,\nu_a) = \frac{N_f-1}{2}\,. }
For generic $\nu_a$ the Higgs branch of this theory is lifted.
For odd $N_f$, the theory has no Coulomb branch because an effective theta angle $\theta=\pi$ is generated, and the $y \rightarrow 1$ limit of the elliptic genus agrees with the counting of vacua (the vacua are discrete for generic $\nu_a$).  However, for $N_f$ even, there is an unlifted non-compact Coulomb branch for any $\nu_a$, and this apparently contributes $-\frac{1}{2}$ to the elliptic genus (rather than a divergence as above).  As $\nu_a \to 0$ there are two non-compact branches in the moduli space; we will discuss this situation in Section $4.5$ below.

Another interesting case is the elliptic genus of theories with an asymptotically cylindrical target space, such as a sigma model on a cigar.
These theories have a continuum of states even after twisting by flavor symmetries. As a result, the usual cancellation between bosons and fermions does not hold precisely, leading to a non-holomorphic result \refs{\TroostUD,\HarveyNHA}.  

The discussion above implies that the elliptic genus is in some ways a more refined observable of a CFT than the $\S^2$ partition function, since it gets contributions from different branches including vacua at infinity, and it also seems sensitive to some features of the asymptotic K\"ahler potential on non-compact branches of the theory. We will see below examples of putative dual pairs of $2d$ theories, whose $\S^2$ partition functions match, but whose $\T^2$ partition functions do not, and we will discuss the implications.

\subsec{IR dualities}

Let us now consider various IR dualities between different ${\cal N}=(2,2)$ theories, some of which were discussed in the previous section. Several different behaviors are possible.

The simplest case is when two different theories flow at low energies to the same compact CFT. This case is similar to higher dimensional IR dualities; the speed of the flow is controlled by the dimension of the lowest irrelevant operator consistent with the symmetries of the problem.

When the two theories are non-compact the situation is more complicated. 
In such a case the spectrum is generally continuous. The continuous operators are not normalizable (they are delta-function normalizable) so they will not be generated during a renormalization group flow, but the classically-marginal metric still has a non-trivial flow. We expect the low-energy description to be given by a sigma model on the space of classical vacua, or on a space where we have a superpotential whose derivative vanishes at infinity. Namely, it is given by some (twisted) chiral superfields, with some effective superpotential which can be computed exactly, and with some metric (K\"ahler potential).

One option is that the asymptotic regions could have angular directions which approach a finite size at infinity. The sigma model description is valid when the curvature of their metric is small. Such a space can be a fixed point of the renormalization group flow (or an approximate one) when the asymptotic compact directions are Ricci-flat; the simplest case is a cylinder, with an asymptotic circle. When this is the form of the metric at high energies, it seems likely that it will remain so also at low energies, namely that the theory will flow to a fixed point describing a sigma model of the same asymptotic form; the only things that can change are normalizable metric deformations near the origin, or a linear dilaton may be generated along the radial direction \HoriAX. These are typically irrelevant operators in the IR, suggesting that as long as the asymptotic metric is the same, IR dualities are plausible. Examples are Liouville theory -- a chiral superfield $X$ with a periodic identification of its imaginary part, a $W = \mu e^{-X}$ superpotential, and a metric which is flat at $X \to \infty$ -- and sigma models on spaces with the topology and asymptotic geometry of a cigar.

In \HoriAX\ an IR duality was suggested between this supersymmetric Liouville theory and a specific gauge theory. The gauge theory is a $U(1)$ gauge theory coupled to two chiral superfields, on one of which the gauge transformation acts in the standard way $\Phi \to e^{i\alpha} \Phi$, while the other is twisted by it, $P \to P + i \alpha$. All the fields have canonical metrics at high energies. It was shown in \HoriAX\ that the low-energy description of the gauge theory, below the scale set by the gauge coupling, is a sigma model on a space which is asymptotically a cylinder. This is the same asymptotic metric as in the Liouville theory, and it has finite angular directions at infinity. A detailed renormalization group analysis showed that the two theories flow to the same theory, which from the point of view of the sigma model may be viewed as a cigar. This is an example of an IR duality between two UV theories with different metrics, that flow to the same non-compact IR CFT.

The other option is that all the angular directions could open up at infinity, including the simplest case of free fields. In this case the renormalization group flow can be much more complicated. We argued in \AharonyJKI\ that a typical renormalization group flow can change the asymptotic form of the metric, so that even two theories whose high-energy metric is different asymptotically on the moduli space can flow to the same theory at low energies. More precisely, the renormalization group flow leads to a change in the metric, first near the origin of the moduli space, and then gradually it flows out towards infinity. In such a duality, any computations that are localized in the interior of the moduli space would be the same on both sides at low energies, though it takes a longer and longer renormalization group time for this to be true as one goes out on the moduli space (and, even in the interior, the presence of many classically marginal operators means that one approaches the fixed point quite slowly). The metric and observables at infinity stay different for any finite renormalization group time. Two examples of this type were discussed in detail in \AharonyJKI, and we will mention further examples below.

As a typical example, consider a duality proposed in \HoriKT\ between a $U(1)$ theory with $N_f$ chiral superfields of charges $Q_i$ ($i=1,\cdots,N_f$, $\sum_{i=1}^{N_f} Q_i=0$) and an FI parameter $t$, and a theory of $N_f$ twisted chiral superfields $Y_i$, subject to a constraint $\sum_{i=1}^{N_f} Q_i Y_i = t$ and a periodicity $Y_i \equiv Y_i + 2\pi i$, with a twisted superpotential 
\eqn\ysuppot{\cW = \mu \sum_{i=1}^{N_f} e^{-Y_i}.}
 For generic values of $t$ the gauge theory flows to some sigma model on its non-compact $(N_f-1)$-dimensional Higgs branch, and the theory of the twisted chiral superfields is also described by a sigma model on a similar space, with the superpotential \ysuppot\ implying that all vacua are at infinity. 

Our considerations above make it clear that in order to state the duality we need to specify not just the superpotential of the $Y_i$'s, but also their K\"ahler potential at high energies. The gauge theory is asymptotically free and its metric for the chiral superfields is flat at high energies. The Higgs branch may be described in terms of gauge-invariant variables formed from these chiral superfields; the metric formed from such variables is generally no longer flat, and it does not have any angular directions that do not grow at infinity. In the dual theory, it is natural to choose the UV metric for the $Y_i$ to be flat. However, in that case the metric at infinity is $(N_f-1)$ copies of a cylinder, with angular directions that do not grow. As described above, we do not expect the renormalization group to change the behavior of such compact directions at infinity, so with this UV metric the two theories are not dual to each other. If we choose a different metric in which the angular metric of the space of $Y_i$ grows at infinity, it is possible that the two theories would be dual to each other; in order to give more evidence for this one would need to analyze the RG flow for any specific value of the high-energy metric. Note that the (related) model of \HoriAX\ does not suffer from these problems.

\subsec{$3d$ theories on a circle}

$3d$ ${\cal N}=2$ gauge theories can have both Coulomb branches, labeled by scalars in vector multiplets, and Higgs branches, labeled by scalars in chiral multiplets. The geometry of the Coluomb branch is asymptotically cylindrical, with the angular directions coming from the dual photons on the moduli space. On the other hand, on the Higgs branch all the angular directions grow towards infinity. In the $3d$ theory, the metric and its asymptotic behavior play no role in the low-energy dynamics.

When we put these theories on a circle, the moduli space and its asymptotic metric generally retain the same form. Let us define $\gamma = g^2 r$, where $g$ is the $3d$ gauge coupling. The asymptotic angular directions on the Coulomb branch can now be described either using the $3d$ dual photons, which have (when canonically normalized) a periodicity $\sqrt{\gamma}$ (up to constants that we will not be interested in), or using the holonomies of the $U(1)$ gauge fields on the circle, which are T-dual to the dual photon variables, and have periodicity $1/\sqrt{\gamma}$. The first description is in terms of chiral multiplets, and the second in terms of twisted chiral multiplets, whose zero modes can be viewed as two dimensional vector multiplets.

In the theory on a circle the metric and its asymptotic behavior play a crucial role in the low-energy dynamics. As mentioned above, we do not expect the RG flow of the metric to change the asymptotic behavior when the angular directions remain finite at infinity. Thus, the IR theories related to the Coulomb branch will be different for the same theory with different values of $g^2 r$; note that this is true already for a free $3d$ vector multiplet on a circle, whose IR limit includes a compact scalar with a radius depending on $\gamma$. Clearly, in such cases with non-compact Coulomb branches we cannot expect to have any IR dualities for the theory on a circle, given that even the same gauge theory has different IR limits depending on its gauge coupling. Note that the low-energy theory on Higgs branches is expected to be independent of $\gamma$, so this is not a problem for dualities between Higgs branch CFTs, and we believe that these can still hold in $2d$ (when both Higgs and Coulomb branches exist, they decouple in the IR, as we discuss in Section $4.5$).

When we take the $r\to 0$ limit, the behavior of the low-energy theory on Coulomb branches will depend on how we scale $g^2$ in this limit. If $\gamma$ remains fixed we retain a finite circle at infinity also in the $2d$ limit. If $\gamma$ goes to zero, and in particular when $g_{2d}^2=g^2 / r$ remains fixed, corresponding to reducing to a $2d$ gauge theory, the valid description is in terms of the holonomy, which sits in a $2d$ vector multiplet (or twisted chiral multiplet) with an asymptotically flat metric. If $\gamma$ goes to infinity the $2d$ limit is best expressed in terms of the same chiral multiplet parameterizing the Coulomb branch in $3d$, whose angular direction now grows at infinity.   When we have two IR dual $3d$ theories, there is always one scaling of the $3d$ metrics that are dual, which corresponds to first flowing to the CFTs in $3d$ and only then reducing on a circle, but with this scaling the $2d$ theory may be singular.  In many cases choosing the canonical UV metric on both sides will not lead to an IR duality, and in some cases there is even no choice of the UV metric on both sides that will give the same non-singular theory in the 2d limit

As a first example consider $3d$ ${\cal N}=2$ pure gauge theories on a circle, starting with $G=SU(2)$. The $3d$ ${\cal N}=2$ theory has a runaway superpotential $1/Y$ in the monopole variable $Y$ \AffleckAS, which is a $3d$ chiral superfield labeling the Coulomb branch. Its asymptotic metric is a cylinder of radius $g$ in the coordinate $X\equiv \log(Y)$, and it has no supersymmetric vacuum. The classical moduli space has a $\Z_2$ identification coming from the Weyl group of $SU(2)$, but the variable $X$ does not have any $\Z_2$ action ($Y$ is the square of the naive $U(1)$ monopole). In the natural dimensionless asymptotic coordinate, $X$, the superpotential goes as $W = \mu e^{-X}$ for some scale $\mu$.

On a circle, at energies below $1/r$, we get asymptotically a sigma model on a cylinder of radius $\sqrt{\gamma}$, still with the superpotential $W = \mu e^{-X}$. It is natural to guess that this flows to a $2d$ Liouville theory in the IR, which has the same asymptotic metric and superpotential (even though the precise metric coming from the gauge theory is different). Note that the central charge of this IR theory depends on $\gamma$.

As we decrease $\gamma$ it is natural to T-dualize the asymptotic circle, such that we have a sigma model in terms of a twisted chiral superfield $\Sigma$ whose imaginary part is asymptotically the T-dual of ${\rm Im}(X)$ (and has periodicity $1/\sqrt{\gamma}$). The asymptotic metric of this sigma model is a cylinder, and there is no twisted superpotential (for any radius of the circle). The T-duality tranformation is not well-defined close to the origin of $\Sigma$, but $\Sigma$ itself (containing the holonomy on the circle) is well-defined, and it is natural to guess that we can describe the theory by a smooth sigma model in this variable, and that the theory flows to a sigma model on a cigar, which is dual to the Liouville theory of the previous paragraph. This is qualitatively similar to the flow  described in \HoriAX. 

If we take the $2d$ limit with $\gamma \to 0$ the radius of the cigar goes to infinity, so that we obtain in this limit a free twisted chiral superfield.
This agrees with the low-energy description of the $2d$ pure $SU(2)$ theory, that was discussed in \AharonyJKI. In terms of the $2d$ twisted chiral multiplet $\Sigma$ (equivalent to the $2d$ vector field)  the moduli space is asymptotically $\R^2/\Z_2$, and we conjectured in \AharonyJKI\ that at low energies the singularity is smoothed out near the origin in the variable ${\tilde Y} = {\rm tr}(\Sigma^2)$, so that the theory flows (after an infinite renormalization group time) to a free twisted chiral superfield ${\tilde Y}$. Note that in this description the metric flows in a complicated way under the renormalization group flow, and even changes its asymptotic form. This seems different from our picture of the previous paragraph where the asymptotics is unchanged. However, the precise relation between $\Sigma$ and ${\tilde Y}$ is complicated (even in the asymptotic region); in particular in the $3d$ theory there is a $\Z_2$ action taking ${\rm Re}(\Sigma) \to -{\rm Re}(\Sigma)$ and only the region of positive ${\rm Re}(\Sigma)$ is kept, while in the $2d$ $\Sigma$ variable the identification is $\Sigma \to -\Sigma$ and there is no identification on the coordinate ${\tilde Y}$. But at low energies for any finite distance on the moduli space we flow to a free twisted chiral superfield in either of these variables, so they become equivalent.

A similar picture is presumably valid for general $G$. The theory on a circle may be described by ${\rm rank}(G)$ chiral superfields with a runaway superpotential for each of them, leading to ${\rm rank}(G)$ copies of Liouville theory with some identifications relating them. For a small circle it is more natural to T-dualize this to ${\rm rank}(G)$ twisted chiral superfields with a complicated metric, and as $\gamma \to 0$ this presumably flows to ${\rm rank}(G)$ free twisted chiral superfields. This is also the expected low-energy description for the $2d$ pure SYM theory.

\subsec{Additional examples}

\item{a.} {\it The duality of $U(1)_{-1/2}$ to single chiral superfield}

In the first example of Section $3.1$ we discussed the reduction on a circle of the duality between the $3d$ $U(1)_{-1/2}$ theory and the $3d$ theory of a free chiral superfield. We saw that for a generic FI parameter (mass on the free chiral side) this leads to the $2d$ Hori-Vafa duality \HoriKT\ between the Landau-Ginzburg model \efw\ and a free $2d$ chiral superfield.
 
However, when $m\to 0$ we obtain on both sides a non-compact theory with a continuous spectrum, and we need to be careful about the K\"ahler potential.
If we start from a finite $3d$ gauge coupling $g$ and reduce to two dimensions, the $3d$ $D$-term ${1\over g^2} |\Sigma|^2$ becomes, in terms of $X$ defined in Section $2.3$,
\eqn\efD{  \int d^2x \int d^4 \theta \frac{1}{4 \pi^2 g^2 r} | X|^2. }
If we define a dimensionless parameter $\gamma=g^2 r$ and hold this finite as $r\to 0$, then this action, together with the twisted superpotential \efw, gives (for $\zeta=0$) the ${\cal N}=2$ supersymmetric Liouville theory, in which the asymptotic metric is a cylinder. (This is also what we get at low energies for finite $r$.) This theory depends on $\gamma$, and for all finite values of $\gamma$ it is not dual to a free chiral, but rather to an $SL(2,\R)/U(1)$ coset model at level $k=2\pi^2 \gamma$ \refs{\GiveonPX,\HoriAX}, or equivalently, to a sigma model on a cigar with asymptotic radius proportional to $\frac{1}{\sqrt{\gamma}}$. For any other scaling of $g$ we do not get a finite K\"ahler potential for $X$ in the limit in which its superpotential \efw\ is finite. On the other hand, on the free chiral side there is no problem in scaling the K\"ahler potential to remain finite as $r\to 0$.

Let us see what went wrong with the naive argument. We started with the $3d$ theory at finite $g$, which is not exactly dual, but only IR dual, to a free chiral.  The operations of flowing to the IR and reducing on $\S^1_r$ evidently do not commute. In $3d$, for ${\rm Re}(\Sigma) \gg g$ the metric is flat in $\Sigma$ and not in the natural Coulomb branch coordinate $V_+$ which maps to the free chiral field, and it has the geometry of a cigar with asymptotic radius $g$. In three dimensions this is still a free field theory in the IR at any point on the Coulomb branch, because the $3d$ metric on the moduli space is irrelevant in the IR. However, when we reduce to $2d$, for any value of $r$ we get a $2d$ sigma model on this cigar rather than a free scalar field, since the $2d$ metric is classically marginal and must be included in the IR description, consistent with the discussion above. 
The bottom line is that the reduction from $3d$ does not directly lead to the $2d$ Hori-Vafa duality \HoriKT, for any choice of the $3d$ K\"ahler potential. It may or may not be possible to choose a $2d$ K\"ahler potential for \efw\ that would lead to a $2d$ duality.

One way to rectify this is to use the proposal of \KapustinHA\ for a $3d$ dual of Theory $A$ (the $U(1)_{-1/2}$ theory) at finite $g$, namely along the full renormalization group flow, not just at the IR fixed point.  Then, as shown in \AganagicUW, carefully reducing the all-scales dual theory $B$ with some fixed value of $\gamma$, one finds the description proposed in \HoriAX\ for a theory which flows to the $SL(2,\R)/U(1)$ coset model at level $k=2\pi^2 \gamma$, which is known to be dual to the ${\cal N}=2$ Liouville theory.

\vskip 20pt

\item{b.} {\it $SU(N_c)$ and $USp(2N_c)$ dualities}

In Section $3.4$ we discussed the reduction of the $3d$ duality between $SU(N_c)$ theories with $N_f$ fundamental and $N_a$ anti-fundamental flavors to $2d$, and we argued that for $N_a < N_f-1$ the reduction implies that the same duality should hold also in $2d$. The reduction indeed implies a duality of the mass-deformed theory with discrete vacua, and a matching of the $\S^2$ partition functions, but, as discussed above, it does not necessarily imply a $2d$ duality of the massless theories.

The matching of the elliptic genus gives an independent check of the $2d$ duality, as it does not come from a limit of a $3d$ partition function. We have computed the elliptic genera on the two sides and found agreement only when
\eqn\suntwoddualitycondition{  N_a< N_c < N_f-N_a \,.}
The failure of the duality for other values of $N_a$ is directly related to the issues with non-compact branches that we discussed above. 

The $3d$ theories that we started from, with $0 \leq k < (N_f-N_a)/2$, have no Coulomb branches, since there is always an effective CS term on these branches. Note that the $2d$ $SU(N_c)$ theories do have non-compact Coulomb branches, but since these branches do not arise from a reduction of the $3d$ Coulomb branches, the problems mentioned above with the reduction of dualities with non-compact Coulomb branches do not arise.  However, when (say) $N_a \geq N_c$, the $3d$ $SU(N_c)$ theory has a non-compact branch on which anti-baryonic ${\tilde Q}^{N_c}$ operators obtain vacuum expectation values. The D-terms imply that such vacuum expectation values must come together with expectation values for the vector multiplet scalars, such that these branches are really mixed Higgs-Coulomb branches, and the asymptotic metrics on these branches lead to problems after the reduction, just like in the examples discussed above. Thus, we believe that the duality only holds in $2d$ when the additional condition \suntwoddualitycondition\ is satisfied; in other cases it does not work because of different asymptotic behaviors of the metric on this non-compact branch. This is consistent with the behavior of the elliptic genera.

A similar discussion applies for the $USp(2N_c)$ theories discussed in Section $3.4$.  In $3d$, for $2N_f$ odd, the Coulomb branch is lifted, and so we expect the duality to reduce to two dimensions.  Indeed, we saw above that this reduces to the duality of \HoriPD, and one can check that the elliptic genus matches across this duality.  However, for $2N_f$ even, the situation is more subtle.  If $k=0$, there is an unlifted Coulomb branch, which maps across the duality to a singlet chiral field, and as above we expect the metrics on the two sides of the duality to differ when we put the theories on a circle and go to two dimensions. And for $k>0$ we found that the reduction of the $2d$ duality includes the Hori-Vafa duality which has the issues discussed above. Thus, the $3d$ duality does not imply a $2d$ duality, and indeed, the elliptic genera fail to match.  However, since we expect the Higgs and Coulomb branches to flow to two decoupled CFTs (as we discuss in more detail below), we may still conjecture a duality between the Higgs branches, even when $2N_f$ is even.  It would be interesting to gather further evidence for this duality, \eg\ by isolating the contribution of each branch to the elliptic genus.

\subsec{Theories with multiple non-compact branches}

Additional issues arise in theories which have more than one non-compact branch in their moduli space.
When there are two (or more) non-compact unlifted branches that classically intersect
(say) at a point at the origin, there are two possibilities. One is that the branches join
together smoothly, and that there is a single CFT describing all the
non-compact regions. The other is that the branches split into a direct sum of separate
CFTs at low energies, and that a semi-infinite throat develops 
in the target space of each of these CFTs near the
origin due to quantum effects. This possibility is believed to happen in ${\cal N}=(4,4)$
supersymmetric gauge theories \refs{\AharonyTH,\WittenYU}, and it can happen also in
${\cal N}=(2,2)$ theories; a famous example is the gauged linear sigma model for the conifold. In such a case we obtain a discrete sum of different CFTs at low energies, and IR dualities may relate some or all of these different CFTs. Note that this discrete sum which arises in $2d$ theories should be distinguished from the one discussed above that arises in $r\to 0$ limits of $d=3$ theories, though the practical consequences of the two situations are similar.

In the case of a theory which flows to an IR fixed point with multiple branches appearing in a direct sum, the $\S^2$ partition function will, a priori, be a sum of contributions from these summands.  However, because of the prefactor $(l \Lambda_{UV} )^{c/3}$ in \stwok, as we take the UV cutoff to infinity only the contribution from the term (or terms) with the largest value of $c$ survives.

The elliptic genus will also be a sum of contributions from these summands.  Unlike the $\S^2$ partition function, since there is no divergent prefactor in the elliptic genus, these may contribute with equal weight. In the previous section we saw that in the $y \to 1$ limit the elliptic genus of $SU(2)$ with an even number of flavors receives an integer contribution from the Higgs branch, plus $-\frac{1}{2}$ from the Coulomb branch, whenever it exists; we expect the two branches to decouple in the IR when both exist. In general it may be difficult to separate the elliptic genus into contributions from the various summand theories.

As in ${\cal N}=(4,4)$ theories, one way to argue for an IR splitting of
branches is by computing the central charge of the CFT associated
with each branch, using its asymptotic region (in particular this can
easily be done when the asymptotic metric is flat). If different branches
have different central charges then they cannot be part of a single
CFT. 
Note that for non-compact CFTs there is no $c$-theorem that can be used to control these flows. 

Another argument involves the $U(1)_R$ currents; an IR SCFT
must have left-moving and right-moving $U(1)_R$ currents that are
part of the superconformal algebra. These currents must be well-defined
local operators, so they cannot rotate angular directions that grow in asymptotic regions
of the moduli space. Assuming that the $U(1)_R$ symmetries are visible
in the UV, this can be used to identify the $U(1)_R$ symmetries in the
IR SCFTs describing the different branches, and if these differ between
branches then again each branch must belong to a separate IR CFT.
This argument is less rigorous since it can be bypassed by additional
R-symmetries appearing in the IR. In addition, as we discussed above, the RG flow can change the asymptotic form of the metric, though we do not expect it to take angular directions that grow at infinity to directions that do not grow at infinity.

Splittings of branches are common in gauge theories that have more
than one non-compact branch, but they can also happen in
Landau-Ginzburg models. As an example, consider the
theory of three massless chiral superfields with a $W=XYZ$ superpotential.
The classical supersymmetric vacua are at $X=Y=0$ for any $Z$
and permutations of this, giving three one-complex-dimensional
branches that meet at the origin $X=Y=Z=0$. The physics near
the origin is strongly coupled at low energies, but far from the
origin we have at low energies just one free massless field, so
we expect each branch to describe a one-complex-dimensional
sigma model on an asymptotically flat space. One may think
that at low energies there would be a single CFT describing all
three branches. However, the UV theory has one vector-like $U(1)_R$
symmetry and two additional vector-like $U(1)$ global symmetries
(there is also an axial $U(1)_R$ symmetry that does not act on
the bottom components of the chiral superfields). On
each branch one can find combinations of these symmetries
that give a $U(1)_R$ that does not act by rotating the
asymptotic region, but there is no $U(1)_R$ that does not
act on the asymptotic regions of all the branches. Thus, assuming that there are
no accidental vector-like symmetries arising that do not act on any
of the branches, and that the RG flow does not significantly change the asymptotic form of the metric, one expects
each branch to flow at low energies to a separate SCFT, namely
the strong coupling physics near the origin separates
the three branches.  This behavior was also observed in other LG model examples in \WittenZH.

As discussed in Section $3.2$, a naive reduction from $3d$ would imply that this $2d$ theory is dual to an LG model of a single twisted chiral superfield with the superpotential \newwds. This duality holds for finite values of $\zeta$ and $m$, but the discussion above implies that it must fail when both $\zeta$ and $m$ are taken to zero and the $XYZ$ model develops three separate branches. Even when just a single non-compact branch arises, the duality already has the K\"ahler potential problem discussed in the first example of the previous subsection, but the problem becomes even worse when both $\zeta$ and $m$ vanish.

As an example of a $\cN=(2,2)$ gauge theory which exhibits a splitting of its branches, consider the $2d$ SQED theory with
$N_f$ massless chiral superfields $Q$ of charge $(+1)$ and $N_f$
massless chiral superfields ${\tilde Q}$ of charge $(-1)$, with $N_f>1$ (for $N_f=1$ this theory was analyzed in \AharonyJKI).
There is a specific
value of the theta angle, $\theta = \pi N_f$, for which the Coulomb branch of this
theory is not lifted (for a vanishing FI parameter), and it can then
be parameterized by the twisted chiral
superfield $\Sigma$. The theory also has a Higgs branch
of dimension $(2N_f-1)$. Both branches are asymptotically
flat in appropriate coordinates (the coordinate $\Sigma$ for
the Coulomb branch, and the coordinates
$Q$ and ${\tilde Q}$ divided by complex gauge transformations
for the Higgs branch). The RG flow in this case is believed to significantly change the asymptotic form of the metric on the Higgs branch, but not the number of non-compact dimensions and the symmetries acting on them. Assuming this, it implies that if the gauge theory flows to
conformal field theories then the Coulomb branch CFT has
${\hat c}=1$ and the Higgs branch CFT has ${\hat c}=2N_f-1$.
As in \WittenYU, the fact that these central charges are
different means that the two SCFTs must form a direct sum,
with an infinite throat connecting them, as in the ${\cal N}=(4,4)$ SQED
theory.
	
By analogy with the ${\cal N}=(4,4)$ case, a simple conjecture
for the theory in the `throat' is that it is a sigma model on a
semi-infinite cylinder. This is suggested by a one-loop computation
of the metric on the Coulomb branch, which gives a behavior
near the origin of $|d\Sigma|^2 / |\Sigma|^2$, though higher
order corrections that modify this form are generally expected (the
perturbative expansion on the Coulomb branch is in
powers of $e^2 / \Sigma^2$). One argument for this form
is that it also arises as an effective theory near the origin of
the Higgs branch by integrating out the charged chiral
multiplets in the region $\Sigma \to \infty$, as in \AharonyDW, and in that description there are no
higher order corrections to the metric (the perturbative
expansion in this effective theory near the origin of 
the Higgs branch is in powers of $d\Sigma / \Sigma^2$).
As in the ${\cal N}=(4,4)$ case, one expects to have different
linear dilatons in the `throat' from the point of view of the
Higgs and Coulomb branches, that give rise to the appropriate
central charges. In other words, the energy-momentum
tensors in the different throats are related by an `improvement
term'. In this case, from the point of view of the Coulomb branch
there is no linear dilaton in the throat, while in the Higgs branch with $N_f>1$
one expects the dilaton to grow towards the origin.

Upon turning on an FI term, the Coulomb branch is lifted, and
one expects also the `throat' of the Higgs branch to be lifted.
In the $\Sigma$ variable this is described by turning on a
Liouville-like twisted superpotential that removes the region
near the origin (the twisted superpotential is $\cW = t \Sigma$,
but since the variable in the `throat' that has a canonical metric is proportional to $\log(\Sigma)$,
this is a Liouville superpotential). The natural variables describing the Higgs
branch are chiral superfields $M = Q {\tilde Q}$, and in these
variables the same deformation is described by smoothing out
the geometry of the Higgs branch near the origin; in the
`throat' these variables are T-dual to the $\Sigma$ variable,
so this is a special case of the cigar/Liouville duality.

Similarly, when turning on masses for the chiral multiplets, 
the Higgs branch is lifted
and so is the Coulomb branch `throat', leading to a smooth
sigma model on the Coulomb branch. In this case there is
again a Liouville twisted superpotential in this `throat' near the origin in the
$\Sigma$ variable. The twisted superpotential (which can be computed) now behaves for small $\Sigma$ as $\Sigma^2$, and the canonically normalized variable is
now proportional to $(-\log(\Sigma))$.

Above we argued that $3d$ dualities with non-compact Higgs branches could lead to $2d$ dualities if their asymptotic metrics have the same IR limit, but that this does not happen for non-compact Coulomb branches. In the cases where both exist, it seems plausible that the Higgs branch CFTs would still be dual, but compared to our previous discussion there is also the extra requirement that the asymptotic metrics in the throats also have the same IR limit, and not just the asymptotic metrics at infinity.

\vskip 20pt
\noindent {\bf Acknowledgments:}

We would like to thank Eran Avraham, Chris Beem, Francesco Benini, Oren Bergman, Cyril Closset, Abhijit Gadde, Zohar Komargodski, and Itamar Yaakov for useful discussions over the last four years.  We would especially like to thank Nathan Seiberg for collaboration throughout a large portion of this project and for many useful discussions.
The work of OA and SSR was supported in part  by the I-CORE program of the Planning and Budgeting Committee and the Israel Science Foundation (grant number 1937/12). The work of OA was supported in part by an Israel Science Foundation center for excellence grant, by the Minerva foundation with funding from the Federal German Ministry for Education and Research, and by the ISF within the ISF-UGC joint research program framework (grant no.\ 1200/14). OA is the Samuel Sebba Professorial Chair of Pure and Applied Physics.  
SSR is  a Jacques Lewiner Career Advancement Chair fellow. The research of SSR was also supported by Israel Science Foundation under grant no. 1696/15. SSR would like to thank the Simons Center for Geometry and Physics and the Aspen Center for Physics for hospitality during final stages of the project. BW was supported in part by the National Science Foundation under Grant No. NSF PHY11-25915.

\appendix{A}{Background and definitions}

In this appendix we review general rudimentary facts about two dimensional and three
dimensional theories with four supercharges, which we will use in this paper, in order to fix our notations and conventions.

\subsec{Two dimensional ${\cal N}=(2,2)$ supersymmetric field theories}

Our conventions follow those of \WittenYC. 
Supersymmetric Lagrangians include the usual K\"ahler potential
for chiral superfields and for twisted chiral superfields. For Abelian theories the vector superfield can be rewritten in terms of a gauge-invariant twisted chiral superfield $\Sigma$. In particular,
the integration over superspace of $-{1\over 4e^2} \Sigma {\bar \Sigma}$
gives the standard kinetic terms for the vector multiplet. One can also
write down superpotentials for chiral superfields, and similarly twisted
superpotentials for twisted chiral superfields. 
A special case is a twisted superpotential that is linear in $\Sigma$,
with a coefficient $t = {\hat r} + i \theta$. In the two dimensional action $\hat{r}$
multiplies $D$ and is the Fayet-Iliopoulos (FI) parameter, while $\theta$
multiples $F_{01}$ and it is a two dimensional periodic theta angle.

Many of the $2d$ ${\cal N}=(2,2)$ theories we will analyze will be gauge theories (Abelian or
non-Abelian), whose low-energy dynamics (at least in the region of
large expectation values for the scalars in the vector multiplets) may
be described by an effective twisted superpotential for the twisted
chiral superfields $\Sigma_i$ related to the vector superfields
(including superfields in the Cartan of any non-Abelian groups).
This effective twisted superpotential may be computed exactly at one-loop order.
If the theory has chiral superfields $\Phi_a$ with charges $Q_a^i$
under the $i$'th $U(1)$ gauge symmetry, and twisted masses $M_a$ which can be viewed as the scalars in background vector multiplets coupled to global symmetries, then
the effective twisted superpotential is
\eqn\efftwistpot{\cW = \sum_i t_i \Sigma_i + \sum_a \cW^{(2d)}_\chi(\sum_i Q_a^i \Sigma_i + M_a), }
where the first term is the contribution of $2d$ FI parameters, which are only included for Abelian factors of the gauge group, and the contribution of a single charged chiral multiplet is:
\eqn\twodchi{ \cW_\chi^{(2d)}(\Sigma) = \Sigma (\log (\Sigma/\mu) - 1)\,, 
}
where $\mu$ is a dynamical scale.  Note that charged vector
superfields do not contribute to \efftwistpot.\foot{More precisely, the vector superfields may contribute an effective FI parameter to $\cW$ \HoriEWA.  For the theories considered in this paper, this term is either trivial, or may be absorbed into a redefinition of the UV FI parameter.  We expect it may be relevant, however, when one considers $so(N)$ gauge theories.  We thank O. Bergman for discussions on this point.}  Also, note that $\Sigma$ is not an unconstrained twisted chiral field, because its $F$-term contains the field strength, $F_{\mu \nu}$, whose flux is quantized.  As discussed for instance in \BeemMB, we may account for this by summing over all branches of the twisted superpotential related by $\cW \rightarrow \cW + 2 \pi i n^i \Sigma_i$, $n_i \in \Z$.  Note that this makes \twodchi\ well-defined, as the branch cuts of the logarithm simply take us to another branch of $\cW$.

Discrete supersymmetric vacua are then given by the solutions of
\eqn\vacua{{\partial {\cW} \over {\partial \Sigma_i}} = t_i + \sum_a Q^i_a \log(\sum_j Q_a^j \Sigma_j + M_a) = 0 \; ({\rm mod}\; 2 \pi i \Z), }
where the modulo corresponds to a change of branch.  This can be conveniently expressed as
\eqn\expvac{ \exp \bigg({\partial {\cW} \over {\partial \Sigma_i}}\bigg)  =e^{t_i} \prod_a (\sum_j Q_a^j \Sigma_j + M_a)^{Q_a^i}  = 1, }
which gives a system of polynomial equations in the $\Sigma_j$.  Following \refs{\HoriDK,\AharonyJKI}, when the gauge group is non-Abelian, we exclude solutions which are not acted on freely by the Weyl symmetry, as these lead to supersymmetry breaking configurations.  

The twisted superpotential can also be naturally defined for higher dimensional systems compactified to effectively $2d$ $\cN=(2,2)$ systems.  In the main text we discuss the case of $3d$ $\cN=2$ theories on a circle.  See \refs{\NekrasovXAA,\ClossetARN,\ClossetZGF} for more examples and details.

There are two interesting types of supersymmetric partition
functions in two dimensions that we use in this paper -- the elliptic genus
and the sphere partition function.

Let's assume that a supersymmetric theory has a left-moving
$U(1)_R$ symmetry $J_L$, and additional Abelian global
symmetries with charges $K_j$ (these may be charges in the
Cartan of non-Abelian global symmetries). The elliptic genus
of the theory is defined as a trace over the Ramond-Ramond
sector (the sector with periodic boundary conditions for the
fermions) of the Hilbert space of the theory on  a circle as
\eqn\ellgen{Z(\tau, z, u) = {\rm Tr}_{RR} \left[ (-1)^F
e^{\pi i \tau (H + i P)} e^{-\pi i {\bar \tau} (H - i P)}
e^{2\pi i z J_L} \prod_j e^{2\pi i u_j K_j} \right],}
where $H$ is the Hamiltonian and $P$ is the momentum along the circle.
Alternatively, it may be defined as a twisted path integral of
the Euclidean theory on a torus with complex structure $\tau$.
When viewed (through the state/operator correspondence) as
a sum over operators, only right-handed-chiral operators
contribute to the sum.

The elliptic genus \ellgen\ does not change under a
renormalization group flow, so if we have some gauge theory
that flows to a superconformal theory, we can compute the
elliptic genus of this SCFT in the free high-energy theory. The only subtle
issue is that to do this we have to identify the correct
$U(1)_R$ symmetry that becomes part of the superconformal
algebra at low energies; assuming that there are no accidental
R-symmetries, this can be done by $c$-extremization \BeniniCZ. 
Once this is known, the elliptic genus of a gauge theory can be 
computed in many cases \BeniniNDA; it is essentially given by a
one-loop computation (from the point of view of the partition
function on a torus).

Another type of supersymmetric partition function which was
studied more recently is the partition function on a two-sphere $\S^2$.
There are several ways to compactify an ${\cal N}=(2,2)$ theory
on $\S^2$ while preserving some supersymmetry. These all
require the existence of an R-symmetry; some require a vector-like
$U(1)_R$, and others (related by mirror symmetry) an
axial $U(1)_R$. We will only use the coupling that utilizes the vector-like $U(1)_R$, and without $R$-symmetry flux,\foot{The second one was discussed in \refs{ \DoroudXW, \DoroudPKA,\GerchkovitzGTA}.  In addition, one may consider the topological twist ($A$- or $B$-model), where one turns on an $R$-symmetry flux.} since
we will be interested in theories that have (at high energies)
a vector $U(1)_R$ but not necessarily an axial $U(1)_R$. In particular, this coupling appears
when reducing the $3d$ $\S^2\times \S^1$ supersymmetric partition function on a circle.
In this method, in order to define the theory on a round $\S^2$ one turns on only one of the 
scalars in the background supergravity multiplet (in addition to the $\S^2$
metric), while on a squashed $\S^2$ (which gives the same partition
function) one turns on also a background $U(1)_R$
field, but with zero total flux.
If we consider a (non-)Abelian gauge theory coupled to charged chiral
superfields, its partition function on $\S^2$ can be computed via localization
as shown in \refs{\BeniniUI,\DoroudXW,\GomisWY}. 

The $\S^2$ partition function includes a contribution from all chiral multiplets in the theory, as well as from the twisted superpotential.  To treat these in a unified way, it is useful to define functions
\eqn\wpmdef{ \cW^{(\pm)}_{\S^2}(\Sigma_i,m_a,Y_\alpha,X_\beta), }
where $\Sigma_i$ and $m_a$ are (the bottom components of) the gauge multiplets and of the vector multiplets coupled to the global symmetries, respectively, and $Y_\alpha$ and $X_\beta$ are, respectively, dynamical and background twisted chiral fields.  The $\S^2$ partition function, normalizing the radius of $\S^2$ to one, is then a function of the background fields, and is given by an integral over constant values of the dynamical fields:
\eqn\stwowpm{\eqalign{& \cZ_{\S^2}(m_a,X_\beta) =\cr &\qquad\, \frac{1}{|W|}\int \prod_i \frac{d\Sigma_i d\bar{\Sigma}_i}{\pi} \prod_\alpha\frac{dY_\alpha d\bar{Y}_\alpha}{\pi} \exp \bigg( \cW_{S^2}^{(+)}(\Sigma_i,m_a,Y_\alpha,X_\beta) -\cW_{S^2}^{(-)}(\bar{\Sigma}_i,\bar{m}_a,\bar{Y}_\alpha,\bar{X}_\beta) \bigg). }}
Here $|W|$ is the order of the Weyl group of the gauge group.  More precisely, the gauge multiplets, such as $\Sigma_i$, contain the field strengths, and so are subject on $\S^2$ to a quantization condition, $\Sigma_i + \bar{\Sigma}_i \in  \Z$, so the notation  $\int d \Sigma_i d\bar{\Sigma}_i$ refers to a real integral and a sum over $\Z$.

It remains to provide the definition of the functions $\cW_{\S^2}^{(\pm)}(\Sigma,m,Y,X)$.  These receive contributions from the various ingredients defining the UV theory.  The bare, UV twisted superpotential, $\cW_{UV}(\Sigma_i,m_a,Y_\alpha,X_\beta)$, contributes as:
\eqn\tscontstwo{ \cW_{\S^2}^{(\pm)} = \cdots + \cW_{UV}(\Sigma_i,m_a,Y_\alpha,X_\beta).}
We also  have the option to turn on a supersymmetric dilaton background, which is a function, $\Omega_{UV}(\Sigma,m,Y,X)$, of the twisted chiral fields, and contributes to $\cW_{S^2}^{(\pm)}$ with opposite signs:
\eqn\dilcontstwo{ \cW_{\S^2}^{(\pm)} = \cdots \pm \Omega_{UV}(\Sigma_i,m_a,Y_\alpha,X_\beta).}
Note that under a change of twisted chiral variables, $Y \rightarrow Y'(Y)$, a Jacobian is induced which contributes to the dilaton background as:
\eqn\diljac{ \Omega_{UV} \rightarrow \Omega_{UV} - \log (\frac{\partial Y'}{\partial Y}).}
Finally, a single chiral multiplet of R-charge $\Delta$ coupled to a gauge multiplet $\Sigma$ (including dynamical and background fields) contributes:
\eqn\chistwocont{  \cW_{\S^2}^{(\pm)}  = \cdots - \log\bigg (\Gamma (\frac{1}{2} + \Sigma \pm \frac{1-\Delta}{2} )\bigg). }
We can think of $\cW_{\S^2}^{(\pm)}$ as a curved space analogue of the flat space effective twisted superpotential.  In particular, note that if we take $\Sigma \rightarrow \infty$, so that the scale of $\Sigma$ is much larger than the scale of the curvature of $\S^2$, we find:
\eqn\chistwolimit{ - \log \bigg(\Gamma (\frac{1}{2} + \Sigma \pm \frac{1-\Delta}{2} )\bigg) \;\;\; \sra \;\;\; \Sigma(\log (\Sigma)-1)  \pm \frac{1-\Delta}{2} \log (\Sigma) + \cdots , }
reproducing the flat space effective twisted superpotential \twodchi\ of a chiral multiplet, as well as a contribution to the dilaton background.

\subsec{Three dimensional theories with ${\cal N}=2$ supersymmetry}

Our conventions for three dimensional theories with ${\cal N}=2$ supersymmetry in flat space follow those of
\AharonyBX.  
In addition to the usual $4d$ multiplets, the theory can include linear multiplets $\tilde{\Sigma}$,\foot{We write these with a tilde to distinguish them from the twisted chiral field strength multiplet $\Sigma$ in two dimensions.} satisfiying ${\cal D}^2 \tilde{\Sigma} = {\bar {\cal D}}^2 \tilde{\Sigma}=0$, which contain conserved currents.  An important example is the field strength multiplet, $\tilde{\Sigma}_V=\epsilon^{\alpha \beta} {\cal D}_\alpha {\bar {\cal D}}_\beta \Tr(V)$, which contains the conserved current $\star \Tr(F)$.  Such a conserved current arises for each $U(1)$ factor in the gauge group, and we refer to this global symmetry as a $U(1)_J$ topological symmetry.

There are two types of parameters appearing in the action that will be important when we consider reduction to two dimensions.  First, for each simple or $U(1)$ factor of the gauge group we choose a gauge coupling $g$.  Second, for each global symmetry, we can couple a background vector multiplet $\tilde{\Sigma}_{BG}$ for which we turn on a constant VEV $M$ for the real scalar $\sigma_{BG}$. We refer to these as real mass parameters.  For example, for a free chiral $\Phi$ and the $U(1)$ symmetry acting on it, this enters as a mass term (after integrating out the auxiliary fields):
\eqn\rmdef{
{\cal L}_m = \int d^4 \theta \Phi^\dagger e^{M \theta \bar{\theta}} \Phi  = M^2 |\phi|^2 + i M \epsilon^{\alpha \beta} \psi^\dagger_\alpha \psi_\beta . 
}
The real mass corresponding to a $U(1)_J$ global symmetry is a Fayet-Iliopoulos term for the corresponding $U(1)$ gauge group factor.
We can also add a Chern-Simons (CS) term for each vector multiplet, 
whose normalization is quantized in order to ensure invariance under large gauge transformations.

\

\item{a.}{\it The three dimensional index}

In three dimensions a useful partition function which we will utilize in our computations is the supersymmetric index, which is identical to a partition function on $\S^2\times \S^1$.
The $3d$ supersymmetric index \refs{\BhattacharyaZY\KimWB\ImamuraSU-\KapustinJM}  for a $3d$ ${\cal N}=2$ theory is defined as
\eqn\indthreedef{{\cal I}(\{\bu_a,\,\bm_a\};\bq)=
\Tr_{{{\cal H}}_{{\{ { m}_a\}}}} \left[ (-1)^{2J_3} \bq^{\frac12\left(E+J_3\right)}\;\prod_a \bu_a^{F_a} \right]\,.
}
Here ${\cal H}_{\{\bm_a\}}$ is the Hilbert space of the theory on $\S^2$ with background fluxes for global symmetries, parametrized by $\bm_a$, on the $\S^2$.  $E$ is the energy of the state on $\S^2$ (which, in the conformal case, is the same as the dimension of the corresponding operator), $R$ is the R-charge, $J_3$ is the Cartan generator of $SO(3)$ rotations of $\S^2$, and $F_a$ are charges for global symmetries.

This index can be evaluated by performing the path integral on $\S^2\times \S^1$, which can be computed by localization. In this interpretation the $\bu_a$ are the holonomies around the circle for background gauge fields for the global symmetries, and the index is an integral over holonomies $\bz_i$ for the dynamical gauge fields, and a sum over their fluxes $\bn_i$. For convenience the radius of $\S^2$ will be set to one, and we will denote the radius of $\S^1$ in these units (namely, the ratio of the two radii) as $\tau$. The radius $\tau$ enters explicitly in the expressions for the partition functions through $\bq=e^{-\tau}$.

It will be useful to write the index in a way which makes the connection to the $\S^2$ partition function more transparent.  In particular, we may think of the $\S^2 \times \S^1_\tau$ partition function as a particular example of the $\S^2$ partition function, where we happen to be studying a $2d$ theory with infinite towers of fields, namely, the $3d$ theory compactified on $\S^1_\tau$.  Then, as in \stwowpm, we may write the index as:
\eqn\stwowpm{ \cZ_{\S^2 \times \S^1_\tau}(m_a) = \frac{1}{|W|}\int \prod_i \frac{d\Sigma_i d\bar{\Sigma}_i}{\pi} \exp \bigg( \cW^{(+)}_{\S^2 \times \S^1_\tau}(\Sigma_i,m_a)-\cW^{(-)}_{\S^2 \times \S^1_\tau}(\bar{\Sigma}_i,\bar{m}_a) \bigg)\,. }
Note the following differences from the $2d$ case.  First, there are no twisted chiral fields apart from the vector multiplets $\Sigma$ and $m$.  Also, large gauge transformations around $\S^1_\tau$ now identify $\Sigma \sim \Sigma+ 2 \pi i \tau^{-1}$.  To relate this to the usual $\bz,\bn$ that are summed over in the index, we write:\foot{The $e^{i\pi}$ factor here is related to the subtleties with definition of $(-1)^F$ in presence of magnetic fluxes. We use the notations of \AharonyDHA. }
\eqn\indparrel{
e^{ \tau \Sigma} = (e^{\pi i}\bq)^{-\bn/2} \bz, \;\;\;\;e^{\tau \bar{\Sigma}} =(e^{-\pi i} \bq)^{-\bn/2} \bz^{-1}\,,
}
where the quantization condition of $\Sigma+{\bar{\Sigma}}\in \Z$ imposes $\bn \in \Z$.  A similar mapping holds for the background fields for global symmetries; below we include these inside $\Sigma$.

As in $2d$, the functions $\cW^{(\pm)}_{\S^2 \times \S^1_\tau}$ are built from the various ingredients in the UV action.  A Chern-Simons term contributes:\foot{In addition, contact terms involving the R-symmetry contribute to the dilaton background in a similar way.}
\eqn\threedcsindex{ \cW^{(\pm)}_{\S^2 \times \S^1_\tau} = \cdots + \frac{1}{2} k \tau \Sigma(\Sigma+\frac{2 \pi i}{\tau}),} 
while a chiral multiplet of R-charge $\Delta$ contributes:
\eqn\threedchiralindex{ \cW^{(\pm)}_{\S^2 \times \S^1_\tau} = \cdots - \log \bigg((e^{\tau \Sigma} \bq^{1 \mp (1-\Delta)};\bq)\bigg), }
where $(\bz;\bq) \equiv \prod_{j=0}^\infty (1- \bz \bq^j)$.

As with the $\S^2$ partition function, we may think of $\cW^{(\pm)}_{\S^2\times \S^1_{\tau}}$   as a curved space analogue of the effective twisted superpotential of a $3d$ theory on $\R^2 \times \S^1_\tau$.  Namely, we see that the Chern-Simons contribution \threedcsindex\ agrees with that of the flat space theory \threedwdef\ for $\tau = 2\pi r$, and if we take $\Sigma$ large in \threedchiralindex\ then, using \BeemMB:  
\eqn\limpoch{
(\bq z;\bq) \sim \exp \bigg( -\sum_{\ell=0}^\infty \frac{\tau^{\ell-1} B_\ell}{\ell!} Li_{2-\ell}(z)\bigg)\,,
}
where $B_\ell$ are the Bernoulli numbers, we find:

\eqn\indexchirallimit{ {\cW^{(\pm)}_{\S^2 \times \S^1_\tau}}^{3d \; chiral}(\Sigma)  \sra \frac{1}{\tau} {\rm Li}_2(e^{-\tau \Sigma})  \pm \frac{1-\Delta}{2} \log (1 - e^{ \tau \Sigma})  + \cdots. }
This reproduces the flat space effective twisted superpotential of a $3d$ chiral multiplet, with an additional contribution to the effective dilaton.

Finally, we note that the index of a $3d$ chiral multiplet can be written as the $\S^2$ partition function of an infinite number of $2d$ chiral multiplets, corresponding to the KK modes. Properly regulating and normalizing the infinite product over the KK modes this can be written as (denoting ${\bf z}=e^{i\tau\zeta}$ for the $U(1)$ background field):
\eqn\KKtemp{
\eqalign{
&{\cal I}_\Delta({\bf z},{\bf m};{\bf q})=\left(\bq^{\frac{1-\Delta}2}\bz^{-1}\right)^{\bm/2}
\frac{( \bz^{-1}\bq^{+ {\bm}/2+1-\frac\Delta2};\bq)}{( \bz \bq^{ +{\bm}/2+\frac\Delta2};\bq)}=
\frac{e^{-\frac2\tau \left({\rm Li}_2\left(e^{-i \tau  \left(\zeta +\frac{\Delta -1}{2} i \right)}\right)-\frac{\pi ^2}{6}\right)}}
{i^m e^{-\pi  \left(\zeta +\frac{\Delta -1}{2} i \right)-\frac{1}{2} \tau  \left(\zeta +\frac{\Delta -1}{2} i \right)^2}}\times\cr
&\qquad \prod_{\ell=1}^\infty  
i^m e^{\Delta-1 -2 i \zeta -\frac{4 i \pi  \ell}{\tau }} \left(\frac{\Delta -1}{2} i +\zeta +\frac{2 \pi  \ell}{\tau }\right)^{1-\Delta +2 i \zeta +\frac{4 i \pi  \ell}{\tau }}
Z(\Delta,\zeta+\frac{2\pi\,\ell}\tau,m)\cr 
&\qquad\; \prod_{\ell=0}^\infty 
(-i)^m e^{\Delta -1 -2i\zeta +\frac{4 \pi i  \ell}{\tau }} \left(-\frac{\Delta -1}{2} i -\zeta +\frac{2 \pi  \ell}{\tau }\right)^{1-\Delta  +2i\zeta -\frac{4 \pi i  \ell}{\tau }}
Z(\Delta,\zeta-\frac{2\pi \,\ell}\tau,m)
\,.
}
} Here ${\cal I}_\Delta$ is the index of a chiral field of R-charge $\Delta$ in $3d$,\foot{Note that here for brevity insice the Pochhammer symbols we dropped the factors of $(-1)^\bm$ which are needed with proper definition of the fermion number \DimoftePY\AharonyDHA.} and the $\S^2$ partition function of a chiral field of R-charge $\Delta$ in two dimensions is $$Z(\Delta,\gamma,m)=\frac{\Gamma(\frac\Delta2-i \gamma -\frac{m}2)}{\Gamma(\frac{2-\Delta}2+i \gamma-\frac{m}2)}\,.$$ $\gamma$ is a twisted mass parameter for the $U(1)$ under which the field is charged, and $m$ is its flux.

\appendix{B}{Details of reductions}

In this appendix we spell out the details of some of the reductions of $3d$ theories and dualities discussed in Sections $2$ and $3$.

\subsec{Reduction of $USp(2N_c)$ theories }

Consider the $3d$ $USp(2N_c)$ theory with $N_f$ flavors ($2N_f$ fundamentals) and Chern-Simons level $k$, where the quantization condition is $k+N_f \in \Z$, and we take $k \geq 0$.  One can write the twisted superpotential for this theory similarly to the $U(N_c)$ case discussed in Section $2.4$, and one finds that the vacuum equations are:
\eqn\spthreed{ \prod_{a=1}^{2N_f}(x_j \nu_a - 1) = {x_j}^{2k}  \prod_{a=1}^{2N_f}(\nu_a - x_j) \,.}
We must solve this for each $x_j$, $j=1,\cdots,N_c$.  The Weyl symmetry acts as permutations of the $x_j$ and flips $x_j \rightarrow x_j^{-1}$.  We must pick the $x_j$ so that none of them are related by this Weyl-symmetry, nor fixed by it.  The polynomial equation above has $2(k+N_f)$ solutions.  Note that $x_j=\pm 1$ are always solutions, and we should exclude these.  The remaining solutions come in pairs related by $x_j \rightarrow {x_j}^{-1}$, so there are $k+N_f-1$ Weyl-inequivalent solutions.  Thus the number of vacua is:
\eqn\spvac{ N_{vac} = \pmatrix{k+N_f-1 \cr N_c}. }

Now let us consider the reduction to $2d$ with fixed masses.  As in the $U(N_c)$ case, if we take $r\Sigma_j$ finite, we find, to leading order, the twisted superpotential for the $2d$ $USp(2N_c)$ with $N_f$ flavors.  The vacuum equation in $2d$ is:
\eqn\sptwodeq{ \prod_{a=1}^{2N_f}(\Sigma_j + m_a) =  \prod_{a=1}^{2N_f}(-\Sigma_j+m_a)\,. }
This has $2N_f$ solutions when $2N_f$ is odd, but only $2N_f-1$ when $2N_f$ is even, since the leading term in the polynomial cancels.  One of these solutions is zero, which we exclude as it leads to residual gauge symmetry, and the remaining ones come in pairs related by a sign. So for odd $2N_f$ we have $(2N_f-1)/2$ solutions, and for even $2N_f$, $N_f-1$ solutions.

However, in the $3d$ theory on a circle we found $(k+N_f-1)$ physical solutions to the corresponding equation,
so we must find the remaining ones.  If we assume that $x_j$ is a finite distance from $1$ as $r \rightarrow 0$, then \spthreed\ becomes approximately:
\eqn\spextravac{ (x_j-1)^{2N_f} = {x_j}^{2k}  (1-x_j)^{2 N_f} \,.}
This has $2N_f$ solutions near $x_j=1$, and $2k$ additional solutions with ${x_j}^{2k}=(-1)^{2N_f}$:
\eqn\splarge{ x_j = \left\{ \matrix{ e^{2 \pi i n/(2k)},& 2N_f \;\rm{even}\cr
e^{2 \pi i (n-\frac{1}{2})/(2k)},&  2N_f  \;\rm{odd} } \right. \;\;\; n=0,1,2,\cdots,2k-1\;\;\;. }

Let us consider the behavior when $2N_f$ is odd or even separately.

\

\item{a.}{\it $2 N_f$ odd}

For $2N_f$ odd, one of the additional $2k$ solutions is the unphysical $x=-1$, and the others come in pairs, so we have $(2k-1)/2$ extra solutions with $\Sigma_j$ of order $r^{-1}$. These combine with the $(2N_f-1)/2$ solutions above, with $\Sigma_j$ finite, to give
give all $(k+N_f-1)$ physical solutions.  
As in the $U(N_c)$ case, we can choose $(N_c-\ell)$ of the eigenvalues from the additional $(2k-1)/2$ solutions, which contribute a trivial decoupled sector, and the remaining $\ell$ contribute an $USp(2\ell)$ theory.  Thus we find that the $2d$ theory is given by a direct sum:
\eqn\spodddecompmt{ USp(2N_c)_k , \; 2N_f \; \rm{odd}\;\;\; \rightarrow \;\;\; \bigoplus_{\ell={\rm max} (0,N_c-k+\frac{1}{2})}^{{\rm min}(N_f-\frac{1}{2},N_c)} \pmatrix{{k-\frac{1}{2}}\cr{N_c-\ell}} USp(2 \ell)_{N_f}}
\

\item{b.}{\it $2N_f$ even }

In the case $k=0$, all of the valid solutions to \spthreed\ descend to solutions of \sptwodeq, and so the $3d$ $USp(2N_c)$ theory reduces to the $2d$ $USp(2N_c)$ theory without any additional sectors.  For $2N_f$ even and $k>0$, of the $2k$ solutions \splarge\ we should not count the one with $n=0$ as it violates our assumption that $x_j$ is not near $1$, and the one at $x_j=-1$ is also unphysical, so we have $(k-1)$ additional physical solutions.  Recall that we only found $(N_f-1)$ physical solutions to \sptwodeq\ in this case.  Thus we are missing a physical solution to \spthreed, which approaches $1$, but is not captured by \sptwodeq.  We claim that the solution scales as $\Sigma_j \sim r^{-1/2}$.  Let us write $X_j=\sqrt{2 \pi r} \Sigma_j$, and expand the effective twisted superpotential to leading order in $r$.  We find:
\eqn\splg{ \cW = k {X_j}^2 + 2\bigg( \sum_{a=1}^{2N_f} m_a \bigg) \log (X_j)\,. }
Thus we have two equivalent vacua sitting at $X_j = \pm i \sqrt{\frac{ \sum_a m_a }{k}}$, or:
\eqn\spsqrtsol{ \Sigma_j = \pm i \sqrt{\frac{ \sum_a m_a }{2 \pi r k}} \,.}
We can define $X_j=\frac{1}{\sqrt{k}} e^{Y_j/2}$, and we find (up to a constant):
\eqn\spwy{ \cW = e^{Y_j} + Y_j \sum_{a=1}^{2N_f} m_a \,.}
This twisted LG model is related by the Hori-Vafa duality \HoriKT\ to a chiral superfield with mass $\sum_{a=1}^{2N_f} m_a$.

To summarize, as $r \rightarrow 0$, the physical solutions  for $\Sigma_j$ behave as:

\nlb $(N_f-1)$ solutions have $\Sigma_j$ finite, given by the solutions of \sptwodeq;
\nlb $(k-1)$ solutions have $\Sigma_j$ of order $r^{-1}$, the top line in \splarge\ for $n \neq 2k$;
\nlb The remaining solution sits at $\Sigma_j$ of order $r^{-1/2}$, given by \spsqrtsol.

\

Thus when we choose $N_c$ eigenvalues, we have the following ways of distributing them:


\nlb $(N_c-\ell)$ are among the $k-1$ solutions of order $r^{-1}$, and the remaining $\ell$ are finite.  This gives an $USp(2 \ell)$ theory plus a decoupled sector.
\nlb $(N_c-\ell-1)$ are among the $k-1$ solutions of order $r^{-1}$, one is of order $r^{-1/2}$, which gives the LG model in \spwy, and the remaining $\ell$ are finite.  

\

This leads to the following direct sum of theories in $2d$:
\eqn\spevendecompmtapp{ \eqalign{USp(2N_c)_k , \; 2N_f \; \rm{even}\;\;\; \rightarrow \;\;\; \bigg(\bigoplus_{\ell={\rm max} (0,N_c-k+1)}^{{\rm min}(N_f-1,N_c)} \pmatrix{k-1\cr N_c-\ell} USp(2 \ell) , \;N_f \bigg) \cr
\bigoplus \bigg(\bigoplus_{\ell={\rm max} (0,N_c-k)}^{{\rm min}(N_f-1,N_c)} \pmatrix{k-1\cr N_c-\ell-1}  USp(2\ell) , \;N_f + \; {\rm LG \; model}\bigg)\,.  }} 

As discussed in the main text, this only holds at the level of the massive theory, and we must be more careful when we consider the precise low-energy CFTs.  

\subsec{Duality appetizer}

Let us consider in more detail the reduction of the gauge theory appearing in the duality appetizer, discussed in Section $3.4$.  Theory A is an $SU(2)$ theory with an adjoint chiral multiplet $\Phi$ and CS level $1$.  The vacuum equation is:
\eqn\DAvac{ (x^2 \nu -1)^2 = x^2 (\nu-x^2 )^2, }
where $x=e^{2 \pi r \Sigma}$ and $\nu=e^{2 \pi r m}$ are the gauge and flavor symmetry parameters, respectively.  This has solutions $x=\pm 1$, which we exclude, as well as:
\eqn\DAvacsol{ x^2 = \frac{1}{2} (  \nu^2 + 2 \nu - 1 \pm (\nu+1) \sqrt{ (\nu+3)(\nu-1)} ). }
The choices of $\pm$ are related by $x \rightarrow x^{-1}$, and so are Weyl-equivalent, but we also have two choices of sign in taking the square root of the whole expression.  These are related by a gauge transformation which is single-valued in $SO(3)$ but not $SU(2)$, and so these are physically inequivalent.  

Recall that the duality appetizer relates this to theory B that is a free chiral $M$, which we identify with ${\rm Tr}(\Phi^2)$, tensored with a decoupled topological sector, $U(1)_{k=2}$.  This decoupled sector contributes exactly the two states we need to match the $SU(2)$ side.

Next let us study the behavior of this theory upon reduction to two dimensions.  If we hold the twisted mass $m$ of the adjoint chiral finite, so that $\nu=e^{2 \pi r m}$ approaches $1$, we find the solutions behave as:
\eqn\DAscale{ \Sigma \approx \pm i \sqrt{ \frac{m}{2\pi r}}, \;\;\; \frac{i}{2r} \pm i \sqrt{\frac{m}{2 \pi r}}. }
The choices of $\pm$ are related by Weyl-symmetry, but the shift by $\frac{i}{2 r}$ gives two physically distinct vacua.  

Let us focus on the first vacuum; the second is essentially the same.  We claim that we find a regular $2d$ limit provided that we rescale $\Sigma$ by $r^{1/2}$ to describe the theory in a regular way.  Thus we define $X = \sqrt{2 \pi r} \Sigma$, and we can derive the effective twisted superpotential for $X$ to leading order in $r$ as:
\eqn\DAWtwod{ W = 2 m \log (X) + X^2. }
If we define $X=e^{Y/2}$,\foot{This change of variables can be justified by studying the limit of the effective dilaton, as described in Section $2.5$.} this becomes:
\eqn\DAWtwodredefine{ W = m Y + e^Y. }
This can be Hori-Vafa-dualized \HoriKT\ into a chiral multiplet of mass $m$.  In addition to the dual of $Y$, there is another chiral multiplet, coming from the Cartan component of the adjoint chiral, which also has mass $m$.  The above analysis is not sufficient to determine their interaction, but it should be such as to restrict the naive $U(1)^2$ symmetry which acts on them to a single $U(1)$, under which they both have charge $1$ (note that this is difficult to arrange with a non-singular superpotential).
In the second vacuum of the $SU(2)$ theory, one finds an identical $2d$ theory upon reduction.  

In the reduction of the dual theory, each of the two CS states is tensored with a free chiral of mass $2m$, so one finds a direct sum of two copies of the free chiral.  Thus we expect each copy of the above theory of two chirals of mass $m$, with the superpotential coupling implicitly described above, to map to the free chiral of mass $2m$.  The matching of the twisted superpotential across this duality was also discussed in \ClossetZGF.  One can also check that the $\S^2$ partition functions of the resulting theories match, which follows from the reduction of the supersymmetric index.  However, for this duality to hold precisely, we will need to include a certain $\Z_2$ orbifold, which the $\S^2$ partition function and twisted superpotential are insensitive to (being defined on the simply connected spaces $\S^2$ and $\R^2$, respectively), but which we can see by studying the elliptic genus.

Before doing this, it will be useful to first make a digression to mention a closely related $2d$ duality.  Consider a pair of chirals $Y,Z$ with superpotential $W=Z Y^2$.  This has a $\Z_2$ symmetry acting on $Y$.  Consider also a free chiral, $X$, which also has a $\Z_2$ symmetry acting as $X \rightarrow -X$.  Then it is claimed in \HoriPD\ that the orbifold of these theories under the corresponding $\Z_2$'s are dual, with $Z$ mapping to $X^2$.\foot{A similar duality also appears in \DiFrancescoTY, where adding the field $Y$ with the superpotential $Z Y^2$ is interpreted as a Jacobian from a change of variables from $X$ to $X^2=Z$.}  More precisely, by gauging this $\Z_2$ with various choices of discrete torsion, we find several slightly different versions of the duality, as we describe in more detail below.  For any version, both theories have a $U(1)$ symmetry, which acts on $X,Y,Z$ with charges $1,-1,$ and $2$, respectively.

To describe this $\Z_2$ orbifold concretely, let us compute the elliptic genus of the dual pair.
The elliptic genus of a chiral field of $R$-charge $\Delta$ is given by
\eqn\chiell{
I_\Delta(z)= y^{-\frac12}\frac{\theta(q\,z y^{\frac\Delta2-1};q)}{\theta(q y^{\frac\Delta2}z;q)}\,,\qquad\quad \theta(z;q)\equiv (z;q)(q \,z^{-1};q)\,,
} 
where $z$ is a fugacity for the $U(1)$ flavor symmetry, and $y$ is a fugacity for the right-moving $R$-symmetry.  Let us define,
\eqn\defIs{
I_{ab} = y^{-\frac{b}2}I_{\frac14}((-1)^a q^{\frac{b}2}z)\,,\qquad\quad
\widetilde I_{ab}=y^{-\frac{b}2}I_{\frac12}(q^{  b}   z^2)\,I_{\frac34}((-1)^aq^{-\frac{b}2} z^{-1})\,.
} Then the elliptic genus of the orbifolded theories on the two sides of the duality is
\eqn\orbiF{
\frac12\sum_{a,b=0}^1 (-1)^{ac+be} I_{ab} =(-1)^{1+(c+1)(e+1)}\;
\frac12\sum_{a,b=0}^1 (-1)^{ac+be} \widetilde I_{ab}\,.
}  This equality is true.
Here $c,d$ are the holonomies of the ``quantum'' $\Z_2$ symmetry, which we will denote $\hat {\Z}_2$, that one acquires after orbifolding 
(the analogue of $U(1)_J$ in $3d$). The term $(-1)^{1+(c+1)(e+1)}$ is a background term for these
holonomies and is an analogue of the contact terms in $3d$. It is crucial for the duality to work.  The background term makes the mapping of the $\Z_2$'s on the two sides of the duality non-trivial.
 Setting $c=e=0$, ``unrefining'' the $\hat{\Z}_2$, we obtain the following simple equality:
\eqn\soeq{
\frac12\left(I_{00}+I_{01}+I_{10}+I_{11}\right)=
\frac12\left(\widetilde I_{00}+\widetilde I_{01}+\widetilde I_{10}+\widetilde I_{11}\right)\,.
} This is the familiar form of the $\Z_2$ orbifold elliptic genus summing over all the twisted sectors.
Fourier transforming the holonomy around the temporal cycle,
\eqn\defYs{
Y_{e,m}=\sum_{a=0}^1 (-1)^{ae}\,I_{am}\,,\qquad
\widetilde Y_{e,m}=\sum_{a=0}^1 (-1)^{ae}\, \widetilde I_{am}\,,
} the identities we get take an even simpler form,
\eqn\summarY{
Y_{e,m}=
(-1)^{e}
\widetilde Y_{e,(e+m+1){\rm\ mod\ }2}\,.
}

We can now obtain a dual of the free chiral field $X$ by gauging the ``quantum'' $\hat{\Z}_2$. Note that on one side of the duality we just get the free chiral ``ungauging'' the orbifold, as desired, however on the other side of the duality  that's not true due to the background term
\eqn\sudual{
I_{ab}=\frac12\sum_{c,d=0}^1 (-1)^{(c+a+1)(d+b+1)} \widetilde I_{cd}\,.
} For example 
\eqn\sudualZ{
I_{00}=\frac12\left(-\widetilde
I_{00}+\widetilde
I_{01}+\widetilde
I_{10}+\widetilde
I_{11}\right)\,.
} By putting the background term on the other side of the duality we can undo the orbifold on side (B) (the $Z Y^2$ side) at the cost of having a non-trivial theory on side (A). The fact that the background terms have such a huge effect is reminiscent of the $U(N)$ dualities being simple in $3d$ while $SU(N)$ dualities are complicated~\AharonyDHA. There the singlet fields dual to monopole operators play a crucial role when ungauging the $U(1)_B$ symmetry by gauging $U(1)_J$. 
Using the $Y$'s defined in \defYs\ the three dualities above can be written as
\eqn\dualswithY{
\eqalign{
Y_{00}+Y_{01}&=\widetilde Y_{00}+\widetilde Y_{01}\,,\cr
 Y_{00}+Y_{10}& =-\widetilde Y_{10}+\widetilde Y_{01}\,,\cr
 -Y_{10}+ Y_{01}&=\widetilde Y_{0,0}+\widetilde Y_{10}\,.
}
} These three dualities were discussed in~\HoriPD\ and referred to as $O_-(1)\leftrightarrow O_-(1)$ and $SO(1)\leftrightarrow O_+(1)$ dualities. These dualities are very reminiscent of the 
distinctions between different global properties of dualities with $so(N_c)$ gauge Lie algebras
in $4d$ \refs{\readinglines,\RazamatOPA} and in $3d$~\AharonyKMA.
 
After this digression, let us now return to the reduction of the duality appetizer to two dimensions.  Let us start from the duality above with
both sides of the duality being orbifolds, and then put the background terms on side (A) and
orbifold the ``quantum'' $\hat{\Z}_2$. The theory (A) then has a chiral field $X$  orbifolded by $\Z_2$
and by the quantum $\hat{\Z}_2$, with the dual side (B) being the model with fields $Z$ and $Y$ with superpotential $W=ZY^2$ and no orbifolds.
We add on side  (B) a field $\Omega$ coupled to the rest through a $W=Z Y^2+\Omega Y$ superpotential.
On side (A) we then have superpotential $W=\Omega \widetilde Y$ where $\widetilde Y$ is the operator dual to $Y$ on side (B), which we expect to be a certain twist operator. On side (B) the equations of motion imply that $\Omega=Y=0$ and the theory is just a free field $Z$ of charge $2$. On the other side we have a model of fields $\Omega$ and $X$ with the superpotential which assigns them both charge $1$, and the orbifold. This is the duality giving us the original duality found by reducing the appetizer above.  We expect that the $\Z_2$ orbifold can be derived from $3d$ by starting from the Weyl symmetry of $SU(2)$, and after performing the Hori-Vafa duality, this should map to the ``quantum'' $\hat{\Z}_2$ symmetry above, but we leave the details to future investigation.

The duality we get is similar to the one in $3d$ in
some respects. We have a free field $X$ dual to a gauge theory of a field $Z$ with the identification $X\sim Z^2$. The gauge group is 
discrete in $2d$ and $SU(2)$ in $3d$. 
 Note that this duality is rather non-trivial 
as it involves on one side of the duality a
superpotential  built from composite (twist) operators. Such superpotentials are reminiscent of, though
not directly related to, the monopole superpotentials one has to consider in $3d$ in order to make some of the dualities work~\AharonyGP. 

Finally, one can ask how we should understand physically the 
identity with the LG sector before performing a mirror symmetry transformation to the theory of chirals, \ie, the identity we get most directly from the reduction of the $3d$ duality.  On one side of this identity we have contributions which can be
interpreted as contributions from a chiral field and a dynamical twisted chiral field with twisted chiral superpotential. Again these two fields should be coupled since the symmetries under which they are charged and their charges are correlated. One way to couple them is by coupling the defect operator 
of the twisted chiral field, \ie, the winding modes of field $Y$ which are chiral, to the chiral field
through a superpotential. Such a coupling will not alter the sphere partition function.

\listrefs
\end